\documentclass[12pt,a4paper]{article}
\usepackage{amsfonts}
\usepackage{amssymb}
\usepackage{graphicx}
\usepackage{setspace}
\usepackage[sectionbib]{natbib}
\usepackage{bibunits}
\defaultbibliographystyle{chicago}
\defaultbibliography{fss}
\usepackage{amsmath}
\usepackage{amsthm}
\usepackage{palatino}
\usepackage{paralist}
\usepackage{subcaption}
\usepackage{xcolor}
\usepackage{multirow}
\usepackage{booktabs}
\usepackage{dsfont}
\usepackage{indentfirst}
\usepackage{enumerate}
\usepackage{longtable}
\usepackage{caption}
\usepackage{comment}
\usepackage{algpseudocode} 
\usepackage{tikz}
\usetikzlibrary{arrows}
\usetikzlibrary{positioning}
\usetikzlibrary{calc}
\usetikzlibrary{positioning,arrows.meta}
\newdimen\nodeDist
\graphicspath{ {Figures/} }

\usepackage[hyperindex,breaklinks]{hyperref}
\hypersetup{colorlinks=true,       
	linkcolor=red,       
	citecolor=blue,        
	filecolor=magenta,      
	urlcolor=cyan           
}  

\usepackage{algorithm,algpseudocode}
\usepackage{booktabs, longtable}


\newcommand{\SR}{\text{SR}}

\newcommand{\PC}{\text{PC}}

\newtheorem{theorem}{Theorem}

\newtheorem{lemma}{Lemma}

\newtheorem{assumption}{Assumption}

\usepackage[toc,page]{appendix}
\usepackage[figuresright]{rotating}
\usepackage{amsmath,float}

\usepackage[left = 1in, right = 1in, top = 1in, bottom = 1in]{geometry}
\usepackage{titlesec}
\titleformat*{\section}{\large\bfseries}
\titleformat*{\subsection}{\normalsize\bfseries}
\titleformat*{\subsubsection}{\normalsize\bfseries}
\titleformat*{\paragraph}{\normalsize\bfseries}

\titlespacing{\section}{0pt}{*1.5}{*1.5}
\titlespacing{\subsection}{0pt}{*1.2}{*1.2}
\titlespacing{\subsubsection}{0pt}{*1.2}{*1.2}
\titlespacing{\paragraph}{0pt}{*1.2}{*1.2}

\usepackage{pdflscape}
\begin{document}

\title{\large{
Selecting and Testing Asset Pricing Models: A Stepwise Approach}\thanks{\scriptsize{The paper was previously circulated as ``Anomaly or Risk Factor? A Stepwise Evaluation''.
The authors thank the department editor (Lukas Schmid), the associate editor, and three anonymous referees for their insightful suggestions.
The authors also thank Lilun Du, Damir Filipovic, Jingyu He, Fuwei Jiang, Christian Julliard, Bin Li, George Milunovich, Xiao Qiao, Raman Uppal, Dacheng Xiu, Zhixin Zhou, and participants at seminars and conferences hosted by City University of Hong Kong, Eastern China Normal University, Greater China Area Finance Conference (Xiamen University), Macquarie University, Wuhan University, Southern University of Science and Technology, and the 2024 Oxford-Man Institute Machine Learning in Quantitative Finance Conference for their valuable comments and discussions.
Feng (E-mail: \texttt{gavin.feng@cityu.edu.hk}) is at City University of Hong Kong; 
Wang (E-mail: \texttt{hansheng@gsm.pku.edu.cn}) is at Peking University; 
Lan (E-mail: \texttt{lanwei@swufe.edu.cn}) is at Southwestern University of Finance and Economics; and Zhang (E-mail: \texttt{junzhang2025@seu.edu.cn}) is at Southeast University.
}}
}

\author{ \small{Guanhao Feng} \and \small{Wei Lan} \and \small{Hansheng Wang} \and \small{Jun Zhang}}

\date{ \small{First draft: Apr. 2023; this draft: Jan. 2026 \\ 
\vspace{0.5cm}
Forthcoming at \textit{Management Science}}} 

\pagenumbering{gobble} 

\maketitle
\vspace{-1cm}

\begin{abstract}
\small
\noindent 
The asset pricing literature emphasizes factor models that minimize pricing errors but overlooks unselected candidate factors that could enhance the performance of test assets. 
This paper proposes a framework for factor model selection and testing by (i) selecting the optimal model that spans the joint efficient frontier of test assets and all candidate factors, and (ii) testing pricing performance on both test assets and unselected candidate factors. 
Our framework updates a baseline model (e.g., CAPM) sequentially by adding or removing factors based on asset pricing tests. 
Ensuring model selection consistency, our framework utilizes the asset pricing duality: minimizing cross-sectionally unexplained pricing errors aligns with maximizing the Sharpe ratio of the selected factor model.
Empirical evidence shows that workhorse factor models fail asset pricing tests, whereas our proposed 8-factor model is not rejected and exhibits robust out-of-sample performance.

\medskip


\end{abstract}

\newpage
\pagenumbering{arabic}
\setcounter{page}{1}

\begin{bibunit}

\section{Introduction}
Asset pricing research aims to develop factor models that can price cross-sectional returns, thereby ensuring zero cross-sectional pricing errors.
From the factor mean-variance efficiency perspective \citep{fama1996multifactor}, the goal is to identify a small set of traded factors $f_t$ that span the efficient frontier of all candidate factors $F_t$ and test assets $r_t$.\footnote{Except for the market factor, risk factors are typically long-short portfolios based on firm characteristics, capturing systematic undiversifiable risk. Test assets are characteristics-managed portfolios used to evaluate the model's cross-sectional pricing performance.}
According to \citet*{barillas2017alpha}, this involves comparing the maximal squared Sharpe ratio ($\SR^2$) of the mean-variance efficient portfolio, $\SR^2(f_t)$, with that of the joint efficient frontier of all assets, $\SR^2(F_t, r_t)$.
We define efficient models as those where $\SR^2$ values show no significant difference, indicating that the factor model has reached the joint efficient frontier.
This can also be evaluated in the spanning regression framework, where test asset returns $r_{i,t}$ are regressed on the factor model $f_t$ for each $i$, testing if all intercepts ("alphas") are jointly zero.
Building on this, \citet*{gibbons1989test} provide an asset pricing test framework to evaluate whether a factor model spans the efficient frontier jointly formed by factors and test assets.

Despite the proliferation of factors \citep{cochrane2011presidential}, traditional factor models \citep[e.g.,][]{fama1993common,fama2015five} fail asset pricing tests for standard test assets. This underscores the empirical challenge of identifying efficient factor models.\footnote{\citet*{bryzgalova2023forest} and \citet*{cong2025growing} emphasize that testing asset pricing models requires well-diversified test assets that approximate the efficient frontier spanned by individual stocks. Due to high-dimensional challenges, researchers can only identify locally efficient models based on the given test assets and all candidate factors. \citet*{chernov2025test} introduce a high-dimensional test framework to construct extensive test asset sets.}
Existing studies on factor selection \citep[e.g.,][]{feng2020taming, chib2024winners} focus on statistical fit or investment performance of traded factors, often neglecting their cross-sectional pricing performance in asset pricing tests.
To address these gaps,  we propose a framework for factor model selection and testing by (i) selecting the optimal model that spans the joint efficient frontier of test assets and all candidate factors, and (ii) testing pricing performance on both test assets and unselected candidate factors.

One can only identify the locally optimal model that spans the joint efficient frontier, because empirical datasets of test assets and factors may not fully span the ultimate efficient frontier constructed from individual stocks.
If the joint efficient frontier coincides with the entire asset universe, the globally optimal factor model provides a linear representation of the stochastic discount factor (SDF). Importantly, this model is unique and, by construction, incorporates all risk factors that define the SDF.

Our framework updates a baseline model sequentially by adding or removing factors based on asset pricing tests. 
The selected model, $f_t^{SEL}$, achieves the joint efficient frontier of test assets and all candidate factors, such that $\SR^2(f_t^{SEL}) = \SR^2(r_t, F_t)$. 
Our approach contrasts with traditional factor selection studies, which implicitly assume $\SR^2(r_t) = \SR^2(r_t, f_t^{UNS})$. This assumption overlooks the potential benefits of including unselected candidate factors $f_t^{UNS}$ as additional test assets.
Though unselected candidate factors may be anomalies rather than risk factors, they could enhance test asset performance, enabling more rigorous evaluation.

Our paper tackles two core issues by introducing a novel statistical approach for factor model selection. 
(i) If a factor model fails asset pricing tests, it suggests omitted factors \citep{he2023diagnostics} and may warrant expansion.
When a difficult-to-price factor shows significant alpha relative to the model, its inclusion can enhance the efficient frontier.
(ii) If a factor model is not rejected but overly large, can it be reduced while maintaining efficiency? Removing a factor with an insignificant alpha relative to the reduced model preserves the efficient frontier.
We propose a systematic framework for updating factor models, which combines stepwise factor selection with rigorous asset pricing tests. By stepwise updating models, it constructs an optimal factor model for test assets and all candidate factors.

When adding or removing factors stepwise, multiple model comparisons are required as each update generates alternative models. \cite{barillas2017alpha} suggest evaluating these comparisons using the model $\SR^2$. Testing the improvement in $\SR^2$ from adding a factor is equivalent to assessing the new factor's alpha via factor-spanning regression, underscoring the asset pricing duality in factor model evaluation.
Expanding the model to maximize the $\SR^2$ increment corresponds to reducing pricing errors, as measured by the GRS value \citep{gibbons1989test}. Conversely, removing a factor to minimize the $\SR^2$ decrease aligns with reducing the model while limiting the rise in GRS value. Comparing factor models using $\SR^2$ provides a systematic framework for stepwise factor inclusion or exclusion.
We conduct an asset pricing test to assess the model's mean-variance efficiency and determine the optimal stopping point in the stepwise procedure, guiding model expansion or reduction.

\subsection{Our Methodology}
Our stepwise evaluation framework consists of forward stepwise evaluation (FSE) and backward stepwise evaluation (BSE). 
The FSE step expands a baseline model by sequentially adding candidate factors within the factor spanning regression framework.
Those excluded candidate factors are treated as left-hand side (LHS) test assets. If the factor model is rejected, some candidate factors may exhibit nonzero alphas. 
FSE sequentially moves these difficult-to-price LHS factors to the right-hand side (RHS) of the factor model.
FSE prioritizes adding factors that yield the largest improvement in $\SR^2$ and the greatest reduction in the GRS statistic. Intuitively, a LHS factor that is difficult to price but contributes to improving $\SR^2$ may be included.

We conduct asset pricing tests to determine the stopping point in the forward stepwise procedure. The statistical power of the standard GRS test diminishes as the number of LHS test assets increases. 
To address this limitation, we employ the high-dimensional alpha (HDA) test proposed by \cite{pesaran2023testing}. 
If the HDA test is rejected, the FSE process may still incorporate a difficult-to-price LHS factor into the RHS model. This stepwise inclusion continues until the HDA test fails to reject, aiming to maximize the $\SR^2$ of the RHS factor model while minimizing the GRS statistic for the LHS test assets and unselected candidate factors.

To reduce an efficient but large model, the BSE step is designed to sequentially move potentially redundant factors from the RHS selected model to the LHS. 
BSE removes factors that contribute minimally to the $\SR^2$ while causing only a small increase in the GRS value, ensuring that the reduced model remains efficient.
Intuitively, factors that are easily priced and add minimal marginal contribution to $\SR^2$ are excluded to reduce the model.
Again, the HDA test serves as a statistical tool to evaluate asset pricing models and determine optimal model size.
Additionally, BSE serves as an exit mechanism for the FSE step, mitigating its tendency toward overfitting by systematically removing redundant factors to maintain model efficiency.
In summary, after deriving an efficient factor model via FSE, the BSE process refines the model by removing redundant factors, yielding an optimal model.

\paragraph{Contribution and comparison with existing methods.}
The development of our method is driven intrinsically by the challenges inherent in asset pricing.
Adapted from the seminal work of forward stepwise model selection \citep{wang2009forward}, our FSE method optimally searches for factors that maximize the $\SR^2$ sequentially. 
It stops when the model is efficient and not rejected by the asset pricing test, serving as an economically guided stop sign.
Our new BSE method builds upon forward regression by incorporating an exit mechanism, providing new empirical insights for removing redundant factors.
We prove the consistency of variable screening and selection: the FSE method includes all risk factors, and the BSE method excludes redundant factors.

In the recent literature, Bayesian approaches \citep[e.g.,][]{chib2020factors, chib2025beta} are limited to a small number of factors unless principal component analysis (PCA) is incorporated \citep[e.g.,][]{chib2024winners}.
Our framework complements latent factor models by allowing inclusion or exclusion of observable factors that are documented to represent specific risks, while offering an economic rationale for stepwise changes.\footnote{As suggested by an anonymous associate editor, our stepwise evaluation method extends to latent factors. Standard methods select latent factors based on eigenvalues or variance contribution. In contrast, our approach ranks latent factors using $\SR^2$, which may yield a different order. The number of selected factors is determined via asset pricing tests.}
Furthermore, most methods, including shrinkage approaches \citep[e.g.,][]{bryzgalova2022bayesian, cui2023time}, depend on separate test assets, which are sensitive to model evaluation and selection \citep{lewellen2010skeptical}. 
Our framework employs unselected candidate factors to improve the (optional) test asset set, building on prior studies for more rigorous evaluation.

Our work is related to \citet*{kozak2020shrinking} and \citet*{bryzgalova2023forest}, which examine regularized portfolio optimization under mean-variance utility.
Portfolio regularization helps in asset or factor selection, providing a practical method to achieve the efficient frontier.
Our goal is to identify an optimal model using test assets and all candidate factors for the joint efficient frontier, while ensuring pricing across the cross section of test assets and unselected candidate factors. 
Leveraging the duality between asset pricing and the efficient frontier, our framework delivers an optimal factor model.
Within our framework, factors excluded during the FSE step add minimal incremental pricing performance to the model.
In the BSE step, redundant factors are removed based on asset pricing tests, providing an alternative to statistical regularization and reducing the risk of overfitting in large models.

Our work is also related to \cite{harvey2021lucky} (hereafter "HL2021"), who use the forward stepwise method to select factors for pricing test assets. 
First, HL2021 selects factors based on test assets and removes unselected candidate factors. Our model utilizes the unselected candidate factors as test assets, demonstrating a different modeling design.
Second, HL2021 uses a bootstrap approach for multiple hypothesis testing, while we propose a stepwise framework to identify the parsimonious model.
Third, we establish theoretical foundations for forward stepwise evaluation in asset pricing and introduce backward stepwise evaluation with theoretical guarantees. This framework enhances the special case of HL2021.

\subsection{Empirical Findings}
We analyze monthly U.S. equity returns from 1973 to 2021, using seven different factor models and a comprehensive list of 97 factors. 
First, all these models are inefficient to price the factor zoo jointly and are rejected by the asset pricing test.
We aim to build better economic models by expanding (and reducing) a baseline model.
FSE adds at least five additional factors to these inefficient baseline models, indicating that more factors are needed. 
For example, the expanded efficient model based on market factor (\texttt{MKT}) contains eight factors, adding factors in order:
expected growth (\texttt{REG}), 
post-earnings-announcement-drift (\texttt{PEAD}), 
refined value factor (\texttt{HMLM}),
short-term reversal (\texttt{STR}),
industry lead-lag effect in prior returns (\texttt{ILR}), 
size (\texttt{SMB}),
earnings predictability (\texttt{EPRD}).
The early FSE-added factors have a greater impact on improving the model's performance.
For instance, when \texttt{REG} is added to the baseline model CAPM, the annualized Sharpe ratio increases from 0.47 to 1.76.

Second, BSE removes factors from those expanded efficiency models, suggesting the existence of redundant factors.
For the final reduced models on different baseline models, we find the same eight factors (\texttt{MKT}, \texttt{REG}, \texttt{PEAD}, \texttt{HMLM}, \texttt{STR}, \texttt{ILR}, \texttt{SMB}, and \texttt{EPRD}) that explain the unselected candidate factors, not rejected by the asset pricing test.\footnote{The study by \cite{barillas2018comparing} evaluates 10 factors, 9 of which are analyzed here. They identify a 6-factor model (\texttt{MKT}, investment factor (\texttt{IA}), profitability factor (\texttt{ROE}), \texttt{SMB}, \texttt{HML}, and momentum factor (\texttt{UMD})). 
\cite{avramov2023integrating} investigate 14 factors, 12 of which overlap with our analysis. They propose a 3-factor model (\texttt{PEAD}, quality-minus-junk (\texttt{QMJ}), and intermediary capital risk (\texttt{IMD})).}
Furthermore, BSE removes some baseline factors and late FSE-added factors, confirming that early FSE-added factors are more important.
Our 8-factor model, which provides a sparse model and explains the unselected factor zoo jointly, consistently excels in asset pricing and investment analysis.
The proposed model outperforms benchmark models, achieving an annualized Sharpe ratio of 2.73 over five decades and 1.53 in the recent 16-year out-of-sample (OOS) period.

Finally, our framework can be adapted to evaluate individual factors and determine if a factor can withstand the inclusion and exclusion process.
We observe extremely high selection rates for our eight factors (with \texttt{REG} and \texttt{PEAD} both selected at 100\%), and low selection rates for the remaining candidate factors.
To address potential overfitting concerns, we conduct OOS model comparisons, which underscore the robust performance of the 8-factor model.
For robustness, we incorporate transaction costs into factor returns analysis, identifying annual rebalancing factors as the most resilient.

\subsection{Other Related Literature}
This paper adds to the growing literature on the factor zoo.
Recent studies include replicating or testing candidate factors \citep[e.g.,][]{harvey2016and, green2017characteristics, chordia2020anomalies, jensen2023there}.
Latent factor modeling includes various PCA methods \citep[e.g.,][]{kozak2018interpreting,kelly2019characteristics,lettau2020estimating,lettau2020factors,kim2021arbitrage,huang2022scaled} and deep neural networks \citep[e.g.,][]{gu2021autoencoder,chen2024deep,feng2022deep}. 
We propose a customized approach to optimize a model based on observable factors, distinct from methods targeting other objectives.

We address limitations in factor selection and evaluation methods, particularly their sensitivity to the choice of test assets. 
Many studies compare factor models by different criteria \citep[e.g.,][]{ahmed2019best,hou2019factors}, whereas others compare models by considering transaction costs \citep[e.g.,][]{detzel2023model,li2023comparing}.
\cite{barillas2017alpha} and \cite{barillas2020model} propose the $\SR^2$ as a summary statistic of model quality without using test assets, which also appear in \cite{barillas2018comparing} and \cite{fama2018choosing}.
We introduced a framework that integrates model comparison and testing into the stepwise evaluation and selection process.

Our study contributes to the growing literature on testing high-dimensional factor models. 
Recent advancements have introduced asset pricing tests tailored for high-dimensional settings in the large-$N$, large-$T$ framework \citep[e.g.,][]{fan2015power,feng2022high,zhang2025testing, chernov2025test, zhang2025testingOOS}.
In particular, \cite{pesaran2023testing} enhance the GRS test by proposing a quadratic test statistic for alphas, accommodating a large number of test assets.
Our paper utilizes their HDA test for the stepwise factor selection stop sign, which helps determine the model size.

Finally, this paper contributes to the statistical literature on variable selection by examining SURE independence screening \citep{fan2008sure}. 
Forward regression, a greedy algorithm, has well-established theoretical properties \citep{barron2008approximation, chen2008extended}.
\cite{wang2009forward} develops a forward regression approach for feature screening in ultra-high-dimensional linear models, demonstrating its SURE screening property.
Our approach mitigates the limitations of forward regression by integrating a backward stepwise elimination stage within a financial application.

The structure of the paper is as follows. Section \ref{sec:methodology} details the methodology employed in the study. Section \ref{sec:data} provides a comprehensive description of the dataset. Section \ref{sec:empirical} develops efficient factor models, while Section \ref{sec:Reduced_result} introduces the optimal factor model. Additional applications and robustness checks are discussed in Section \ref{sec:empirical_robust}. Finally, Section \ref{sec:conclusion} concludes the paper. Technical details, simulation evidence, and supplementary empirical results are provided in the Appendix.

\section{Methodology: Stepwise Evaluation} \label{sec:methodology}
\subsection{Efficient Factor Models and the Optimal Model} \label{sec:known_model}
Before introducing the method, we define efficient factor models and the optimal model.
Let $N$ be the total number of candidate factors, $f_{i,t}$ be the return of the $i$-th factor observed at time $t$ for $1 \leq i \leq N$, $1\leq t\leq T$, and $ \textbf{F}_t = (f_{1,t}, \cdots, f_{N,t})^\top  \in \mathbb{R}^{N}$ be the return vector at time $t$. 
We use a generic notation $\mathcal{M} \subset \mathcal{F}$ to denote an arbitrary candidate model, where $\mathcal{F} = \{ 1, \cdots, N\} $ is the full model containing all candidate factors.
Define $\textbf{F}_{(\mathcal{M}) t} = \left\lbrace f_{i,t}: i \in \mathcal{M}\right\rbrace \in \mathbb{R}^{|\mathcal{M}|}$ as the return vector of the model $\mathcal{M}$ at time $t$,
where $|\mathcal{M}|$ denotes the number of factors contained in model $\mathcal{M}$ (i.e., the model size).
Similarly, let
$\textbf{F}_{(\widetilde{\mathcal{M}}) t} = \left\lbrace f_{i,t}: i \in \widetilde{\mathcal{M}}\right\rbrace \in \mathbb{R}^{N-|\mathcal{M}|}$,
$\widetilde{\mathcal{M}} = \{\mathcal{F} \backslash \mathcal{M} \} $ is the complement of $\mathcal{M}$.
For any model $\mathcal{M}$, one can run a factor-spanning regression that regresses $\textbf{F}_{(\widetilde{\mathcal{M}}) t}$ on $\textbf{F}_{(\mathcal{M}) t}$, where the unselected candidate factors $\textbf{F}_{(\widetilde{\mathcal{M}}) t}$ serve as the LHS test asset returns. 
We admit the following spanning regression form:
\vspace{-0.2cm}
\begin{equation}\label{eq: true_model}
\underbrace{\textbf{F}_{(\widetilde{\mathcal{M}})t}}_{\mbox{LHS}}  = \pmb{\alpha}_{(\widetilde{\mathcal{M}}) }  +  \pmb{\beta}_{(\widetilde{\mathcal{M}}) }^\top \underbrace{\textbf{F}_{(\mathcal{M})t} }_{\mbox{RHS}} + \pmb{\varepsilon}_{(\widetilde{\mathcal{M}})t},
\end{equation}
where
$\pmb{\alpha}_{(\widetilde{\mathcal{M}}) }$ is the vector of alphas 
and $\pmb{\beta}_{ ( \widetilde{\mathcal{M}}) }$ is the matrix of betas.
Following \cite{barillas2017alpha}, we define efficient factor models as those where $\SR^2$ values show no significant difference, indicating the model reaches the ultimate mean-variance efficient frontier.
Accordingly, if model $\mathcal{M}$ is efficient within the spanning regression framework (\ref{eq: true_model}), its intercept terms $\pmb{\alpha}_{(\widetilde{\mathcal{M}})}$ are jointly zero, indicating no cross-sectional pricing errors.
Otherwise, 
$\pmb{\alpha}_{(\widetilde{\mathcal{M}}) }$ should be different from zero.
While multiple efficient models may exist, the optimal model $\mathcal{T}$ is defined as the one that is the smallest.
Importantly, the optimal model $\mathcal{T}$ is uniquely defined and integrates all risk factors underlying the SDF.
We aim to develop efficient models and derive the optimal model from a baseline by analyzing alpha's properties.

\subsection{Stepwise Evaluation Framework} \label{sec:2.2}

The stepwise evaluation involves two key components: selecting factors to add or remove from the model and determining the stopping point. Factor selection criteria guide the inclusion or exclusion of factors, while stopping criteria assess the fitness of asset pricing models. 
At each step, models are evaluated, and the best model is selected by modifying a single factor. The process iterates until the model achieves efficiency.
Detailed methods and algorithms are provided in Appendix \ref{sec:method_detail}.

\subsubsection{Factor Selection Criterion: Maximal $\SR^2$} \label{sec:2.2.1}
\paragraph{Two non-nested model comparisons.}
\cite{gibbons1989test} quantify the gains from adding test assets to a set of factors using the maximal squared Sharpe ratio increase. 
In model (\ref{eq: true_model}), the improvement in the $\SR^2$ from adding unselected candidate factors $\bf{F}_{(\widetilde{\mathcal{M}})} $ to the investment universe $\bf{F}_{(\mathcal{M})} $ is a quadratic form in the test-asset alphas (\citealp{barillas2018comparing}):
\vspace{-0.2cm}
\begin{equation*} \label{eq2.5}
\pmb{\alpha}_{(\widetilde{\mathcal{M}}) }^\top  \pmb{\Sigma}_{\pmb{\varepsilon}}^{-1} \pmb{\alpha}_{(\widetilde{\mathcal{M}})} = {\SR^2\{ \bf{F}_{(\widetilde{\mathcal{M}})},\bf{F}_{(\mathcal{M})} \} } - \SR^2 \{\bf{F}_{(\mathcal{M})} \} = {\SR^2\{\bf{F}\} } - \SR^2 \{\bf{F}_{(\mathcal{M})}\},
\vspace{-0.2cm}
\end{equation*}
where $\pmb{\Sigma}_{\pmb{\varepsilon}}$ be the covariance matrix of residuals. $\SR^2\{x\} = \pmb{\mu}_x ^\top \pmb{\Sigma}_x ^{-1}  \pmb{\mu}_x$ is the maximal squared Sharpe ratio delivered by the assets in vector $x$, $\pmb{\mu}_x$ and $\pmb{\Sigma}_x$ are the means and invertible covariance matrices of the assets in vector $x$. 
\cite{barillas2017alpha} argue that when comparing two non-nested factor models, the superior model should be able to span the investment opportunities provided by the test assets and the factors in the other model.
In particular, let us consider two models with factors $\mathcal{M}_1$ and $\mathcal{M}_2$, the model size $|\mathcal{M}_1| = |\mathcal{M}_2|$. Model $\mathcal{M}_1$ is better than model $\mathcal{M}_2$ if\footnote{
If we consider additional test assets $\textbf{R}$, the conclusion remains the same: $ \SR^2 \{\textbf{R}, \mathbf{F}_{(\widetilde{\mathcal{M}}_1)},  \mathbf{F}_{(\mathcal{M}_1)} \} - \SR^2 \{ \mathbf{F}_{(\mathcal{M}_1)} \} <  \SR^2 \{\textbf{R}, \mathbf{F}_{(\widetilde{\mathcal{M}}_2)},  \mathbf{F}_{(\mathcal{M}_2)} \}-  \SR^2 \{ \mathbf{F}_{(\mathcal{M}_2)} \}$. This inequality is equivalent to $\SR^2 \{ \mathbf{F}_{(\mathcal{M}_1)} \} > \SR^2 \{ \mathbf{F}_{(\mathcal{M}_2)} \}$ and the test assets $\textbf{R}$ are irrelevant for model comparison.
}
\vspace{-0.2cm}
\begin{equation} \label{eq:BS_SR}
    \SR^2 \{\mathbf{F}_{(\widetilde{\mathcal{M}}_1)},  \mathbf{F}_{(\mathcal{M}_1)} \} - \SR^2 \{ \mathbf{F}_{(\mathcal{M}_1)} \} <  \SR^2 \{\mathbf{F}_{(\widetilde{\mathcal{M}}_2)},  \mathbf{F}_{(\mathcal{M}_2)} \}-  \SR^2 \{ \mathbf{F}_{(\mathcal{M}_2)} \}.
    \vspace{-0.2cm}
\end{equation}

In particular, they explain the two sides of inequality (\ref{eq:BS_SR}) measure the misspecification of models $\mathcal{M}_1$ and $\mathcal{M}_2$, and thus, model $\mathcal{M}_1$ is considered better (less misspecified) than model $\mathcal{M}_2$ because an investor with access to the factors in model $\mathcal{M}_1$ obtains a lower $\SR^2$ improvement by having access to the unselected candidate factors and the factors in the other model than an investor with access to the factors in model $\mathcal{M}_2$. 
This inequality is equivalent to $\SR^2 \{ \mathbf{F}_{(\mathcal{M}_1)} \} > \SR^2 \{ \mathbf{F}_{(\mathcal{M}_2)} \}$, 
and thus, \cite{barillas2017alpha} demonstrate that test assets are irrelevant for model comparison, and comparing models in terms of $\SR^2$, which assesses the ability of factor models to span the investment opportunity set, is an efficient approach.

Similarly, we can write the GRS value of models $\mathcal{M}_1$ and $\mathcal{M}_2$ as:
\vspace{-0.2cm}
\begin{equation} \label{eq: GSR}
\begin{split}
\mbox{GRS}(\mathcal{M}_1)  &= \frac{T-N}{N - {|\mathcal{M}_1| }} \Big( \frac{1 + \SR^2\{\bf{F}\} } { 1+ \SR^2 \{\bf{F}_{(\mathcal{M}_1)}\} } -1 \Big), \\
\mbox{GRS}(\mathcal{M}_2) 		&= \frac{T-N}{N - {|\mathcal{M}_2| }} \Big( \frac{1 + \SR^2\{\bf{F}\} } { 1+ \SR^2 \{\bf{F}_{(\mathcal{M}_2)}\} } -1 \Big).
\end{split}
\vspace{-0.2cm}
\end{equation}

Under our case, a small $\mbox{GRS}(\mathcal{M})$ value corresponds to a large $ \SR^2\{\bf{F}_{(\mathcal{M})} \}$, which means reducing the mispricing of the factor model is equivalent to increasing the $ \SR^2\{\bf{F}_{(\mathcal{M})} \}$ of the factor model.
Eq. (\ref{eq: GSR}) also shows model $\mathcal{M}_1$ is better than model $\mathcal{M}_2$ if $\mbox{GRS}(\mathcal{M}_1)  < \mbox{GRS}(\mathcal{M}_2)$.

\paragraph{Multiple non-nested model comparisons.}
In each step of the FSE and BSE processes, we consider $J$ possible non-nested models $\mathcal{M}_j $ that need to be compared simultaneously for each $j = 1, \cdots, J$.\footnote{The value of $J$ varies at each step. For instance, in the first step of FSE, $ J = N - |\mathcal{M}_b|$, where $\mathcal{M}_b$ represents the baseline model.} We then calculate the maximal squared Sharpe ratio, $\SR^2 \{\bf{F}_{(\mathcal{M}_j)} \}$,
and the GRS value: 
\vspace{-0.2cm}
 \begin{equation} \label{eq:GRS_j}
    \mbox{GRS}(\mathcal{M}_j) =  \frac{T-N }{{N - |\mathcal{M}_j|} } \Big( \frac{1 + \mbox{SR}^2\{ \textbf{F}  \}  } { 1+ \SR^2 \{\bf{F}_{(\mathcal{M}_j)} \}  }   -1 \Big). 
    \vspace{-0.2cm}
\end{equation}

For each step of FSE and BSE, we can minimize $\mbox{GRS}_j$ or maximize $\mbox{SR}^2_j$ to identify the best and worst factors, respectively.
Accordingly, we can obtain the expanded or reduced model for the current step (adding the best factor for the FSE process or removing the worst factor for the BSE process).
The GRS value is used solely as a model comparison criterion and does not account for the distributional properties of the GRS test statistic.
By leveraging the asset pricing duality between the minimal GRS value for pricing unselected candidate factors and the maximal $\SR^2$ of selected factors, our approach sequentially adds or removes factors while ensuring consistency in selection for all risk factors. This method provides a potential solution for comparing investment results and asset pricing performance between multiple non-nested models.

\subsubsection{Model Stop Criterion: High-dimensional Alpha Test} \label{sec:2.2.2}
We then assess if the expanded or reduced model is efficient and determine when to stop the stepwise inclusion or exclusion process.
In the FSE and BSE processes, 
for any model $\mathcal{M}$,
if $\mathcal{M}=\mathcal{T}$, that is, the model is correctly specified, 
so no intercept exists as displayed in Eq. (\ref{eq: true_model}). If $\mathcal{M}\supset \mathcal{T}$,
we expect that the resulting intercept $\pmb{\alpha}_{(\widetilde{\mathcal{M}}) }$ is close to zero. 
Otherwise, when $\mathcal{M} \not\supset \mathcal{T}$,
$\pmb{\alpha}_{(\widetilde{\mathcal{M}}) }$ should be different from zero. 
To evaluate the performance of the model $\mathcal{M}$, we are then interested in testing the following statistical hypotheses:\footnote{
For further evaluating the model's cross-sectional pricing performance, we incorporate additional test assets ($\textbf{R}_t$).
Let $\widetilde{\pmb{\alpha}}_{(\widetilde{\mathcal{M}}) } = (\pmb{\alpha}_{( \widetilde{\mathcal{M}}) }^\top,  \widetilde{\pmb{\alpha}}^\top)^\top $, where $\widetilde{\pmb{\alpha}}$ be the vector of intercepts from spanning regressions that $\textbf{R}_t$ on $\textbf{F}_{(\mathcal{M})t}$.
This corresponds to testing the null hypothesis: $\mathbb{H}_0 : \widetilde{\pmb{\alpha}}_{(\widetilde{\mathcal{M}}) } = \mathbf{0}$.
}
\vspace{-0.2cm}
\begin{equation} \label{eq: hypotheses}
\mathbb{H}_0 : \pmb{\alpha}_{(\widetilde{\mathcal{M}}) } = \mathbf{0} \mbox{~~vs.~~} \mathbb{H}_1 : \pmb{\alpha}_{(\widetilde{\mathcal{M}}) } \neq \mathbf{0}.
\vspace{-0.2cm}
\end{equation}

To test the hypotheses in Eq. (\ref{eq: hypotheses}), we can use the GRS statistic in Eq. (\ref{eq:GRS_j}). GRS is distributed exactly as a non-central $F$ distribution with degrees of freedom $(N - |\mathcal{M}|)$ and $( T- N)$.\footnote{An important limitation of this result is that it requires that $T > (N - |\mathcal{M}|)$ in our case.
Even when $(N - |\mathcal{M}|) < T$ and $N - |\mathcal{M}|$ is slightly larger, the power of the GRS test may be compromised because it employs an unrestricted sample covariance matrix, which is known to perform badly even for moderate $N - |\mathcal{M}|$; see, for example, \cite{fan2011high}, \cite{fan2015power}, and \cite{zhang2025testingOOS}.} 
The GRS test statistic is commonly used in empirical asset pricing, but it often leads to rejection, limiting its further application \citep{detzel2023model}.
For this reason, this paper introduces a new high-dimensional test to solve this challenge. Recently, \cite{pesaran2023testing} develop a simple quadratic test statistic that ignores off-diagonal elements of the covariance matrix to test the null hypothesis in Eq. (\ref{eq: hypotheses}), defined as the high-dimensional alpha (HDA) test hereafter:{\footnote{We exclude finite-sample adjustments from their test statistic for simplicity and clarity.}}
\vspace{-0.2cm}
\begin{equation} \label{eq: FSE}
\mbox{HDA}(\mathcal{M}) = \frac{T \widehat{\pmb{\alpha}}_{(\widetilde{\mathcal{M}}) }^\top  \mbox{Diag}(\widehat{\pmb{\Sigma}}_{\pmb{\varepsilon}})^{-1}\widehat{\pmb{\alpha}}_{(\widetilde{\mathcal{M}}) } \{{1+ \SR^2 \{\mathbf{F}_{(\mathcal{M})}\} } \}^{-1} - N_1 }{ \sqrt{2N_1 \{ 1+(N_1-1)\widehat{\rho}^2_{NT}\}}} ,
\vspace{-0.2cm}
\end{equation}
where $\widehat{\pmb{\Sigma}}_{\pmb{\varepsilon}}$ is the sample covariance matrix estimator and
$N_1 = N - |\mathcal{M}|$, 
$\widehat{\rho}^2_{N T}$ is a correction estimator term related to the sparsity of $\pmb{\Sigma}_{\pmb{\varepsilon}}$.
Under the null hypothesis, \cite{pesaran2023testing} demonstrate $\mbox{HDA} \overset{d}{\rightarrow} N(0,1 )$ with certain conditions.
Consequently, for any given significance level $\lambda$, we can reject the null hypothesis if  $\mbox{HDA}>z_{ \lambda}$, where $z_{\lambda}$ denotes the  upper $\lambda$-th quantile 
of the standard normal distribution.
The significance level choices for the asset pricing test can be seen as tuning parameters. We have tested the robustness of our results at different significance levels (1\%, 5\%, and 10\%). Our empirical findings remain consistent across all significance levels, leading us to adopt the widely accepted 5\% level in the main results.

\subsection{Baseline Model and Stepwise Evaluation} \label{sec:unknown_model}
Our framework constructs efficient models and identifies the optimal specification from various baselines.
We start with an arbitrary baseline model $\mathcal{M}_b$, such as the FF3 model, and then we illustrate our idea in two scenarios.
When $ \mathcal{M}_b \supset \mathcal{T}$ and is not rejected by the asset pricing test. Then, $\mathcal{M}_b$ is considered an efficient model, and we can use the BSE directly to obtain the optimal model.
When $\mathcal{M}_b \not\supset \mathcal{T}$, we use the FSE to get an efficient model, then the BSE can be used to obtain the optimal model.
Appendix \ref{sec:SE_Illustration} provides illustrative examples of the entire process of our FSE and BSE methods.

For any inefficient baseline model $\mathcal{M}_b \not\supset 
 \mathcal{T} $, 
we employ the FSE process in Appendix \ref{sec:FSE_detail} to add factors into the baseline model $\mathcal{M}_b$ to make it efficient, and then, $\mathcal{M}_b^F = \{ \mathcal{M}_b,  \mathcal{M}_F \}$, where $\mathcal{M}_b^F$ refers to the expanded efficient model (the final model of FSE process) constructed based on the baseline model $\mathcal{M}_b$, and  $\mathcal{M}_F$ represents the added factors.
The number of $\mathcal{M}_F$ is determined by the HDA test.
Moving the candidate factor from the LHS to the RHS helps explain the cross section of expected returns measured by GRS, so the value should decline.

Note that the theoretical requirement $\mathcal{T} = \mathcal{M}_b^F$ under the FSE is impractical, as this condition is not assured even in fixed-dimension settings. However, ensuring $\mathcal{T} \subset \mathcal{M}_b^F$ is feasible. Failure to do so risks omitting at least one critical factor in $\mathcal{M}_b^F$.
In Appendix \ref{sec:FSE_detail}, Theorem \ref{theorem1} establishes that the FSE is screening consistent, ensuring the expanded model detects all risk factors with probability approaching one, i.e., $P(\mathcal{T} \subset \mathcal{M}_b^{F}) \to 1$. The Appendix \ref{sec:proof_th1} provides the proof of this theorem.

For any efficient baseline model $ \mathcal{M}_b^F \supset \mathcal{T}$, 
we employ the BSE process in Appendix \ref{sec:backward} to remove redundant factors to obtain the optimal model, and then, $\mathcal{M}_b^{F+B} = \{ \mathcal{M}_b^{F} \backslash  \mathcal{M}_B \}$, where $\mathcal{M}_b^{F+B}$ refers to the FSE/BSE reduced models (the final model of BSE process) constructed based on the efficient model $\mathcal{M}_b^{F}$ and $\mathcal{M}_B$ represents the removed factors.
The number of $\mathcal{M}_B$ is determined by the HDA test.

Theorem \ref{theorem2} in the Appendix \ref{sec:backward} states that the BSE can consistently select all risk factors, demonstrating the selection consistency property of the proposed BSE. The probability that the optimal model equals the FSE/BSE reduced model tends to be one; that is, $P(\mathcal{T} = \mathcal{M}_b^{F+B}) \to 1$.
The Appendix \ref{sec:proof_th2} contains the proof of Theorem \ref{theorem2}. 

\subsection{Toy Model for Illustrations} 

We present a toy model to highlight key aspects of the FSE approach (similarly applicable to BSE). The model: (i) implements the forward stepwise evaluation method and (ii) contrasts our approach with existing methods, emphasizing differing assumptions.

Starting from a single factor baseline model $\textbf{F}_{(\mathcal{M}^{*}_0)} = \{ f_{1,t} \}$ (e.g., the market factor). 
In the 1-st forward iteration, we need to expand this baseline model by adding another factor $f_{i,t}$ and denote the expanded 2-factor model as $\textbf{F}_{(\mathcal{M}_i)} = \{f_{1,t}, f_{i,t}\}$ for each $i = 2,\cdots, N$. We consider the following $N-1$ non-nested factor model:
\begin{equation} \label{eq: FSE_model1}
\begin{array}{lll}
& \textbf{F}_{(\mathcal{M}_2)} = \{f_{1,t}, f_{2,t}\}: \quad & \textbf{F}_{(\widetilde{\mathcal{M}}_2)t}  = \pmb{\alpha}_{(\widetilde{\mathcal{M}}_2) }  +  \pmb{\beta}_{(\widetilde{\mathcal{M}}_2) }^\top \textbf{F}_{(\mathcal{M}_2)t}  + \pmb{\varepsilon}_{(\widetilde{\mathcal{M}}_2)t} \\
&    &\vdots \\
& \textbf{F}_{(\mathcal{M}_i)} = \{f_{1,t}, f_{i,t}\}:  \quad   & \textbf{F}_{(\widetilde{\mathcal{M}}_i)t}  = \pmb{\alpha}_{(\widetilde{\mathcal{M}}_i) }  +  \pmb{\beta}_{(\widetilde{\mathcal{M}}_i) }^\top \textbf{F}_{(\mathcal{M}_i)t}  + \pmb{\varepsilon}_{(\widetilde{\mathcal{M}}_i)t} \\
&   &\vdots \\
&  \textbf{F}_{(\mathcal{M}_N)} = \{f_{1,t}, f_{N,t}\}: \quad     & \textbf{F}_{(\widetilde{\mathcal{M}}_N)t}  = \pmb{\alpha}_{(\widetilde{\mathcal{M}}_N) }  +  \pmb{\beta}_{(\widetilde{\mathcal{M}}_N) }^\top \textbf{F}_{(\mathcal{M}_N)t}  + \pmb{\varepsilon}_{(\widetilde{\mathcal{M}}_N)t} \\
\end{array}
\end{equation}

As introduced in Section \ref{sec:2.2}, stepwise evaluation involves two key steps under Eq. (\ref{eq: FSE_model1}): (I) selecting factors to include and (II) determining when to stop.

\textbf{(I) Factor Selection:} We use the maximum squared Sharpe ratio of the expanded 2-factor model (${\SR}^2\{f_{1,t}, f_{i,t}\}$) to decide which factor to select. We see that the best factor added is, $ \widehat{a}_1 = \max_{i \in \{2,\cdots, N \}} {\SR}^2\{{f_{1,t}}, f_{i,t}\}$, resulting in the selected model $\mathcal{M}^{*}_1 = \{f_{ \widehat{a}_1 ,t}, f_{1,t}\}$, for example.

To investigate why this factor is added,
we construct the following factor-spanning regression that regression $f_{i,t}$ on $f_{1,t}$ for each $i = 2, \cdots, N$, 
\begin{equation} \label{eq: FSE_model1_1}
    f_{i,t} = \alpha_i + \beta_i f_{1,t} + \varepsilon_{i,t},
\end{equation}
where the intercept term $\alpha_i$ represents the pricing error associated with the $i$-th candidate factors, $\beta_i $ contains the regression coefficient,
$\varepsilon_{it} $ denotes the idiosyncratic error term.
 We assume that $\varepsilon_{i,t}$ is an independent noise term satisfying  
$
\mathbb{E}[\varepsilon_{i,t}]=0, \operatorname{Var}(\varepsilon_{i,t})=\sigma_{\varepsilon_i}^2,
$ and $\varepsilon_{i,t}$ is independent of $f_{1,t}$.
\cite{gibbons1989test} show that $\alpha_i^2/ \sigma_{\varepsilon_i}^2$ is the difference between 
the maximal squared Sharpe ratio one can construct from $f_{1,t}$ and $f_{i,t}$ together and the maximal for $f_{1,t}$ alone,
\[
 \SR^2\{\alpha_i\} = \alpha_i^2/ \sigma_{\varepsilon_i}^2 = 
{\SR}^2\{f_{i,t},f_{1,t}\}  -\SR^2\{f_{1,t}\}.
\]

If $f_{i,t}$ is uncorrelated with $f_{1,t}$ (i.e., $\beta_{i}=0$) under Eq. (\ref{eq: FSE_model1_1}).\footnote{
Latent factors generated by PCA are uncorrelated. For example, following \citet*{kozak2018interpreting} and \cite{lettau2020factors}, the eigendecomposition of the factor covariance matrix 
$ \pmb{\Sigma} = \textbf{Q} \textbf{D} \textbf{Q}^\top, \quad \text{where} \quad \textbf{D} = \text{diag}(d_1, d_2, \ldots, d_N), $
yields eigenvectors $\textbf{Q}$ and eigenvalues $\textbf{D}$, ordered in decreasing magnitude. PCA factors are constructed as:
$ \textbf{PC}_t = \textbf{Q}^\top \textbf{F}_t, $
where $ \textbf{PC}_t = ({\PC}_{1,t}, \ldots, {\PC}_{N,t})^\top $. The maximum Sharpe ratio among PCA factors is given by:
$ \max_{i \in \{1, \ldots, N\}} \SR^2\{\PC_{i,t}\}. $
These latent factors can be incorporated into the candidate set for factor selection.
}
We then obtain $\mathbb{E}(f_{i,t}) =  \alpha_i$ and $\mbox{Var}(f_{i,t}) = \sigma_{\varepsilon_i}^2$. Thus, we have,  $\SR^2\{\alpha_i\}= \SR^2\{f_{i,t}\}$.
In this case, selecting the factor that maximizes ${\SR}^2\{f_{i,t},f_{1,t}\}$ is equivalent to maximizing the marginal single-factor squared Sharpe ratio ${\SR}^2\{f_{i,t}\}$.

Note that when our stepwise evaluation method is applied to uncorrelated latent factors, the factor selection criterion is: $\max_{i} \SR^2\{f_{i,t}\}$.
In contrast, our method allows for correlated candidate factors, and its selection criterion follows a sequential approach: $\max_{i } {\SR}^2\{f_{i,t}, {f_{1,t}}\}$.
We present a simulation study in Appendix \ref{sec:simulation} to illustrate the advantages of our method. 
Our stepwise regression framework effectively addresses factor correlation, a challenge commonly faced by most variable selection techniques.

\textbf{(II) Model Testing:}
We then test the selected model $\mathcal{M}^{*}_1$ to determine whether to continue adding factors.
We conduct asset pricing tests in Section \ref{sec:2.2.2} on the selected model $\mathcal{M}_1^{*}$ by testing  ${\pmb{\alpha}}_{(\widetilde{\mathcal{M}}_1^{*})} = \mathbf{0}$, where ${\pmb{\alpha}}_{(\widetilde{\mathcal{M}}_1^{*})}$ be the vector of intercepts from spanning regressions that $\textbf{F}_{(\widetilde{\mathcal{M}}_1^{*})t}$ on $\textbf{F}_{(\mathcal{M}_1^{*})t}$.
We can introduce additional test assets ($\textbf{R}_t$) into the unselected factor returns to verify whether the selected model $\mathcal{M}^{*}_1$ is efficient in pricing the cross-section asset returns.

Based on the above toy model, we summarize the assumptions underlying our method. 
Beyond conventional conditions on factor returns and error terms, our framework imposes two key assumptions.
First, our stepwise framework selects risk factors included in the optimal model $\mathcal{T}$ based on their ability to reduce pricing errors ($\alpha$). 
We assume the pricing errors of factors in $\mathcal{T}$ are distinguishable from those excluded from $\mathcal{T}$, as established in assumption 3(i) of the Appendix.
Specifically, for any factor in $\mathcal{T}$, its $\alpha$ under any candidate model $\mathcal{M}$ should be large, while for excluded factors, $\alpha$ should be small.
Second, we permit correlation between the LHS unselected candidate factors $\widetilde{\mathcal{M}}$ and RHS selected factors $\mathcal{M}$, provided the correlation remains bounded, as established in assumption 3(ii) of the Appendix.
This flexibility stems from our sequential factor selection approach, where each factor is chosen conditionally on previously selected ones -- a key distinction from single-factor Sharpe ratio rankings, which neglect factor dependencies.
Detailed assumptions are provided in Appendix \ref{sec:assumption}.

\section{Data} \label{sec:data}
We study monthly U.S. equity data from January 1973 to December 2021, considering seven different factor pricing models and a comprehensive list of 97
factors.

\paragraph{Competing models.}

We consider constructing efficient models from seven different factor pricing models.\footnote{The factor model data are downloaded from the author's website.}
The models considered in this study include
(1) the capital asset pricing model (CAPM), which includes only a market factor (\texttt{MKT});
(2) the \cite{fama1993common} 3-factor model, comprising the \texttt{MKT}, size factor (\texttt{SMB}), and value factor (\texttt{HML}), hereafter FF3;
(3) the \cite{fama2015five} 5-factor model, which adds  the profitability factor (\texttt{RMW}) and investment factor (\texttt{CMA}) to the FF3 model, hereafter FF5;
(4) the \cite{fama2018choosing} 6-factor model, which adds the momentum factor (\texttt{UMD}) motivated by the work of \cite{jegadeesh1993returns} to the FF5 model, hereafter FF6; 
(5) the \cite{hou2021augmented} 5-factor model, consisting of the \texttt{MKT}, \texttt{SMB}, investment factor (\texttt{IA}), profitability factor (\texttt{ROE}),  and  expected growth factor (\texttt{REG}), hereafter Q5; 
(6) the \cite{daniel2020short} 3-factor model contains the \texttt{MKT}, post-earnings-announcement-drift (\texttt{PEAD}) factor, and financing factor (\texttt{FIN}), hereafter DHS3; 
(7) the \cite{barillas2018comparing} 6-factor model, consisting of \texttt{MKT}, \texttt{SMB}, and \texttt{UMD} of the FF6 model, the \texttt{ROE} and \texttt{IA} of the Q5 model, and a more ``timely'' version of the value factor (\texttt{HMLM}) by \cite{asness2013devil}, hereafter BS6.

\paragraph{Candidate factors.}
Our empirical analysis considers a comprehensive list of 97 (23+74) factors. 
First, we include factors from popular models, the six factors in FF6, the \texttt{IA}, \texttt{ROE}, \texttt{REG} in Q5, and the \texttt{PEAD} and \texttt{FIN} in DHS3,
 the \texttt{HMLM} in BS6.
Second, we include commonly used investment industry factors, such as betting-against-beta (\texttt{BAB}), and quality-minus-junk (\texttt{QMJ}).
Third, we also include other traded factors: 
market beta (\texttt{BETA}), 
long-term reversal (\texttt{LTR}), 
short-term reversal (\texttt{STR}), 
liquidity (\texttt{LIQ}), 
return volatility (\texttt{VOL}), 
intermediary capital risk (\texttt{IMD}), 
standard unexpected earnings (\texttt{SUE}), 
net share issues (\texttt{NI}), 
and aligned investor sentiment index (\texttt{$S^{PLS}$}).\footnote{We use a regression of the aligned investor sentiment index on stock returns over the past 36 months to create decile portfolios and long-short portfolios, defined as $S^{PLS}$. The aligned investor sentiment index data are downloaded from the author's website \citep*{huang2015investor}.}
Furthermore, we include an additional 74 factors from the exhaustive list compiled by \cite{hou2020replicating}, each with a one-month holding period.\footnote{The data are downloaded from Lu Zhang's website.}
Table \ref{tab:factor_zoo} in the Appendix \ref{sec:description_data} lists factors and their acronyms, along with their monthly average returns and annualized Sharpe ratios.

\paragraph{Test assets.}
We consider 285 (bivariate-sorted) basis portfolios for optional test assets.
The portfolios consist of the following categories: 25 sorted by size and book-to-market ratio, 25 by size and operating profitability, 25 by size and investment, 25 by size and momentum, 25 sorted by size and market beta, 25 by size and short-term reversal, 25 by size and long-term reversal, 25 by size and variance, 25 by size and residual variance, 35 by size and net share issuance, and 25 by size and accruals.\footnote{The data are downloaded from Ken French's website.}

\section{Expanded Factor Models } \label{sec:empirical}

\subsection{Testing Factor Model Efficiency}

Before applying the stepwise evaluation, we show the efficiency of several baseline factor models. 
Panel A of Table \ref{tab:Model_test} shows their performance in pricing the unselected factors.
Both GRS and HDA test statistics highly reject all the baseline models at the 1\% significance level.
Q5 \citep{hou2021augmented} is the best for all these rejected baseline models, as it produces the smallest GRS and HDA values. 
These findings indicate that the baseline models cannot explain the unselected factors, calling for more factors.
We then show how to construct an efficient model using our FSE method based on these inefficient baseline models.

\begin{table}[h!] 
 \caption{Testing Factor Model Efficiency} \label{tab:Model_test}
 \vspace{-0.4cm}
 \begin{center}

\footnotesize{
\setlength{\tabcolsep}{2  em}{
\begin{tabular}{lccccc}
\toprule 
Model     & \# M & GRS   & $p_{\text{GRS}}$ & HDA    & $p_{\text{HDA}}$ \\ \hline 
     \\
      &  \multicolumn{5}{c}{\underline{Panel A: Baseline Models} }  \\
\\
CAPM & 1    & 5.40 & 0.000         & 35.02 & 0.000         \\
FF3  & 3    & 5.35 & 0.000         & 52.51 & 0.000         \\
FF5  & 5    & 4.80 & 0.000         & 36.29  & 0.000         \\
FF6  & 6    & 4.59 & 0.000         & 27.83 & 0.000         \\
Q5   & 5    & \underline{{2.95}} & 0.000         & \underline{{4.55}}  & 0.000         \\
DHS3 & 3    & 3.67 & 0.000       & 7.26  & 0.000         \\
BS6  & 6    & 4.04 & 0.000         & 29.05 & 0.000       \\ 
\\
 &  \multicolumn{5}{c}{\underline{Panel B: Expanded Efficient Models} }  \\
 \\
CAPM$^F$ & 8    & 1.60 & 0.001  & 1.28 & {0.101}  \\
FF3$^F$  & 11   & 1.40 & 0.017  & 1.28 & {0.100}    \\
FF5$^F$  & 11   & 1.57 & 0.002  & 1.57 & 0.058  \\
FF6$^F$  & 12   & 1.51 & 0.004  & \underline{{1.12}} & \underline{{0.131}}  \\
Q5$^F$   & 10   & 1.57 & 0.002  & 1.36 & 0.087  \\
DHS3$^F$ & 9    & 1.61 & 0.001  & 1.30 & 0.097  \\
BS6$^F$  & 12   & 1.49 & 0.006  & 1.41 & 0.079     \\
\bottomrule
\end{tabular}
}}   
 \end{center}
\footnotesize{Notes: 
This table shows the efficiency of the "row" models in pricing the unselected factors (97 candidate factors, excluding the row model).
Panel A includes the baseline factor models, and Panel B includes the expanded efficient models using FSE. 
For each model, we report the GRS test statistic along with its $p$-value, denoted as $p_{\text{GRS}}$, and the HDA test statistic from \cite{pesaran2023testing} along with its $p$-value, represented as $p_{\text{HDA}}$. 
"\# M" represents the number of factors included in the row model.
}
  \vspace{-0.2cm}
\end{table}

Table \ref{tab:sufficient_model} provides expanded efficient models constructed using our FSE with various baseline models (underlined).
FSE adds at least five additional factors to these inefficient baseline models, indicating that more factors are needed. 
For example, the expanded efficient model CAPM$^F$ contains eight factors, adding factors in order (\texttt{REG}, \texttt{PEAD}, \texttt{HMLM}, \texttt{STR}, \texttt{ILR}, \texttt{SMB}, and \texttt{EPRD}).
Another observation is that recently published factors of expected growth \citep{hou2021augmented} and post-earnings announcement drift \citep{daniel2020short}, \texttt{REG} and \texttt{PEAD}, are the first two factors added in all those baseline models, as they were not included in them initially.

\begin{table}[h!]
\caption{
Expanded Factor Models (FSE Process)
} \label{tab:sufficient_model}
\vspace{-0.4cm}
\begin{center}
\footnotesize{
\setlength{\tabcolsep}{0.8em}{
\begin{tabular}{lccccccc}
\toprule
ID           & CAPM$^F$                         & FF3$^F$                          & FF5$^F$                          & FF6$^F$                          & Q5$^F$                           & DHS3$^F$                          & BS6$^F$                           \\ \hline
1            & \underline{MKT} & \underline{MKT} & \underline{MKT} & \underline{MKT} & \underline{MKT} & \underline{MKT}  & \underline{MKT}  \\
2            & REG                              & \underline{SMB} & \underline{SMB} & \underline{SMB} & \underline{SMB} & \underline{PEAD} & \underline{SMB}  \\
             & (0.91$^{***}$)                   &                                  &                                  &                                  &                                  &                                   &                                   \\
3            & PEAD                             & \underline{HML} & \underline{HML} & \underline{HML} & \underline{IA}  & \underline{FIN}  & \underline{IA}   \\
             & (0.64$^{***}$)                   &                                  &                                  &                                  &                                  &                                   &                                   \\
4            & HMLM                             & REG                              & \underline{RMW} & \underline{RMW} & \underline{ROE} & REG                               & \underline{ROE}  \\
             & (0.33$^{**}$)                    & (0.94$^{***}$)                   &                                  &                                  &                                  & (0.55$^{***}$)                    &                                   \\
5            & STR                              & PEAD                             & \underline{CMA} & \underline{CMA} & \underline{REG} & STR                               & \underline{UMD}  \\
             & (0.30$^{**}$)                    & (0.69$^{***}$)                   &                                  &                                  &                                  & (0.73$^{***}$)                    &                                   \\
6            & ILR                              & STR                              & REG                              & \underline{UMD} & PEAD                             & ILR                               & \underline{HMLM} \\
             & (0.70$^{***}$)                   & (0.24$^{*}$)                     & (0.80$^{***}$)                   &                                  & (0.43$^{***}$)                   & (0.24)                             &                                   \\
7            & SMB                              & ILR                              & PEAD                             & REG                              & STR                              & HMLM                              & REG                               \\
             & (0.09)                           & (0.72$^{***}$)                   & (0.69$^{***}$)                   & (0.71$^{***}$)                   & (0.61$^{***}$)                   & (0.45$^{***}$)                    & (0.67$^{***}$)                    \\
8            & EPRD                             & HMLM                             & STR                              & PEAD                             & ILR                              & SMB                               & PEAD                              \\
             & (0.78$^{***}$)                   & (-0.04)                          & (0.32$^{**}$)                    & (0.56$^{***}$)                   & (0.43$^{*}$)                     & (0.39$^{***}$)                    & (0.58$^{***}$)                    \\
9            &                                  & EPRD                             & ILR                              & HMLM                             & HMLM                             & EPRD                              & EPRD                              \\
             &                                  & (1.04$^{***}$)                   & (0.64$^{***}$)                   & (0.32$^{***}$)                   & (0.44$^{***}$)                   & (0.52$^{***}$)                    & (0.94$^{***}$)                    \\
10           &                                  & BAB                              & HMLM                             & EPRD                             & EPRD                             &                                   & STR                               \\
             &                                  & (0.78$^{***}$)                   & (0.01)                            & (0.88$^{***}$)                   & (0.64$^{***}$)                   &                                   & (0.29$^{**}$)                     \\
11           &                                  & SIM                              & EPRD                             & STR                              &                                  &                                   & ILR                               \\
             &                                  & (0.78$^{***}$)                   & (0.99$^{***}$)                   & (0.45$^{***}$)                   &                                  &                                   & (0.62$^{***}$)                    \\
12           &                                  &                                  &                                  & ILR                              &                                  &                                   & TBI                               \\
             &                                  &                                  &                                  & (0.51$^{**}$)                    &                                  &                                   & (0.19)                             \\
             &                                  &                                  &                                  &                                  &                                  &                                   &                                   \\
\# M     & 1                                & 3                                & 5                                & 6                                & 5                                & 3                                 & 6                                 \\
\# M$^F$ & 8                                & 11                               & 11                               & 12                               & 10                               & 9                                 & 12                                \\
\# Added       & 7                                & 8                                & 6                                & 6                                & 5                                & 6                                 & 6       
\\

\bottomrule
\end{tabular}
}
}
\end{center}
\footnotesize{Notes: 
The table provides the FSE process with various baseline models.
 "M" refers to the baseline model, and "M$^F$" refers to the expanded efficient model where FSE stops adding factors.
We report their model sizes and the number of added factors ("\# Added"). 
Those underlined factors are from the baseline models, and the added factors are included in the FSE order. 
We report the model-adjusted alphas in \% for the added factors based on each "column" baseline model with underlined factors, where $^*$, $^{**}$, and $^{***}$ denote significance levels of 10\%, 5\%, and 1\%, respectively.
}
  \vspace{-0.2cm}
\end{table}

Because the forward stepwise process is stopped by the insignificant HDA test, all these expanded models are efficient, as shown in Panel B of Table \ref{tab:Model_test}.
Compared with Panel A of Table \ref{tab:Model_test}, GRS and HDA values have decreased significantly.
The HDA test has higher power in high-dimensional cases, so we mainly use its significance to assess the model's effectiveness.
The HDA statistic at a 5\% significance level does not reject any of the expanded efficient models.
The FF6$^F$ model has the smallest HDA test statistic, though it has added six factors (\texttt{REG}, \texttt{PEAD}, \texttt{HMLM}, \texttt{EPRD}, \texttt{STR}, and \texttt{ILR}) to become a 12-factor model.
These findings indicate the expanded models are efficient in pricing the unselected candidate factors.

We further explore the reasons for adding these factors and conduct a spanning regression using the baseline model to assess its ability to price the added factor in an expanded efficient model.
The numbers in Table \ref{tab:sufficient_model} present the model-adjusted alphas for the added factors based on each baseline model in the column.
Not surprisingly, these baseline models fail to price most added factors. 
For example, the CAPM model fails to price all seven added factors, except for \texttt{SMB}. 
Here, \texttt{SMB} is not a difficult-to-price factor but helps improve $\SR^2$ for the expanded model of the previous six factors. 
As discussed, \texttt{REG} and \texttt{PEAD} are the first two factors added in the forward stepwise process. Both consistently exhibit unexplained alpha with extremely high economic and statistical significance across various baseline models.
These results indicate that we can obtain an efficient model that performs well in pricing the unselected LHS candidate factors by moving a few factors from the LHS to the RHS, which may be difficult to price or significantly improve $\SR^2$ compared to the baseline models.

\subsection{Interpreting Expanded Factor Models}

\begin{figure}[h]
	\caption{Expanded Factor Models (FSE Process) }
\label{fig:FSE_all}
\vspace{-0.4cm}
\begin{center}
	\includegraphics[ width=1\textwidth]{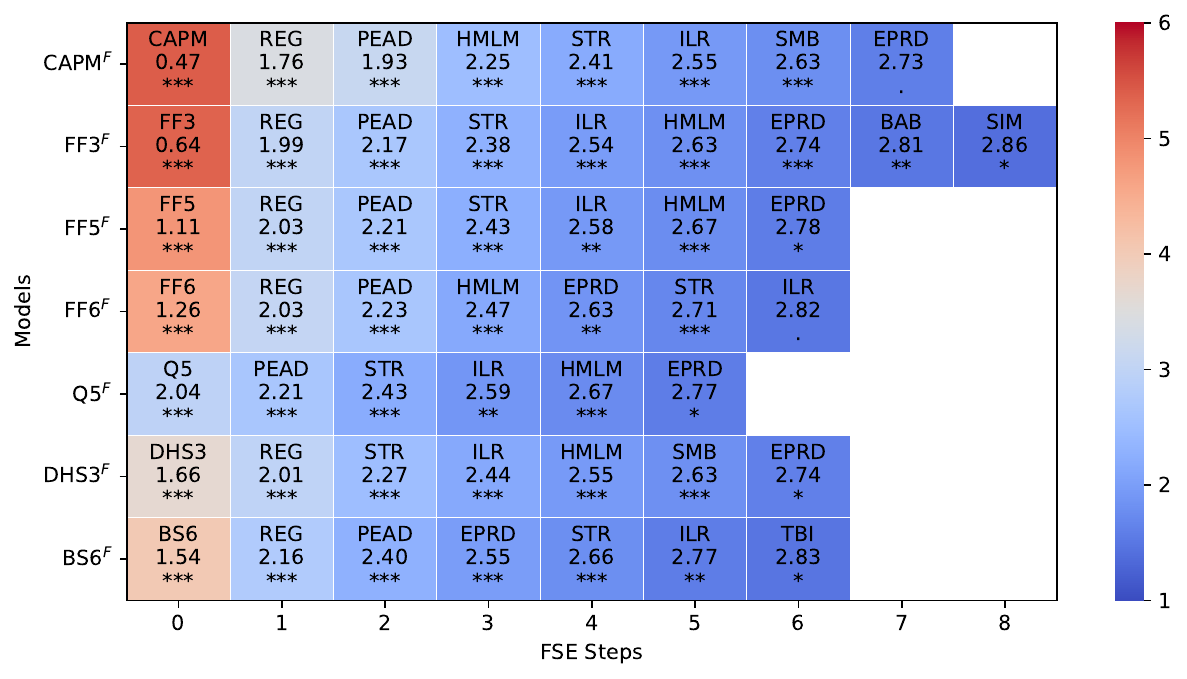}
 \end{center}
\vspace{-0.2cm} 
\footnotesize{Notes: 
This figure illustrates the FSE process for seven baseline models (step 0), displaying the order of factor inclusion that produces the GRS value (color bar) and the annualized Sharpe ratios (numbers) for the expanded model as factors are added one by one over the FSE steps. 
The horizontal axis represents steps forward, while the vertical axis depicts expanded factor models.
The significance code is from the HDA test, where $^.$, $^*$, $^{**}$, and $^{***}$ denote significance levels of greater than 10\%, 10\%, 5\%, and 1\%, respectively.
}
  \vspace{-0.2cm}
\end{figure}

The stepwise factor inclusion is supposed to utilize the asset pricing duality for adding the next factor with the maximal $\SR^2$ increment and the maximal GRS value reduction. 
Figure \ref{fig:FSE_all} illustrates the forward stepwise evaluation process for seven baseline models, displaying the order of factor inclusion, which produces the GRS value (color bar) and the annualized Sharpe ratio (numbers) for each expanded model by adding factors one by one over the FSE steps (from left to right).

First, the baseline models (at step 0) show a high GRS value and a relatively low Sharpe ratio, with the HDA statistic being highly significant across all these models.
Second, incorporating a factor via FSE at step 1 significantly improves the baseline model, lowering the GRS statistic and boosting the Sharpe ratio. The FF3 model exhibits the largest enhancement, with the Sharpe ratio rising from 0.64 to 1.99, a gain of 1.35, followed by the CAPM model. 
Third, as the number of FSE steps increases, the GRS statistic of the expanded model declines, accompanied by a noticeable shift in color (as indicated by the GRS value). However, the incremental improvement in the Sharpe ratio diminishes progressively. 
Early FSE-added factors exert a stronger influence on enhancing model performance.
Fourth, the significance of the HDA test diminishes as more factors are included in the model, eventually becoming insignificant, which suggests that an efficient model has been achieved.
Finally, the expanded efficient models across different baseline models demonstrate comparable GRS values and Sharpe ratios.

\begin{figure}[h]
	\caption{Expanded Factor Models (FF3 Example) }
\label{fig:FSE_FF3}
\vspace{-0.4cm}
\begin{center}
	\includegraphics[ width=1\textwidth]{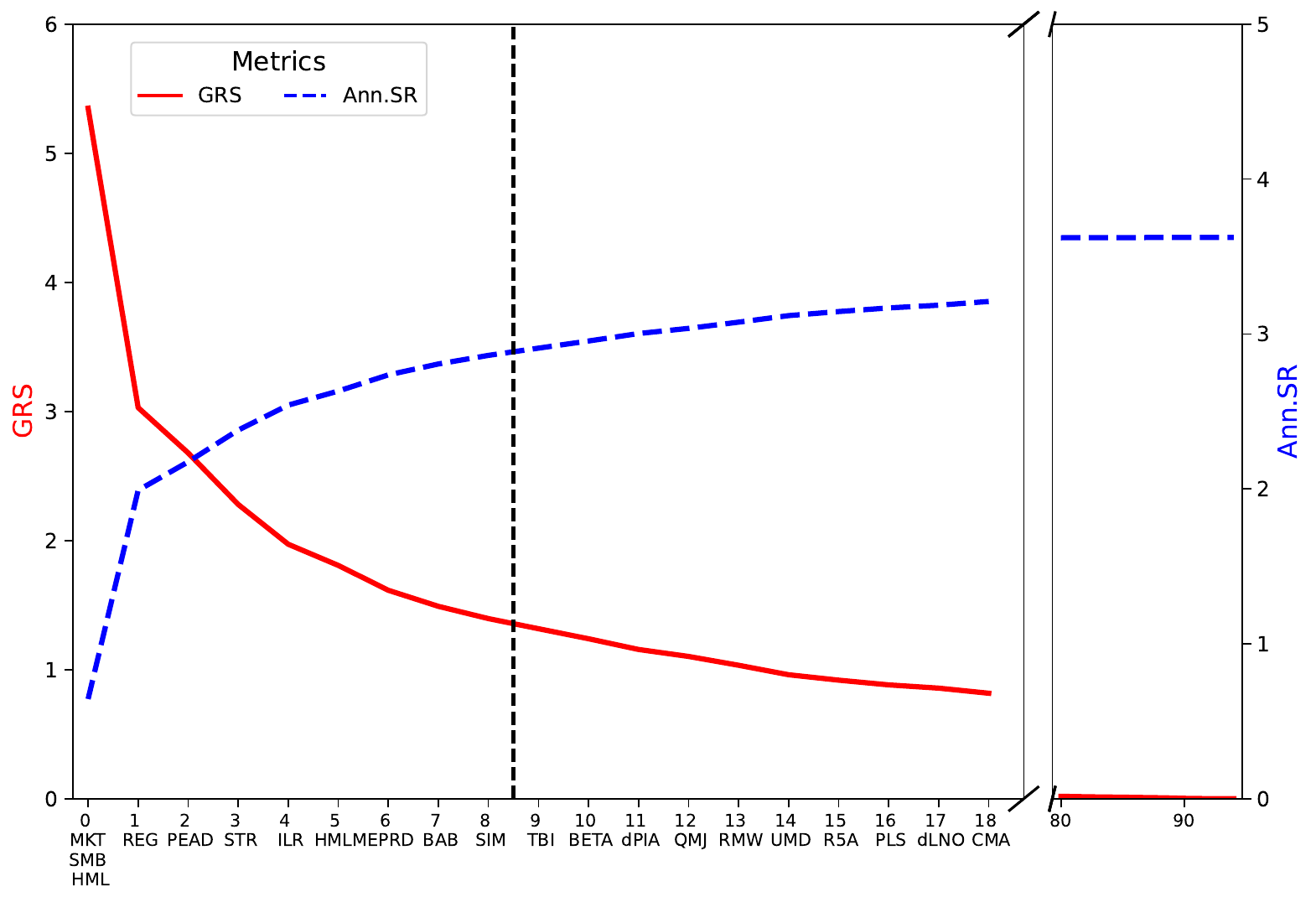}
 \end{center}
\vspace{-0.2cm} 
\footnotesize{Notes: 
This figure illustrates the FSE based on the Fama-French 3-factor model. 
It plots the sequence of minimal GRS values (solid red line, left y-axis) and maximal annualized Sharpe ratios (dashed blue line, right y-axis) across steps. 
The vertical dashed line indicates the stop position from the HDA test.
}
  \vspace{-0.2cm}
\end{figure}

We use the Fama-French 3-factor model as an example to further describe the asset pricing duality objective incorporated within the FSE process. 
Figure \ref{fig:FSE_FF3} displays the order of factor stepwise inclusion that provides the minimal GRS value and the maximal Sharpe ratio while comparing multiple models.
The minimal GRS value (a solid red line on the left y-axis) and the maximal annualized Sharpe ratio (a dashed blue line on the right y-axis) are plotted for each step. 
The figure shows a decreasing trend in the minimum GRS value and a gradual increase in the Sharpe ratio as more factors are added, suggesting improved factor model performance.

In step 1, \texttt{REG} combined with FF3 factors achieves a minimal GRS value of 3.03 and the largest annualized Sharpe ratio of 1.99. Using the HDA test, we assess the efficiency of the expanded model by testing the unexplained alphas of the unselected 93 (97-3-1) candidate factors. The resulting $p$-values indicate high significance, suggesting the need for incorporating additional factors.
In step 2, \texttt{PEAD} appears as the most significant factor among the unselected factors after \texttt{REG}, with a GRS value of 2.68 and a strong rejection in the HDA test. 
During the FSE, \texttt{STR}, \texttt{ILR}, \texttt{HMLM}, \texttt{EPRD}, \texttt{BAB}, and \texttt{SIM} are successively selected until reaching the HDA test with an insignificant $p$-value of 0.1, suggesting the expanded FF3 model (FF3$^{F}$) can explain the unselected factors, and we can stop the factor inclusion.
The unselected factors listed on the x-axis in order are potential factors if we do not consider stopping the FSE process for the baseline model of FF3.
Thus, the final FF3$^{F}$ model is \texttt{MKT} + \texttt{SMB} + \texttt{HML} + 
 \texttt{REG} +\texttt{PEAD} + \texttt{STR} + \texttt{ILR} + \texttt{HMLM}  + \texttt{EPRD} + \texttt{BAB} + \texttt{SIM}.

\subsection{Performance Improvement for Expanded Factor Models} \label{tab:performance_expanded}

\paragraph{Asset pricing performance measures.}
We evaluate the asset pricing performance of each baseline and expanded model using a set of metrics frequently employed in the literature \citep[e.g.,][]{fama2018choosing, ahmed2019best}.
The first measure is the average absolute alphas $A\lvert \alpha \rvert = {N}^{-1}\sum_{i = 1}^{N} |\widehat{\alpha}_{i}|$.
The second measure is the average absolute $t$-values 
$A\lvert t(\alpha) \rvert  = {N}^{-1}\sum_{i = 1}^{N} |t(\widehat{\alpha}_{i})|$.
The third measure is the number of significant alphas with $\lvert t \rvert > 1.96$, $\# \mbox{sign2} = \left[  \sum_{i=1}^{N} I\left\lbrace \lvert t(\widehat{\alpha}_{i}) \rvert > 1.96 \right\rbrace \right]$.
Furthermore, to ensure economic and statistical soundness, we incorporate factor model performance measures \citep[e.g.,][]{kelly2019characteristics, feng2022deep}.
The fourth measure is Total $R^2$:
\vspace{-0.2cm}
\begin{equation}
   \mbox{Total $R^2$}  = 1 - \frac{\sum_{t=1}^T \sum_{i=1}^N (R_{i,t} - \widehat{R}_{i,t})^2} {\sum_{t=1}^T \sum_{i=1}^N (R_{i,t} - \widehat R_{i,t}^{\text{CAPM}})^2 }, 
\vspace{-0.2cm}
\end{equation}
where $\widehat{R}_{i,t} = \widehat{\pmb{\beta}}_i^\top \textbf{f}_t$ and $\widehat R_{i,t}^{\text{CAPM}}$ is the CAPM-implied term. 
Measuring the statistical model fitness, Total $R^2$ represents the fraction of realized return variation explained by the contemporaneous factor realizations, aggregated over all assets and all periods. 
For evaluating asset pricing performance, the last measure is the cross-sectional $R^2$:
\vspace{-0.2cm}
\begin{equation}
\mbox{CS $R^2$}  = 1 - \frac{\sum_{i=1}^N \Big(\bar{R}_i - \widehat{\bar{R}}_i \Big)^2} {\sum_{i=1}^N \Big(\bar{R}_i - \widetilde{ER}_i^{\text{CAPM}} \Big)^2 },
\vspace{-0.2cm}
\end{equation}
where $ \widehat{\bar{R}}_i =  \widehat{\pmb{\beta}}_i^\top \widetilde{\pmb{\lambda}}_f$, the risk premium estimates $\widetilde{\pmb{\lambda}}_f$ are the cross-sectional regression coefficients by regressing $\bar{R}_i$ on $\widehat{\pmb{\beta}}_i$,
and $\widetilde{ER}_i^{\text{CAPM}}$ is the CAPM-implied average return with the constant market risk premium $\widetilde{\pmb{\lambda}}_{\text{MKT}}$.
In short, models with lower values of $A|\alpha|$, $A|t(\alpha)|$, and $\#\text{sign2}$ are considered better, as are models with higher values of Total $R^2$ and CS $R^2$.
To evaluate the performance of mean-variance efficient (MVE) portfolios across different factor models, we report the following investment metrics: 
average returns in \% (AVG), annualized Sharpe ratio (Ann.SR), CAPM alpha in \% ($\alpha_{\text{CAPM}}$), and FF5 alpha in \% ($\alpha_{\text{FF5}}$).

\begin{sidewaystable}
\caption{Performance Improvement for Expanded and Reduced Models}
\label{tab:FSE_performance}
\vspace{-0.4cm}
\begin{center}
\footnotesize{
 \setlength{\tabcolsep}{0.5em}{
\begin{tabular}{lcccccccccccccccccc}
\toprule
\\
         &      &  & \multicolumn{5}{c}{\underline{Asset Pricing (unselected candidate factors)}}                 & \hspace{0.5cm} & \multicolumn{5}{c}{\underline{Asset Pricing (basis portfolios)}}               &  \hspace{0.5cm} & \multicolumn{4}{c}{\underline{Investment}}  \\
         \\
         & \# M &  & $A|{\alpha|}$ & $A|{t(\alpha)|}$ & \# sign2 & Total $R^2$ & CS $R^2$ &  & $A|{\alpha|}$ & $A|{t(\alpha)|}$ & \# sign2 & Total $R^2$ & CS $R^2$ &  & AVG  & Ann.SR & $\alpha_{\text{CAPM}}$ & $\alpha_{\text{FF5}}$ \\ \hline 
\\
& &  & \multicolumn{15}{c}{\underline{Panel A: Baseline Models} }  \\
\\
CAPM     & 1  &  & 0.51 & 3.37 & 81 & 0.0  & 0.0  &  & 0.23 & 1.9  & 127 & 0.0  & 0.0  &  & 0.62 & 0.47 & -            & -        \\
FF3      & 3  &  & 0.47 & 3.44 & 70 & 22.9 & 4.3  &  & 0.15 & 1.81 & 107 & 55.5 & 7.2  &  & 0.82 & 0.64 & 0.38$^{***}$ & -        \\
FF5      & 5  &  & 0.34 & 2.58 & 50 & 31.8 & 12.6 &  & 0.12 & 1.38 & 73  & 60.0 & 71.0 &  & 0.59 & 1.11 & 0.49$^{***}$ & -        \\
FF6      & 6  &  & 0.26 & 2.11 & 37 & 37.6 & 25.9 &  & 0.1  & 1.32 & 67  & 64.2 & 93.8 &  & 0.76 & 1.26 & 0.65$^{***}$ & 0.17$^{***}$         \\
Q5       & 5  &  & 0.17 & 1.11 & 11 & 29.5 & 43.3 &  & 0.11 & 1.16 & 46  & 56.1 & 94.4 &  & 1.18 & 2.04 & 1.12$^{***}$ & 0.83$^{***}$         \\
DHS3     & 3  &  & 0.23 & 1.55 & 29 & 17.8 & 41.2 &  & 0.29 & 2.25 & 168 & 10.7 & 80.0 &  & 1.26 & 1.66 & 1.16$^{***}$ & 0.91$^{***}$         \\
BS6      & 6  &  & 0.28 & 2.15 & 39 & 36.1 & 24.5 &  & 0.12 & 1.49 & 91  & 62.7 & 93.0 &  & 0.9  & 1.54 & 0.81$^{***}$ & 0.46$^{***}$         \\
\\
& &  & \multicolumn{15}{c}{\underline{Panel B: Expanded Efficient Models} }  \\
\\
CAPM$^F$ & 8  &  & 0.15 & 0.93 & 11 & 32.9 & 67.1 &  & 0.1  & 0.96 & 31  & 59.0 & 91.4 &  & 1.16 & 2.73 & 1.12$^{***}$ & 1.02$^{***}$         \\
FF3$^F$  & 11 &  & 0.14 & 0.93 & 8  & 39.9 & 65.3 &  & 0.1  & 1.06 & 40  & 63.9 & 94.6 &  & 1.09 & 2.86 & 1.06$^{***}$ & 0.96$^{***}$         \\
FF5$^F$  & 11 &  & 0.14 & 0.9  & 9  & 43.9 & 68.0 &  & 0.1  & 1.07 & 42  & 65.8 & 96.7 &  & 1.01 & 2.78 & 0.98$^{***}$ & 0.85$^{***}$         \\
FF6$^F$  & 12 &  & 0.13 & 0.86 & 10 & 44.8 & 67.6 &  & 0.1  & 1.09 & 39  & 66.8 & 96.4 &  & 0.94 & 2.82 & 0.91$^{***}$ & 0.79$^{***}$         \\
Q5$^F$   & 10 &  & 0.14 & 0.92 & 9  & 40.9 & 67.8 &  & 0.1  & 1.05 & 43  & 63.5 & 95.6 &  & 1.12 & 2.77 & 1.08$^{***}$ & 0.95$^{***}$         \\
DHS3$^F$ & 9  &  & 0.14 & 0.9  & 8  & 38.9 & 66.1 &  & 0.1  & 1    & 34  & 62.9 & 97.1 &  & 1.16 & 2.74 & 1.12$^{***}$ & 1.01$^{***}$         \\
BS6$^F$  & 12 &  & 0.15 & 0.96 & 9  & 43.7 & 72.5 &  & 0.1  & 1.1  & 47  & 65.1 & 95.5 &  & 1.08 & 2.83 & 1.05$^{***}$ & 0.91$^{***}$         \\
\\
& &  & \multicolumn{15}{c}{\underline{Panel C: 8-Factor Model} }  \\
\\
M8  & 8  &  & 0.15 & 0.93 & 11 & 32.9 & 67.1 &  & 0.1  & 0.96 & 31  & 59.0 & 91.4 &  & 1.16 & \underline{2.73} & \underline{1.12}$^{***}$ & 1.02$^{***}$ \\
\bottomrule
\end{tabular}
}}
\end{center}
\footnotesize{Notes: 
The table below compares different models for asset pricing performance by pricing the unselected factors (out of 97 candidate factors) and 285 basis portfolios, as well as the investment performance by factor tangency portfolio.
We report the asset pricing performance measures in Section \ref{tab:performance_expanded}.
We present the investment performance measures, including average returns in \% (AVG), annualized Sharpe ratio (Ann.SR),  CAPM alpha in \% ($\alpha_{\text{CAPM}}$), and FF5 alpha in \% ($\alpha_{\text{FF5}}$) to assess investment performance.
Panels A, B, and C present the results for the baseline models, the expanded efficient models, and our 8-factor model, respectively.
$^*$, $^{**}$, and $^{***}$ denote significance levels of 10\%, 5\%, and 1\%, respectively.
}
\end{sidewaystable}

Panels A and B of Table \ref{tab:FSE_performance} compare baseline and expanded efficient models in terms of asset pricing and investment performance. 
Expanded models consistently outperform baseline models in asset pricing, reducing unexplained unselected candidate factors and basis portfolio deviations. 
We also find that the absolute levels for unexplained alphas and their $t$-statistics for unselected candidate factors and basis portfolios can be reduced to approximately 0.1\% and 1, which are economically and statistically insignificant.
The CS $R^2$ values, which do not necessarily increase with the addition of more factors, also increase dramatically when expanding the baseline models.
For example, the CAPM$^{F}$ outperforms the baseline model by 67.1\% when adding seven more factors and pricing the unselected candidate factors.

For investment performance, the buy-and-hold strategy, as per the CAPM, has an annualized Sharpe ratio of 0.47 over the past five decades.
Other baseline models, such as Q5, DHS3, and BS6, already exhibit exceptionally high Sharpe ratios and CAPM alphas, with values above 1.5 and 0.8\%, respectively.
Unsurprisingly, adding factors raises the Sharpe ratio; however, all the expanded efficient models generate similarly high Sharpe ratios, ranging from 2.7 to 2.8, and alphas of around 1\%, which are significantly higher than those of the baseline models.

The similar asset pricing and investment performances of different expanded efficient models confirm that the FSE provides robust model selection results.
The expanded efficient models consistently demonstrate positive performance across all asset pricing and investment analysis measures.

\section{Reduced Factor Models} \label{sec:Reduced_result}

The expanded efficient models in Section \ref{sec:empirical} are based on baseline models, such as FF5 and Q5, which may include redundant factors. Therefore, having a scheme for removing potential redundant factors is necessary in efficient but large models and mitigate the greedy nature of FSE against model overfitting.
The proposed BSE can be used to create a reduced factor model from any baseline model or the expanded efficient model. All reduced models presented in the paper are efficient, as factors are removed from the expanded efficient models.

\subsection{Testing the Reduced Factor Models} \label{sec:test_reduced}
Table \ref{tab:Reduced_model} reports the FSE/BSE reduced models constructed using our BSE with various expanded efficient models in Table \ref{tab:sufficient_model}. 
The factors listed in Table \ref{tab:Reduced_model} follow a similar order to those in Table \ref{tab:sufficient_model}.
Note that both types of models are efficient because they have been tested using the HDA test.
The FSE/BSE reduced models can be considered the smallest efficient models for the stepwise evaluation of different baseline models.
The table indicates that the FSE/BSE reduced models maintain the same factors, even after removing various redundant factors. 
BSE can remove redundant factors in different orders from these expanded efficient models, 
suggesting potential redundant factors and model overfitting, this exit mechanism is therefore essential.

For the FSE/BSE reduced models on different baseline models, we find the same eight factors (\texttt{MKT}, \texttt{REG}, \texttt{PEAD}, \texttt{HMLM}, \texttt{STR}, \texttt{ILR}, \texttt{SMB}, and \texttt{EPRD}, hereafter M8) that explain the unselected factors over the past five decades, not rejected by the asset pricing test.
Interestingly, CAPM$^{F+B}$ is the same as CAPM$^{F}$, where no redundant factor is removed for the seven added factors. 
Furthermore, BSE removes some baseline factors and late FSE-added factors, confirming that early FSE-added factors are more important. 
For example, FF3$^{F+B}$ removes three factors, of which \texttt{HML} is excluded, whereas the other two factors (\texttt{BAB} and \texttt{SIM}) are the last two added in the FSE process.
BS6$^{F+B}$ removes four factors, including \texttt{IA}, \texttt{ROE}, and \texttt{UMD} from BS6, and \texttt{TBI} from the last added factor of the FSE process.
The factors removed from the other reduced models are also from the original baseline model. 
In short, these consistent findings show the strength of our approach in selecting a locally optimal model that effectively explains the cross section of unselected candidate factor returns.

\begin{table}[h!]
\caption{Reduced Factor Models (BSE Process)}
\label{tab:Reduced_model}
\vspace{-0.4cm}
\begin{center}
\footnotesize{
\setlength{\tabcolsep}{0.5em}{
\begin{tabular}{lccccccc}
\toprule
ID & CAPM$^{F+B}$ & FF3$^{F+B}$ & FF5$^{F+B}$ & FF6$^{F+B}$ & Q5$^{F+B}$ & DHS3$^{F+B}$ & BS6$^{F+B}$ \\ \hline
1  & MKT      & MKT     & MKT     & MKT     & MKT    & MKT      & MKT     \\
2  & REG      & SMB     & SMB     & SMB     & SMB    & PEAD     & SMB     \\
3  & PEAD     & REG     & REG     & REG     & REG    & REG      & HMLM    \\
4  & HMLM     & PEAD    & PEAD    & PEAD    & PEAD   & STR      & REG     \\
5  & STR      & STR     & STR     & HMLM    & STR    & ILR      & PEAD    \\
6  & ILR      & ILR     & ILR     & EPRD    & ILR    & HMLM     & EPRD    \\
7  & SMB      & HMLM    & HMLM    & STR     & HMLM   & SMB      & STR     \\
8  & EPRD     & EPRD    & EPRD    & ILR     & EPRD   & EPRD     & ILR  \\ 
\\

\# M$^F$ & 8        & 11      & 11      & 12      & 10     & 9        & 12      \\
\# M$^{F+B}$     & 8        & 8       & 8       & 8      & 8     & 8        & 8      \\
\# Removed       & 0        & 3       & 3      & 4       & 2      & 1        & 4
\\

\bottomrule
\end{tabular}
}}
\end{center}
\footnotesize{Notes: 
The table provides FSE/BSE reduced factor models constructed using our BSE based on those expanded efficient models in Table \ref{tab:sufficient_model}. 
"M$^{F+B}$" refers to the FSE/BSE reduced model where BSE stops removing factors based on different large efficient models "M$^F$". 
We also report their model sizes and the number of removed factors ("\# Removed").
}
  \vspace{-0.2cm}
\end{table}

\subsection{Interpreting Reduced Factor Models}

\begin{figure}[h!]
	\caption{Reduced Factor Models (BSE Process) }
\label{fig:BSE_all}
\vspace{-0.4cm}
\begin{center}
	\includegraphics[ width=1\textwidth]{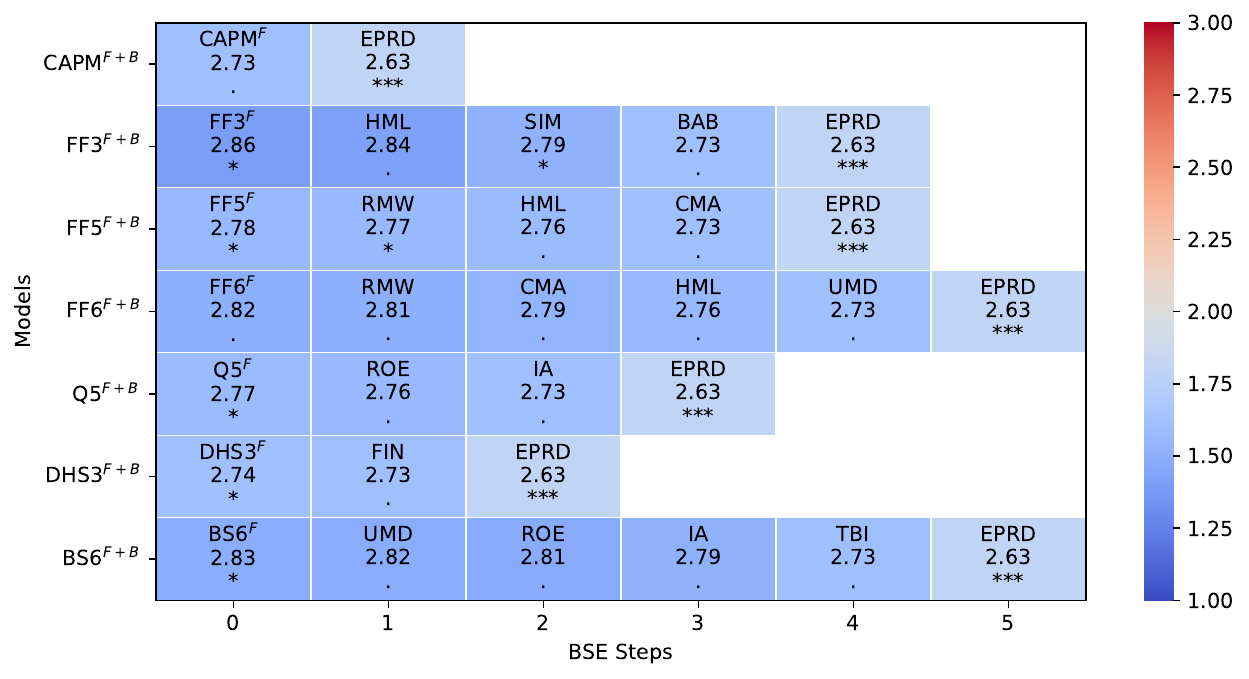}
 \end{center}
\vspace{-0.2cm} 
\footnotesize{Notes: 
This figure illustrates the BSE process for seven expanded efficient models (step 0), displaying the order of factor exclusion, which produces the GRS value (color bar) and the annualized Sharpe ratios (numbers) for the reduced model by removing factors one by one over the BSE steps. 
The horizontal axis displays the number of steps backward, whereas the vertical axis shows the reduced factor models. 
The significance code is from the HDA test, where $^.$, $^*$, $^{**}$, and $^{***}$ denote the significance levels greater than 10\%, 10\%, 5\%, and 1\%, respectively.
}
  \vspace{-0.2cm}
\end{figure}

A smaller but efficient factor model can result in a more interpretable model, even if it slightly decreases performance.
The stepwise factor exclusion uses asset pricing duality to remove the next factor with a minimal reduction in $\SR^2$ and a minimal increase in GRS value.
Figure \ref {fig:BSE_all} shows the BSE process for seven expanded efficient models, displaying the order of factor exclusion, which produces the GRS value (color bar) and the annualized Sharpe ratios (numbers) for the reduced model by removing factors one by one over the BSE steps (from left to right).

First, the expanded efficient models (at step 0) exhibit a comparably low GRS value and a high Sharpe ratio, with the HDA statistic being insignificant across all these models.
Second, in step 1, removing a factor via BSE has a slight negative impact on the model's performance, with an insignificant change in the color (GRS value), indicating that the gradual dropping of the redundant factor does not significantly impact the model's performance. 
The BSE can remove a potential redundant factor except for CAPM$^F$, while its Sharpe ratio is only reduced by 0.01 or 0.02.

Third, as the number of BSE steps increases, the GRS value of the reduced model increases slightly, with a corresponding change in color, and the degree of reduction in the Sharpe ratio gradually increases.
The late BSE-removed factors have a greater impact on reducing the model’s performance.
Fourth, the significance of the HDA test increases as more factors are removed, eventually becoming significant, which suggests that the current factor should not be removed. The HDA test confirms we should not remove \texttt{EPRD} (earnings predictability).
If we remove \texttt{EPRD} from the 8-factor model, the annualized Sharpe ratio decreases from 2.73 to 2.63. 
Finally, the FSE/BSE reduced models across different baseline models reach the same 8-factor model.

\begin{figure}[h]
 \caption{Reduced Factor Models (FF3$^F$ Example)}\label{fig:BSE_FF3}
\vspace{-0.4cm}
\begin{center}
\includegraphics[width=1\textwidth]{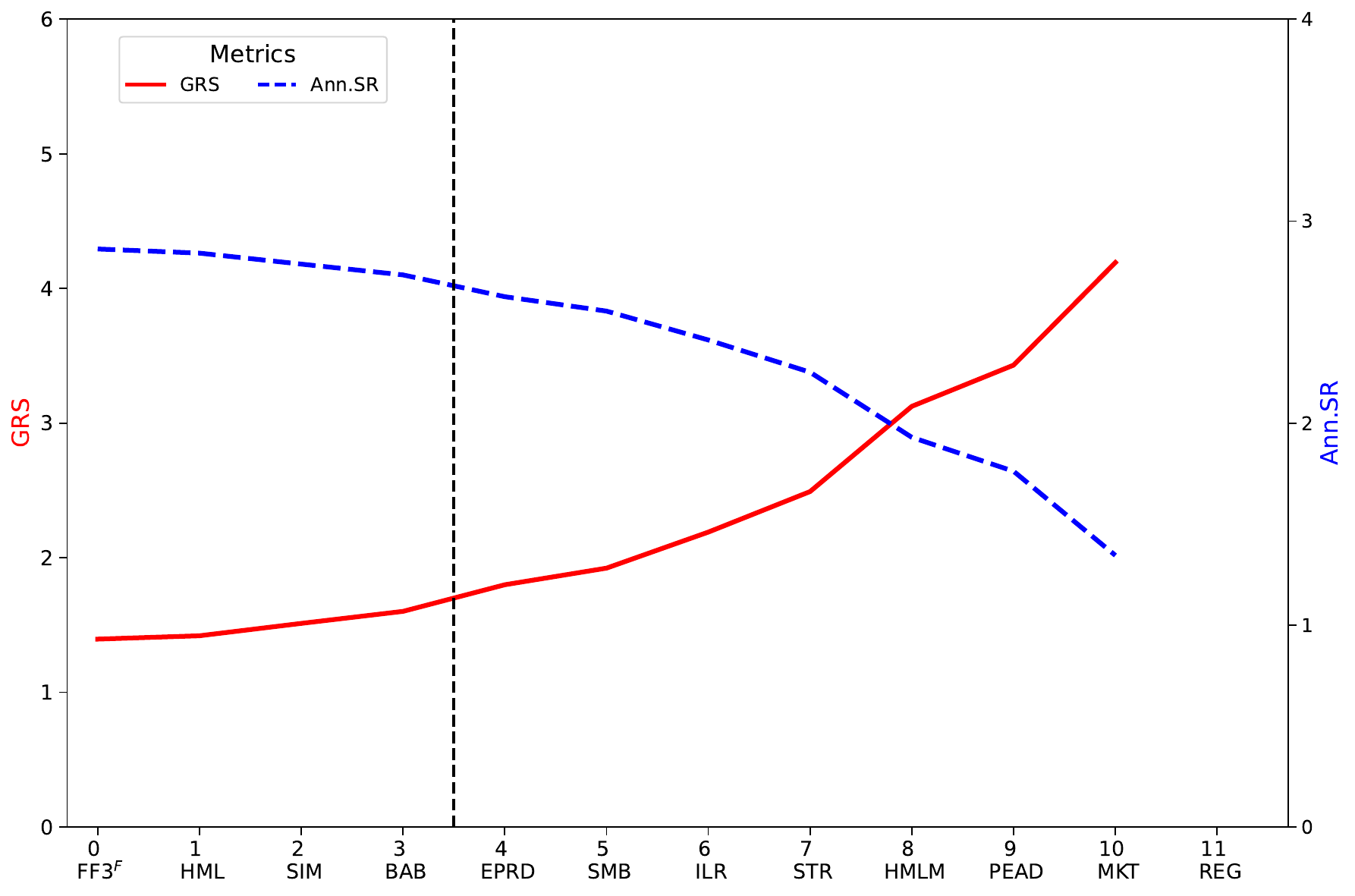}
\end{center}
\vspace{-0.2cm} 
\footnotesize{Notes: 
The figure illustrates the BSE derived from the expanded Fama-French 3-factor model (FF3$^F$). It plots the minimal GRS statistic at each step (solid red line, left y-axis) and the maximal annualized Sharpe ratio (dashed blue line, right y-axis). 
The vertical dashed line indicates the stop position from the HDA test.
}
  \vspace{-0.2cm}
\end{figure}

We again use the expanded FF3 model (FF3$^F$) as an example to further describe the asset pricing duality objective incorporated within the BSE process.
Figure \ref {fig:BSE_FF3} shows the order of factor backward stepwise exclusion that minimizes the GRS value and maximizes the Sharpe ratio when comparing multiple reduced models.
The figure indicates a rising minimum GRS value and a gradual decrease in the Sharpe ratio as more factors are removed, indicating a decline in factor model performance.
In step 1, FF3$^F$ removes the value factor, \texttt{HML}, to achieve a GRS value of 1.42 and a Sharpe ratio of 2.84.
We use the HDA test to evaluate the efficiency of the reduced model by pricing the unexplained alphas of the unselected 87 candidate factors (97-11+1).
The insignificant test result suggests the potential need for removing additional factors. 

While HDA tests are not rejected, \texttt{SIM} and \texttt{BAB} are excluded in steps 2 and 3.
In step 4, excluding \texttt{EPRD} results in rejecting the HDA test at the 5\% level, indicating that \texttt{EPRD} should not be removed, thereby concluding the factor exclusion.
The unselected factors listed on the x-axis are potential factors if we do not consider stopping the BSE process.
Three factors were removed, with \texttt{HML} coming from FF3, whereas \texttt{SIM} and \texttt{BAB} are the last two added in the FSE process. 
The exclusion of \texttt{HML} suggests potential factor proliferation with newly published factors replacing previous ones.
Therefore, the final optimal model FF3$^{F+B}$ is \texttt{MKT} + \texttt{SMB} + \texttt{REG} + \texttt{PEAD} + \texttt{STR} + \texttt{ILR} + \texttt{HMLM} + \texttt{EPRD}.

\subsection{Performance of Reduced Factor Models}

Panel C of Table \ref{tab:FSE_performance} reports the asset pricing and investment performance of the FSE/BSE reduced model (8-Factor Model), which is the same for all baseline models.
The 8-factor model contains the same factors as the CAPM$^F$ model, as shown in Table \ref{tab:Reduced_model}.
The reduced model M$^{F+B}$ has asset pricing and investment performances similar to the expanded model M$^F$, despite having fewer factors.

Our selected 8-factor model achieves an annualized Sharpe ratio of 2.73 and a monthly CAPM alpha of 1.12\%.\footnote{We also evaluate the model-adjusted investment performance based on other benchmark models. The results for these model-adjusted alpha metrics are similar. Our M8 model shows a large and significant alpha on all benchmark models, while the other models have a small alpha on our M8 model, indicating that our M8 model significantly outperforms the benchmark models.}
The study by \cite{barillas2018comparing} examines 10 factors, 9 of which are included in the current analysis. They find a 6-factor model (\texttt{MKT} + \texttt{IA} + \texttt{ROE} + \texttt{SMB} + \texttt{HML} + \texttt{UMD}) that has a Sharpe ratio of 1.35 over the past five decades.
The study by \cite{avramov2023integrating} examines 14 factors, 12 of which are included in the current analysis. They report a 3-factor model (\texttt{PEAD} + \texttt{QMJ} + \texttt{IMD}) with a Sharpe ratio of 1.58.
Our selected 8-factor model's investment performance exceeds that of existing factor selection approaches.

\subsection{Out-of-sample Evaluation}

\begin{sidewaystable}
\caption{In-sample and Out-of-sample Performance}
\label{tab:performance_remaining_oos}
\vspace{-0.4cm}
\begin{center}
\footnotesize{
 \setlength{\tabcolsep}{0.5em}{
\begin{tabular}{lcccccccccccccccccc}

\toprule
\\
            &  & \multicolumn{6}{c}{\underline{Asset Pricing   (unselected candidate factors)}}                            &  & \multicolumn{6}{c}{\underline{Asset Pricing (basis   portfolios)}}                               &  & \multicolumn{3}{c}{\underline{Investment}} \\
            \\
            &  & \multicolumn{2}{c}{\underline{Time 1}} & \multicolumn{2}{c}{\underline{Time 2}} & \multicolumn{2}{c}{\underline{Time 3}} &  & \multicolumn{2}{c}{\underline{Time 1}} & \multicolumn{2}{c}{\underline{Time 2}} & \multicolumn{2}{c}{\underline{Time 3}} &  & \underline{Time 1}  & \underline{Time 2}   & \underline{Time 3}   \\
            \\
            &  & Total $R^2$   & CS $R^2$   & Total $R^2$   & CS $R^2$   & Total $R^2$   & CS $R^2$   &  & Total $R^2$   & CS $R^2$   & Total $R^2$   & CS $R^2$   & Total $R^2$   & CS $R^2$   &  & Ann.SR   & Ann.SR   & Ann.SR   \\ \hline
            \\
            &  & \multicolumn{17}{c}{\underline{Panel A: In-sample}}                 \\
            \\
FF5         &  & 32.3          & 18.9       & 29.8          & 15.1       & 35.7          & 17         &  & 59.1          & 63.8       & 58.8          & 13.4       & 63.9          & 82.4       &  & 1.24     & 1.12     & 1.27     \\
Q5          &  & 30.7          & 34.1       & 29.5          & 41.6       & 31.3          & 46.3       &  & 54.7          & 89.2       & 57.5          & 64.7       & 59.3          & 92.9       &  & 1.86     & 1.95     & 2.45     \\
FF5$^{F+B}$ &  & 29.3          & 54.4       & 22.4          & 53.6       & 37.5          & 63.1       &  & 55.9          & 69.2       & 18.6          & 70.4       & 63.3          & 93.3       &  & 2.25     & 2.26     & 3.4      \\
Q5$^{F+B}$  &  & 29.3          & 54.4       & 36.2          & 65.6       & 32.4          & 64.4       &  & 55.9          & 69.2       & 62.5          & 91.5       & 60.4          & 90.8       &  & 2.25     & 2.58     & 3.42     \\
M8          &  & 32.4          & 60.6       & 34.6          & 62.4       & 37            & 64.5       &  & 57.6          & 72.4       & 60.7          & 50.1       & 62.4          & 90.1       &  & 2.38     & 2.56     & 3.43     \\
\\
            &  & \multicolumn{17}{c}{\underline{Panel B: Out-of-sample}}                                                                                                                                                                        \\
            \\
FF5         &  & 29.1          & 15.6       & 27            & 10.3       & 20.6          & 19.4       &  & 60.5          & 67.7       & 53.9          & 48.4       & 50            & 40.1       &  & 0.59     & 0.96     & 0.66     \\
Q5          &  & 24.2          & 29.2       & 23.4          & 12.3       & 23.4          & 35.4       &  & 57.3          & 92.3       & 47.3          & 95.3       & 46.7          & 36.5       &  & 2.17     & 2.15     & 1.21     \\
FF5$^{F+B}$ &  & 27.5          & 32.9       & 18.7          & 17.2       & 20.6          & 32         &  & 57.8          & 60.8       & 11.3          & 93.6       & 49.7          & 43.3       &  & 2.88     & 2.05     & 1.16     \\
Q5$^{F+B}$  &  & 27.5          & 32.9       & 30.2          & 23.3       & 21.5          & 26.3       &  & 57.8          & 60.8       & 54.4          & 91.4       & 47.2          & 44.3       &  & 2.88     & 2.72     & 1.23     \\
M8          &  & 28.3          & 45.6       & 23.2          & 16.1       & 18.9          & 39.4       &  & 58.6          & 67.9       & 50.7          & 62.4       & 49.1          & 59.1       &  & 3.34     & 2.97     & 1.53          \\

\bottomrule
\end{tabular}
}}
\end{center}
\footnotesize{Notes: 
The table reports asset pricing and investment performance results across various models for both in-sample (INS) and out-of-sample (OOS) analyses. We compare two benchmark models, FF5 and Q5, their reduced variants FF5$^{F+B}$ and Q5$^{F+B}$, and our proposed M8 model. 
The data is split into three consecutive folds, with each fold serving as the test sample, while the remaining two are used for training. Asset pricing performance is evaluated using total and cross-sectional $R^2$ (\%) for unselected candidate factors and basis portfolios. OOS $R^2$ values are derived from INS data. 
Investment performance is assessed through annualized Sharpe ratios of the MVE portfolio, with OOS weights based on INS data. Panels A and B present results for INS and OOS analyses, respectively.
}
\end{sidewaystable}

We conduct an OOS analysis to assess the robustness of our stepwise evaluation and mitigate overfitting concerns. 
Following \cite{feng2022deep}, we split the dataset sequentially into three periods: ``Time1'' (1973/01–1989/04), ``Time2'' (1989/05–2005/08), and ``Time3'' (2005/09–2021/12). The test sample comprises 196 months, while the remaining 392 months are used for model training. Factor sets and portfolio weights are held constant during OOS evaluation. Notably, the factors in reduced models M$^{F+B}$ vary across shorter samples.
 
Table \ref{tab:performance_remaining_oos} presents asset pricing and investment performance results for various models in both in-sample (INS) and OOS analyses. We compare two standard baseline models, FF5 and Q5, their reduced variants FF5$^{F+B}$ and Q5$^{F+B}$, and our proposed M8 model. 
Columns labeled ``Time1'' evaluate the models during ``Time1'', trained using 32 years and 8 months from ``Time2'' and ``Time3''.
Section \ref{tab:performance_expanded} utilizes commonly used metrics from the literature to assess factor model performance, with most being INS measures.
For OOS performance analysis, we report Total and CS $R^2$ for asset pricing, as well as annualized Sharpe ratios for MVE portfolio investment performance.

Our empirical analysis yields three key findings on the asset pricing performance. First, FSE/BSE reduced factor models consistently outperform baseline models, with greater pricing performance for unselected candidate factors and basis portfolios. 
For instance, FF5$^{F+B}$ improves OOS CS $R^2$ from 15.6\% to 32.9\% for pricing unselected candidate factors in the ``Time 1'' period. 
Second, asset pricing models exhibit stronger pricing performance for basis portfolios compared to unselected candidate factors, as reflected in higher Total and CS $R^2$ values. This highlights the models’ effectiveness in capturing cross-sectional variation among basis portfolios.
Finally, the findings underscore the importance of incorporating unselected candidate factors as test assets when optimizing model selection.

Regarding investment performance, we find:
(1) All reduced FSE/BSE models using INS data outperform benchmark models in both INS and OOS analyses. Adding more factors does not consistently enhance OOS performance, underscoring the advantages of our stepwise evaluation process.
(2) The M8 model delivers superior OOS Sharpe ratios compared to other models. Notably, the 8-factor model achieves an OOS annualized Sharpe ratio of 1.53 during the latest fold period. 
Although the M8 model does not consistently yield the highest Total and CS $R^2$, its Sharpe ratio is consistently and significantly superior. Importantly, while the Sharpe ratio remains invariant to test asset selection, $R^2$ is sensitive to it.
As demonstrated by \cite{barillas2017alpha}, the Sharpe ratio serves as a critical metric for determining the relative performance of factor models, providing a definitive comparison of their effectiveness.

Third, the asset pricing and investment performance of reduced FSE/BSE models are comparable both in INS and OOS, consistently incorporating most of the selected eight factors. This underscores the robustness of the two-step forward-backward stepwise selection methodology.
In summary, reduced FSE/BSE models exhibit strong OOS performance across asset pricing and investment metrics.

\section{Further Asset Pricing Applications} \label{sec:empirical_robust}
\subsection{Evaluating Individual Factors} \label{sec:SE_individual}

After evaluating factor models, we can assess the performance of individual factors using a stepwise evaluation. This involves conducting a two-step forward and backward evaluation of the testing factor within a baseline model. 
If the factor is removed during the process, another factor must replace its role. If the factor remains in the final model, it is considered useful. 
Unlike other factor testing methods, our approach provides an enter/exit mechanism for updating controlling factors. Table \ref{tab:bench_row_CAPM} reports the individual factor test. 
With a 2-factor baseline model (\texttt{MKT} + one "row" factor), corresponding to 96 (97-1) models, we use FSE to expand the baseline model to an expanded efficient model and BSE to remove redundant factors.

\begin{table}[h!]
\caption{Testing Individual Factors (Baseline = MKT + Row Factor)}
\label{tab:bench_row_CAPM}
\vspace{-0.4cm} 
    \begin{center}
\footnotesize{\setlength{\tabcolsep}{0.25em}{
\begin{tabular}{ccccccccccclccccc}
\toprule
ID & Factor  & Selected    & Same         & Rate &  & ID & Factor & Selected    & Same         & Rate &  & ID & Factor & Selected    & Same         & Rate \\ \hline
1  & \underline{{MKT}}     & -            & -            & \underline{{0.93}} &  & 34 & dBE    &              & $\checkmark$ & 0.00 &  & 66 & IVFF   &              & $\checkmark$ & 0.00 \\
2  & \underline{{SMB}}     & $\checkmark$ & $\checkmark$ & \underline{{0.84}} &  & 35 & dCOA   &              & $\checkmark$ & 0.00 &  & 67 & IVG    &              & $\checkmark$ & 0.00 \\
3  & HML     &              &              & 0.00 &  & 36 & dFIN   &              & $\checkmark$ & 0.00 &  & 68 & IVQ    &              & $\checkmark$ & 0.00 \\
4  & RMW     &              & $\checkmark$ & 0.00 &  & 37 & dFNL   &              & $\checkmark$ & 0.00 &  & 69 & NDF    &              & $\checkmark$ & 0.00 \\
5  & CMA     & $\checkmark$ &              & 0.01 &  & 38 & dII    &              & $\checkmark$ & 0.00 &  & 70 & NEI    &              & $\checkmark$ & 0.00 \\
6  & UMD     &              & $\checkmark$ & 0.00 &  & 39 & dLNO   &              &              & 0.00 &  & 71 & NOA    & $\checkmark$ &              & 0.01 \\
7  & \underline{{HMLM}}    & $\checkmark$ & $\checkmark$ & \underline{{0.97}} &  & 40 & dNCA   &              & $\checkmark$ & 0.00 &  & 72 & NOP    &              & $\checkmark$ & 0.00 \\
8  & QMJ     &              & $\checkmark$ & 0.02 &  & 41 & dNCO   &              & $\checkmark$ & 0.00 &  & 73 & OA     &              & $\checkmark$ & 0.00 \\
9  & BAB     & $\checkmark$ &              & 0.09 &  & 42 & dNOA   &              & $\checkmark$ & 0.00 &  & 74 & OCA    &              & $\checkmark$ & 0.00 \\
10 & LIQ     &              & $\checkmark$ & 0.00 &  & 43 & DP     &              & $\checkmark$ & 0.00 &  & 75 & OCP    &              & $\checkmark$ & 0.00 \\
11 & \underline{{STR}}     & $\checkmark$ & $\checkmark$ & \underline{{0.98}} &  & 44 & dPIA   &              & $\checkmark$ & 0.00 &  & 76 & OL     &              & $\checkmark$ & 0.00 \\
12 & LTR     &              & $\checkmark$ & 0.00 &  & 45 & dROA   &              & $\checkmark$ & 0.00 &  & 77 & OLE    & $\checkmark$ &              & 0.01 \\
13 & \underline{{REG}}     & $\checkmark$ & $\checkmark$ & \underline{{1.00}} &  & 46 & dROE   &              & $\checkmark$ & 0.00 &  & 78 & OP     &              & $\checkmark$ & 0.00 \\
14 & VOL     &              & $\checkmark$ & 0.00 &  & 47 & DUR    &              & $\checkmark$ & 0.00 &  & 79 & OPA    &              & $\checkmark$ & 0.00 \\
15 & SUE     &              & $\checkmark$ & 0.00 &  & 48 & dWC    & $\checkmark$ &              & 0.01 &  & 80 & OPE    & $\checkmark$ &              & 0.01 \\
16 & IA      & $\checkmark$ &              & 0.02 &  & 49 & EBP    &              & $\checkmark$ & 0.00 &  & 81 & 52W    &              &              & 0.00 \\
17 & ROE     &              & $\checkmark$ & 0.00 &  & 50 & EM     &              & $\checkmark$ & 0.00 &  & 82 & PDA    &              & $\checkmark$ & 0.00 \\
18 & IMD     &              & $\checkmark$ & 0.00 &  & 51 & EP     &              & $\checkmark$ & 0.00 &  & 83 & POA    &              & $\checkmark$ & 0.00 \\
19 & NI      &              & $\checkmark$ & 0.00 &  & 52 & \underline{{EPRD}}   & $\checkmark$ & $\checkmark$ & \underline{{0.90}} &  & 84 & PTA    &              & $\checkmark$ & 0.00 \\
20 & BETA    & $\checkmark$ &              & 0.09 &  & 53 & ETL    &              &              & 0.00 &  & 85 & R1A    &              & $\checkmark$ & 0.00 \\
21 & \underline{{PEAD}}    & $\checkmark$ & $\checkmark$ & \underline{{1.00}} &  & 54 & ETR    &              & $\checkmark$ & 0.00 &  & 86 & R5A    &              &              & 0.00 \\
22 & FIN     &              & $\checkmark$ & 0.00 &  & 55 & GPA    &              & $\checkmark$ & 0.00 &  & 87 & RER    &              & $\checkmark$ & 0.00 \\
23 & $S^{PLS}$ &              & $\checkmark$ & 0.00 &  & 56 & HS     &              & $\checkmark$ & 0.00 &  & 88 & Resid  &              & $\checkmark$ & 0.00 \\
24 & ABR     &              & $\checkmark$ & 0.00 &  & 57 & IG     & $\checkmark$ &              & 0.01 &  & 89 & ROA    &              & $\checkmark$ & 0.00 \\
25 & ACI     & $\checkmark$ &              & 0.01 &  & 58 & ILE    &              & $\checkmark$ & 0.00 &  & 90 & RS     &              & $\checkmark$ & 0.00 \\
26 & ATO     &              &              & 0.00 &  & 59 & \underline{{ILR}}    & $\checkmark$ & $\checkmark$ & \underline{{0.98}} &  & 91 & SG     &              & $\checkmark$ & 0.00 \\
27 & CEI     &              & $\checkmark$ & 0.00 &  & 60 & IM     &              & $\checkmark$ & 0.00 &  & 92 & SIM    & $\checkmark$ &              & 0.04 \\
28 & CIM     &              & $\checkmark$ & 0.00 &  & 61 & IOCA   & $\checkmark$ &              & 0.01 &  & 93 & SP     &              & $\checkmark$ & 0.00 \\
29 & CLA     &              & $\checkmark$ & 0.00 &  & 62 & IR     &              & $\checkmark$ & 0.00 &  & 94 & TA     &              & $\checkmark$ & 0.00 \\
30 & COP     & $\checkmark$ &              & 0.01 &  & 63 & ISFF   &              & $\checkmark$ & 0.00 &  & 95 & TBI    & $\checkmark$ &              & 0.02 \\
31 & CP      &              & $\checkmark$ & 0.00 &  & 64 & ISQ    &              & $\checkmark$ & 0.00 &  & 96 & TV     &              & $\checkmark$ & 0.00 \\
32 & CTO     &              & $\checkmark$ & 0.00 &  & 65 & IVC    &              & $\checkmark$ & 0.00 &  & 97 & VHP    &              & $\checkmark$ & 0.00 \\
33 & DAC     &              & $\checkmark$ & 0.00 &  &    &        &              &              &      &  &    &        &              &              &   \\
\bottomrule
\end{tabular}
}}
\end{center}
\footnotesize{Notes: 
This table presents the individual factor test within the stepwise evaluation framework. With a 2-factor baseline model (MKT + one "row" factor), we use FSE to expand it to an efficient model and BSE to remove redundant factors. The "Selected" column shows the inclusion of the "row" factor after BSE, denoted by a "$\checkmark$". The "Same" column indicates if the reduced model from a different baseline model is consistent with our selected 8-factor model (underlined), with "$\checkmark$" denoting consistency. The "Rate" column also provides the selection frequency of each factor (included in the reduced model) among the 96 models.
}
  \vspace{-0.2cm}
\end{table}

First, the ``Selected'' column has a total count of 21, indicating that only 21.9\% of the 96 models do not remove the row factor. This finding suggests that most individual factors are redundant and can be excluded by the stepwise evaluation.
Some famous factors, such as \texttt{CMA} and \texttt{IA} for investment, \texttt{BETA}, and \texttt{BAB}, remain selected within the 2-factor baseline model. 
Most factor testing methods do not account for the high correlation among factors, which may lead to model misspecification. While our framework does not fully resolve this issue, the enter-exit two-step scheme advances the analysis of factors across baseline models.

Second, in the ``Same'' column, the total count is 76, indicating that 79.2\% of the 96 models are consistent with our 8-factor model in Section \ref{sec:test_reduced}. 
The 8-factor model can be robustly selected regardless of the baseline model used, even if the baseline model is a simple 2-factor model with a redundant factor.
In addition, we consistently find extremely high selection rates for our selected eight factors, with \texttt{REG} and \texttt{PEAD} both at 100\%, and the \texttt{SMB} factor having a lower selection rate of 84\%. All other factors had selection rates close to 0 except for these eight factors.

\begin{table}[!h]
\caption{Testing Individual Factors (Baseline = FF3 + Row Factor)}
\label{tab:bench_row_FF3}
\vspace{-0.4cm}
    \begin{center}
\footnotesize{\setlength{\tabcolsep}{0.2em}{
\begin{tabular}{ccccccccccclccccc}
\toprule
ID & Factor  & Selected    & Same         & Rate &  & ID & Factor & Selected    & Same         & Rate &  & ID & Factor & Selected    & Same         & Rate \\ \hline
1  & \underline{{MKT}}     & -            & -            & \underline{{0.96}} &  & 34 & dBE    &              & $\checkmark$ & 0.00 &  & 66 & IVFF   &              & $\checkmark$ & 0.00 \\
2  & \underline{{SMB}}     & -            & -            & \underline{{0.95}} &  & 35 & dCOA   &              & $\checkmark$ & 0.00 &  & 67 & IVG    &              & $\checkmark$ & 0.00 \\
3  & HML     & -            & -            & 0.15 &  & 36 & dFIN   & $\checkmark$ &              & 0.01 &  & 68 & IVQ    &              & $\checkmark$ & 0.00 \\
4  & RMW     &              & $\checkmark$ & 0.00 &  & 37 & dFNL   & $\checkmark$ &              & 0.01 &  & 69 & NDF    &              & $\checkmark$ & 0.00 \\
5  & CMA     &              & $\checkmark$ & 0.00 &  & 38 & dII    &              & $\checkmark$ & 0.00 &  & 70 & NEI    &              & $\checkmark$ & 0.00 \\
6  & UMD     &              & $\checkmark$ & 0.03 &  & 39 & dLNO   &              & $\checkmark$ & 0.00 &  & 71 & NOA    &              & $\checkmark$ & 0.00 \\
7  & \underline{{HMLM}}    & $\checkmark$ & $\checkmark$ & \underline{{1.00}} &  & 40 & dNCA   & $\checkmark$ &              & 0.01 &  & 72 & NOP    &              & $\checkmark$ & 0.00 \\
8  & QMJ     & $\checkmark$ &              & 0.02 &  & 41 & dNCO   & $\checkmark$ &              & 0.01 &  & 73 & OA     & $\checkmark$ &              & 0.01 \\
9  & BAB     &              & $\checkmark$ & 0.18 &  & 42 & dNOA   & $\checkmark$ &              & 0.01 &  & 74 & OCA    &              & $\checkmark$ & 0.00 \\
10 & LIQ     &              & $\checkmark$ & 0.00 &  & 43 & DP     &              & $\checkmark$ & 0.00 &  & 75 & OCP    & $\checkmark$ &              & 0.01 \\
11 & \underline{{STR}}     & $\checkmark$ & $\checkmark$ & \underline{{0.99}} &  & 44 & dPIA   & $\checkmark$ &              & 0.01 &  & 76 & OL     &              & $\checkmark$ & 0.00 \\
12 & LTR     &              &              & 0.00 &  & 45 & dROA   &              & $\checkmark$ & 0.00 &  & 77 & OLE    &              & $\checkmark$ & 0.00 \\
13 & \underline{{REG}}     & $\checkmark$ & $\checkmark$ & \underline{{1.00}} &  & 46 & dROE   &              & $\checkmark$ & 0.00 &  & 78 & OP     &              & $\checkmark$ & 0.00 \\
14 & VOL     &              & $\checkmark$ & 0.00 &  & 47 & DUR    &              & $\checkmark$ & 0.00 &  & 79 & OPA    &              &              & 0.00 \\
15 & SUE     &              & $\checkmark$ & 0.00 &  & 48 & dWC    & $\checkmark$ &              & 0.01 &  & 80 & OPE    & $\checkmark$ &              & 0.01 \\
16 & IA      &              & $\checkmark$ & 0.00 &  & 49 & EBP    &              & $\checkmark$ & 0.00 &  & 81 & 52W    & $\checkmark$ &              & 0.01 \\
17 & ROE     &              & $\checkmark$ & 0.00 &  & 50 & EM     &              & $\checkmark$ & 0.00 &  & 82 & PDA    &              & $\checkmark$ & 0.00 \\
18 & IMD     &              & $\checkmark$ & 0.00 &  & 51 & EP     &              & $\checkmark$ & 0.00 &  & 83 & POA    & $\checkmark$ &              & 0.01 \\
19 & NI      &              &              & 0.00 &  & 52 & \underline{{EPRD}}   & $\checkmark$ & $\checkmark$ & \underline{{0.99}} &  & 84 & PTA    &              & $\checkmark$ & 0.00 \\
20 & BETA    & $\checkmark$ &              & 0.04 &  & 53 & ETL    &              & $\checkmark$ & 0.00 &  & 85 & R1A    &              & $\checkmark$ & 0.00 \\
21 & \underline{{PEAD}}    & $\checkmark$ & $\checkmark$ & \underline{{1.00}} &  & 54 & ETR    &              & $\checkmark$ & 0.00 &  & 86 & R5A    & $\checkmark$ &              & 0.01 \\
22 & FIN     &              & $\checkmark$ & 0.00 &  & 55 & GPA    &              & $\checkmark$ & 0.00 &  & 87 & RER    &              & $\checkmark$ & 0.00 \\
23 & $S^{PLS}$ &              & $\checkmark$ & 0.00 &  & 56 & HS     &              & $\checkmark$ & 0.00 &  & 88 & Resid  &              & $\checkmark$ & 0.00 \\
24 & ABR     &              & $\checkmark$ & 0.00 &  & 57 & IG     &              & $\checkmark$ & 0.00 &  & 89 & ROA    &              & $\checkmark$ & 0.00 \\
25 & ACI     &              & $\checkmark$ & 0.00 &  & 58 & ILE    &              & $\checkmark$ & 0.00 &  & 90 & RS     &              & $\checkmark$ & 0.00 \\
26 & ATO     &              & $\checkmark$ & 0.00 &  & 59 & \underline{{ILR}}    & $\checkmark$ & $\checkmark$ & \underline{{0.99}} &  & 91 & SG     &              & $\checkmark$ & 0.00 \\
27 & CEI     &              & $\checkmark$ & 0.00 &  & 60 & IM     &              & $\checkmark$ & 0.00 &  & 92 & SIM    &              & $\checkmark$ & 0.01 \\
28 & CIM     &              & $\checkmark$ & 0.00 &  & 61 & IOCA   &              & $\checkmark$ & 0.00 &  & 93 & SP     &              & $\checkmark$ & 0.00 \\
29 & CLA     &              & $\checkmark$ & 0.00 &  & 62 & IR     &              & $\checkmark$ & 0.00 &  & 94 & TA     &              & $\checkmark$ & 0.00 \\
30 & COP     &              & $\checkmark$ & 0.00 &  & 63 & ISFF   & $\checkmark$ &              & 0.01 &  & 95 & TBI    &              & $\checkmark$ & 0.00 \\
31 & CP      &              & $\checkmark$ & 0.00 &  & 64 & ISQ    &              & $\checkmark$ & 0.00 &  & 96 & TV     &              & $\checkmark$ & 0.00 \\
32 & CTO     &              & $\checkmark$ & 0.00 &  & 65 & IVC    & $\checkmark$ &              & 0.01 &  & 97 & VHP    &              & $\checkmark$ & 0.00 \\
33 & DAC     &              & $\checkmark$ & 0.00 &  &    &        &              &              &      &  &    &        &              &              &  \\ 
\bottomrule
\end{tabular}
}
}
\end{center}
\footnotesize{Notes: 
This table presents the individual factor test within the stepwise evaluation framework. With a 4-factor baseline model (FF3 + one "row" factor), corresponding to 94 models, we use FSE to expand it to an efficient model and BSE to remove redundant factors.
The table format follows Table \ref{tab:bench_row_CAPM}.
}
  \vspace{-0.2cm}
\end{table}

Table \ref{tab:bench_row_FF3} presents the individual factor test results with the 4-factor baseline model (FF3 + one "row" factor), corresponding to 94 (97-3) models. 
These results are similar to Table \ref{tab:bench_row_CAPM}, and the selection rates for our selected eight factors are all close to 1.
In particular, the selection rate of \texttt{HML} from FF3 is only 15\%, indicating our method can effectively remove a redundant factor from the baseline model. These findings demonstrate the robustness of our approach in choosing a small set of factors that simultaneously achieve the maximal Sharpe ratio and explain the cross section of unselected candidate factor returns, even when dealing with hundreds of factors.

\subsection{Machine Learning Baseline Models} 
\label{sec:PCA}

The recent machine learning literature employs PCA methods to create latent factors for dimension reduction and shrinkage methods (e.g., LASSO) for selecting observable factors.
To further apply our stepwise evaluation framework, we expand the factor list by incorporating PCA latent factors and further consider two strong baseline models, risk-premia-PCA (RP-PCA) of \cite{lettau2020factors} and reduced-rank approach (RRA) of \citet*{he2023shrinking}, both including high-dimensional portfolios in estimating principal components using the same sample.\footnote{The estimation includes extracting five RP-PCA factors from 285 basis portfolios and extracting five RRA factors from 97 factors using the same 285 basis portfolios.}

Additionally, LASSO is widely used for variable selection. We also consider two shrinkage methods: Lasso adopted by \citet*[FGX2020,][]{feng2020taming} and elastic-net regularization adopted by \citet*[KNS2020,][]{kozak2020shrinking} to obtain alternative baseline models.
Again, our stepwise evaluation can help add or remove factors to make these baseline models efficient.

Table \ref{tab:Reduced_other_method} displays the FSE/BSE reduced models based on PCA and LASSO baseline models, respectively. 
In Panel A, RP-PCA$^{F+B}$ retains all the RP-PCA factors and adds six additional factors to create the smallest adequate model, which includes four from our M8 model.
RRA$^{F+B}$ removes the \texttt{RRA4} factor from the baseline RRA model and adds four additional factors from our M8 model.
Compared to our M8 model in Table \ref{tab:Reduced_model}, RRA$^{F+B}$ does not include the \texttt{MKT} and \texttt{SMB} factors. 
The reason is that this \texttt{RRA1} factor has the largest loadings on the \texttt{MKT} and \texttt{SMB} factors, consistent with \citet{he2023shrinking}. 
These PCA latent factors, which combine information from multiple observable factors, result in FSE/BSE reduced models that differ from the 8-factor model due to changes in the list of factors.

\begin{table}[h!]
\caption{Reduced Models with Other Baseline Models}
\label{tab:Reduced_other_method}
\vspace{-0.4cm}
\begin{center}
\footnotesize{
\setlength{\tabcolsep}{0.15em}{
\begin{tabular}{ccccccccccc}
\toprule
\\
\multicolumn{5}{c}{\underline{Panel A : PCA Baseline Models}}                   &  & \multicolumn{5}{c}{\underline{Panel B : LASSO-based Baseline Models}}                  \\
\\
RP-PCA  & RRA  &  & RP-PCA$^{F+B}$ & RRA$^{F+B}$ & \hspace{1cm} & FGX2020 & KNS2020 &  & FGX2020$^{F+B}$ & KNS2020$^{F+B}$ \\  \hline
RP-PCA1 & RRA1 &  & \underline{RP-PCA1}        & \underline{RRA1}         &  & MKT     & MKT     &  & \underline{MKT}             & \underline{MKT}             \\
RP-PCA2 & RRA2 &  & \underline{RP-PCA2}        & \underline{RRA2}         &  & UMD     & SMB     &  & \underline{UMD}             & \underline{SMB}             \\
RP-PCA3 & RRA3 &  & \underline{RP-PCA3}        & \underline{RRA3}         &  & BAB     & BAB     &  & \underline{BAB}             & REG             \\
RP-PCA4 & RRA4 &  & \underline{RP-PCA4}        & \underline{RRA5}         &  & CIM     & ROE     &  & \underline{CIM}             & STR             \\
RP-PCA5 & RRA5 &  & \underline{RP-PCA5 }       & REG          &  & R5A     & ABR     &  & REG             & PEAD            \\
        &      &  & REG            & PEAD         &  &         & CIM     &  & STR             & HMLM            \\
        &      &  & HMLM           & STR          &  &         & dROE    &  & PEAD            & EPRD            \\
        &     &  & EPRD           & ILR          &  &         &         &  & HMLM            & ILR             \\
        &      &  & PEAD           &              &  &         &         &  & EPRD            &                 \\
        &      &  & HML            &              &  &         &         &  & BETA            &                 \\
        &      &  & 52W            &              &  &         &         &  &                 &                \\
\bottomrule
\end{tabular}
}}
\end{center}
\footnotesize{Notes: 
The table presents reduced models alongside other baseline models created using PCA and LASSO methods. 
Panel A includes the original RP-PCA and RRA models, along with their corresponding reduced models, whereas Panel B includes the FGX2020 and KNS2020 models, along with their corresponding reduced models.
Those underlined factors are from the baseline models.
}
  \vspace{-0.2cm}
\end{table}

In Panel B, the FGX2020 and KNS2020 methods select five and seven factors, respectively.
Through the FSE and BSE processes, FGX2020$^{F+B}$ removes \texttt{R5A} factor from the baseline model of FGX2020 and adds six additional factors. 
Furthermore, KNS2020$^{F+B}$ removes five factors from the baseline model of KNS2020 and adds six factors to obtain our 8-factor model.
These findings provide strong evidence for the robustness of our selected eight factors, as demonstrated by the stepwise evaluation in Table \ref{tab:Reduced_model}.

\begin{table}[h!]
\caption{Performance of Different Reduced Models with Other Methods}
\label{tab:BSE_performance_other_method}
\vspace{-0.4cm}
\begin{center}
\footnotesize{\setlength{\tabcolsep}{0.15em}{
\begin{tabular}{lcccccccccc}
\toprule
\\
                 &  & \multicolumn{5}{c}{\underline{Panel A : PCA Baseline Models }}                      & \multicolumn{4}{c}{\underline{Panel B : LASSO-based Baseline Models}}             \\ 
                 \\
                 &  & RP-PCA       & RRA          & RP-PCA$^{F+B}$ & RRA$^{F+B}$ &  \hspace{1cm}  & FGX2020      & KNS2020      & FGX2020$^{F+B}$ & KNS2020$^{F+B}$ \\ \hline
                 \\
\# Factor        &  & 5            & 5            & 11             & 8            &  & 5            & 7            & 10              & 8               \\
\\
                 & \multicolumn{10}{c}{\underline{Asset Pricing (unselected candidate factors)}}                                                                   \\
                 \\
$A|{\alpha|}$    &  & 0.24         & 0.31         & 0.13           & 0.13         &  & 0.31         & 0.33         & 0.15            & 0.15            \\
$A|{t(\alpha)|}$ &  & 1.9          & 2.49         & 0.87           & 0.85         &  & 2.24         & 2.41         & 0.94            & 0.93            \\
\# sign2         &  & 34           & 49           & 6              & 8            &  & 47           & 50           & 8               & 11              \\
Total $R^2$        &  & 33.8        & 37.0         & 42.3           & 38.8         &  & 15.8        & 28.1         & 38.5           & 32.9           \\
CS $R^2$        &  & 29.4        & 27.3         & 62.7           & 65.8         &  & 25.0        & 19.5        & 62.2           & 67.1           \\
\\
                 & \multicolumn{10}{c}{\underline{Asset Pricing (basis portfolios)}}                                                         \\
                 \\
$A|{\alpha|}$    &  & 0.11         & 0.11         & 0.1            & 0.09         &  & 0.2          & 0.14         & 0.16            & 0.1             \\
$A|{t(\alpha)|}$ &  & 1.53         & 1.38         & 1.26           & 0.95         &  & 1.7          & 1.66         & 1.29            & 0.96            \\
\# sign2         &  & 91           & 66           & 61             & 30           &  & 111          & 94           & 62              & 31              \\
Total $R^2$        &  & 71.9         & 65.4         & 74.2           & 65.2         &  & 13.26        & 56.6        & 41.7           & 59.0           \\
CS $R^2$        &  & 98.2         & 97.4         & 96.5           & 95.8         &  & 96.1        & 78.9        & 90.5           & 91.4           \\
\\
                 & \multicolumn{10}{c}{\underline{Investment}}                                                                                            \\
                 \\
AVG             &  & 1.03         & 0.43         & 0.72         & 0.88         &  & 1.49         & 1.28         & 1.18         & 1.16         \\
Ann.SR          &  & 1.49         & 1.17         & 2.93         & 2.6          &  & 1.33         & 1.48         & 2.8          & 2.73         \\
$\alpha_{\text{CAPM}}$ &  & 0.92$^{***}$ & 0.36$^{***}$ & 0.70$^{***}$ & 0.85$^{***}$ &  & 1.30$^{***}$ & 1.15$^{***}$ & 1.15$^{***}$ & 1.12$^{***}$ \\
$\alpha_{\text{FF5}}$  &  & 0.57$^{***}$ & 0.15$^{***}$ & 0.61$^{***}$ & 0.75$^{***}$ &  & 1.01$^{***}$ & 0.88$^{***}$ & 1.04$^{***}$ & 1.02$^{***}$   \\
\bottomrule
\end{tabular}
}}
\end{center}
\footnotesize{Notes: 
The table below compares different models to showcase their asset pricing and investment performance.
Panel A provides the performance results with PCA baseline models, whereas Panel B reports the results with LASSO-based baseline models.
The table format follows Table \ref{tab:FSE_performance}.}
  \vspace{-0.2cm}
\end{table}

Table \ref{tab:BSE_performance_other_method} compares the asset pricing and investment performance of the different models presented in Table \ref{tab:Reduced_other_method}.
All reduced models outperform baseline models in asset pricing, offering greater explanatory power with unselected candidate factors and portfolios.
For example, the RRA$^{F+B}$ model has three more factors than the RRA model. The absolute levels for unexplained alphas and their $t$-statistics for unselected factors can be shrunk to about 0.13\% and 0.85, which are economically and statistically insignificant. The CS $R^2$ values increase from 27.3\% to 65.8\%.
The LASSO-based model in Panel B exhibits comparable results.
In summary, our stepwise evaluation is effective for adopting other dimension reduction or factor selection techniques to develop an economically and statistically meaningful model.

\subsection{Out-of-sample Model Comparisons}

To address overfitting, we evaluate model performance using OOS data. Following \citet{fama2018choosing}, we employ a bootstrap procedure to compare the differences between INS and OOS Sharpe ratios across models.
The bootstrap process splits the 588-month sample period (Jan. 1973 to Dec. 2021) into 294 consecutive month pairs: (1, 2), $\cdots$, (587, 588). Each bootstrap run randomly assigns one month from each pair to INS and the other to OOS, drawing 294 pairs with replacement. For each bootstrap iteration, we calculate the INS Sharpe ratio ($\SR^2$) and apply the corresponding MVE portfolio weights to the unused months to estimate the OOS Sharpe ratio for the INS tangency portfolio.

\begin{table}[h!]
\caption{
Bootstrapped Maximal Squared Sharpe Ratios}
\label{tab:model_comparison_sr_boot}
\vspace{-0.4cm}
\begin{center}
\footnotesize{
\setlength{\tabcolsep}{0.2em}{
\begin{tabular}{lccccccccccccccc}
\toprule
\\
 & \multicolumn{15}{c}{\underline{Panel A: In-sample Bootstrap Results}}                                                               \\
\\
 &          &  & \multicolumn{11}{c}{\underline{Probability (\%) that the Row Performs Better than the Column}} &  &       \\
\\
        & M-SR$^2$ &  & FF3 & FF5  & RRA  & FF6  & FGX2020 & KNS2020 & RP-PCA & BS6  & DHS3 & Q5  & M8 &  & Best \\ \hline
CAPM    & 0.02  &  & 0   & 0    & 0    & 0    & 0       & 0       & 0      & 0    & 0    & 0   & 0  &  & 0    \\
FF3     & 0.04  &  &     & 0    & 0    & 0    & 0       & 0       & 0      & 0    & 0    & 0   & 0  &  & 0    \\
FF5     & 0.11  &  & 100 &      & 30.6 & 0    & 8.9     & 0.9     & 0.7    & 0.1  & 0.2  & 0   & 0  &  & 0    \\
RRA     & 0.13  &  & 100 & 69.4 &      & 15.8 & 12.4    & 1       & 0      & 0    & 0    & 0   & 0  &  & 0    \\
FF6     & 0.15  &  & 100 & 100  & 84.2 &      & 31.1    & 5.1     & 2.4    & 0.1  & 1.1  & 0   & 0  &  & 0    \\
FGX2020 & 0.16  &  & 100 & 91.1 & 87.6 & 68.9 &         & 9.2     & 16.9   & 11.1 & 4.4  & 0   & 0  &  & 0    \\
KNS2020 & 0.2   &  & 100 & 99.1 & 99   & 94.9 & 90.8    &         & 55.1   & 39.5 & 18.4 & 0.2 & 0  &  & 0    \\
RP-PCA  & 0.2   &  & 100 & 99.3 & 100  & 97.6 & 83.1    & 44.9    &        & 33.5 & 17.4 & 0   & 0  &  & 0    \\
BS6     & 0.21  &  & 100 & 99.9 & 100  & 99.9 & 88.9    & 60.5    & 66.5   &      & 25.6 & 0.1 & 0  &  & 0    \\
DHS3    & 0.24  &  & 100 & 99.8 & 100  & 98.9 & 95.6    & 81.6    & 82.6   & 74.4 &      & 0.9 & 0  &  & 0    \\
Q5      & 0.36  &  & 100 & 100  & 100  & 100  & 100     & 99.8    & 100    & 99.9 & 99.1 &     & 0  &  & 0    \\
M8      & 0.65  &  & 100 & 100  & 100  & 100  & 100     & 100     & 100    & 100  & 100  & 100 &    &  & 100  \\
\\
  & \multicolumn{15}{c}{\underline{Panel B: Out-of-sample Bootstrap Results} }                                                            \\
                     \\
 &          &  & \multicolumn{11}{c}{\underline{Probability (\%) that the Row Performs Better than the Column}} &  &       \\ 
                     \\
        & M-SR$^2$ &  & FF3 & FF5  & RRA  & FF6  & FGX2020 & KNS2020 & RP-PCA & BS6  & DHS3 & Q5  & M8 &  & Best \\ \hline
CAPM    & 0.02  &  & 22.2 & 0.6  & 0.7  & 0.3  & 0.2  & 0.2  & 0.1    & 0    & 0       & 0       & 0   &  & 0    \\
FF3     & 0.03  &  &      & 0.4  & 0.3  & 0.2  & 0.2  & 0.1  & 0      & 0    & 0       & 0       & 0   &  & 0    \\
FF5     & 0.08  &  & 99.6 &      & 31.3 & 9.8  & 15.2 & 4.2  & 1.9    & 0.3  & 0       & 0       & 0   &  & 0    \\
RRA     & 0.09  &  & 99.7 & 68.7 &      & 22.3 & 15   & 4.5  & 0.2    & 0.2  & 0       & 0       & 0   &  & 0    \\
FF6     & 0.11  &  & 99.8 & 90.2 & 77.7 &      & 32.1 & 10.6 & 0.7    & 0.3  & 0.1     & 0       & 0   &  & 0    \\
FGX2020 & 0.12  &  & 99.8 & 84.8 & 85   & 67.9 &      & 16.8 & 10.1   & 8.6  & 1       & 0.1     & 0   &  & 0    \\
KNS2020 & 0.15  &  & 99.9 & 95.8 & 95.5 & 89.4 & 83.2 &      & 34.1   & 26   & 2.3     & 0.1     & 0   &  & 0    \\
RP-PCA  & 0.16  &  & 100  & 98.1 & 99.8 & 99.3 & 89.9 & 65.9 &        & 37.8 & 7.5     & 0.2     & 0   &  & 0    \\
BS6     & 0.17  &  & 100  & 99.7 & 99.8 & 99.7 & 91.4 & 74   & 62.2   &      & 9.1     & 0       & 0   &  & 0    \\
DHS3    & 0.22  &  & 100  & 100  & 100  & 99.9 & 99   & 97.7 & 92.5   & 90.9 &         & 2.2     & 0   &  & 0    \\
Q5      & 0.32  &  & 100  & 100  & 100  & 100  & 99.9 & 99.9 & 99.8   & 100  & 97.8    &         & 0.2 &  & 0.2  \\
M8      & 0.56  &  & 100  & 100  & 100  & 100  & 100  & 100  & 100    & 100  & 100     & 99.8    &     &  & 99.8
\\
\bottomrule
\end{tabular}
}}
\end{center}
\footnotesize{Notes: 
This table presents bootstrap comparisons of squared Sharpe ratios across models. The 588-month sample period (Jan. 1973 - Dec. 2021) is divided into 294 adjacent pairs: months (1, 2), $\cdots$, (587, 588). Each bootstrap run randomly selects one month from each pair as INS and the other as OOS, drawing 294 pairs with replacement. 
INS $\SR^2$ is calculated using these paired months, with portfolio weights applied to unused months to estimate OOS Sharpe ratios for the INS tangency portfolio. The first column, "M-SR$^2$" reports the average $\SR^2$ from 1,000 INS and OOS simulation runs. The next eleven columns, labeled by model names, show the percentage of bootstrap simulations where the Sharpe ratio of the row model exceeds that of the column model. The final column, "Best", displays the percentage of simulations in which the row model achieves the highest squared Sharpe ratio among all models.
Panel A reports in-sample results, while Panel B presents out-of-sample results.
}
\vspace{-0.2cm}
\end{table}

This analysis yields three key metrics: average squared Sharpe ratios, pairwise outperformance probabilities, and frequencies of ``Best" performance. By integrating ridge regularization with rigorous OOS testing, this procedure mitigates overfitting and enables statistically robust comparisons across models. The approach ensures reliable rankings and robustness in model selection.

Table \ref{tab:model_comparison_sr_boot} presents the bootstrap comparison of squared Sharpe ratio differences across models. 
The M8 model emerges as the top performer, with a 99.8\% probability of outperforming Q5 and a 100\% probability against all other models. In contrast, traditional factor models, such as the CAPM, FF3, and FF5, consistently rank among the weakest performers, with negligible probabilities of being the best-performing model.
Machine learning approaches, including Lasso-based and PCA-based models, significantly outperform early factor models, such as the CAPM, FF3, and FF5. While not as dominant as M8, newer models such as DHS3 and Q5 demonstrate strong performance, with probabilities nearing unity for outperforming both Lasso-based and PCA-based models. Notably, Q5 retains a minimal 0.02\% probability of being the top performer across simulations.
These results highlight the positive and robust performance of the M8 model to span the investment opportunity set in OOS analyses.

\subsection{Robustness Analysis with Transaction Costs}
Portfolio analysis must account for transaction costs to enhance the robustness of investment evaluation. We conduct a straightforward exercise, subtracting transaction costs directly from factor returns to evaluate their impact.
Following \cite{kelly2023modeling}, we incorporate one-way trading costs based on rebalancing frequency. 
For annually rebalanced factors, a deduction of 2 basis points (bps) is applied to monthly returns. 
Monthly rebalanced factors incur trading costs of either 12 or 24 bps.
One can anticipate that both the LHS unselected candidate factors and the RHS factor models incur a fixed cost, increasing the likelihood of annual rebalanced factors entering the RHS model.
In the M8 model, \texttt{MKT} is a buy-and-hold factor with no transaction costs, \texttt{SMB} and \texttt{EPRD} are rebalanced annually, while the remaining factors (\texttt{REG}, \texttt{PEAD}, \texttt{HMLM}, \texttt{STR}, \texttt{ILR}) are rebalanced monthly.

\begin{table}[h!]
\caption{Reduced Factor Models with Transaction Costs (CAPM Example)} \label{tab:cost}
\vspace{-0.4cm}
\begin{center}
\footnotesize{\setlength{\tabcolsep}{1.5em}{
\begin{tabular}{ccccccc}
\toprule
\\
\underline{Panel A: 0 bps} &  & \multicolumn{2}{c}{\underline{Panel B: 12 bps}}   &  & \multicolumn{2}{c}{\underline{Panel C: 24 bps}}    \\
\\
M8               &  & FF5$^{F+B}$       & Q5$^{F+B}$       &  & FF5$^{F+B}$       & Q5$^{F+B}$        \\ \hline
MKT              &  & MKT               & MKT              &  & MKT               & MKT               \\
REG              &  & \underline{SMB}  & \underline{SMB} &  & \underline{SMB}  & \underline{SMB}  \\
PEAD             &  & \underline{CMA}  & \underline{IA}  &  & \underline{CMA}  & \underline{IA}   \\
HMLM             &  & REG               & REG              &  & REG               & ROE               \\
STR              &  & \underline{EPRD} & PEAD             &  & \underline{EPRD} & REG               \\
ILR              &  & R5A               & STR              &  & LTR               & \underline{EPRD} \\
\underline{SMB}              &  & PEAD              & ILR              &  & \underline{LIQ}  & R5A               \\
\underline{EPRD}             &  & HMLM              &                  &  & \underline{OCP}  & LTR               \\
                 &  &                   &                  &  & \underline{dWC}  & \underline{OCP}  \\
             \bottomrule
\end{tabular}
}}
\end{center}
\footnotesize{Notes: 
The table presents an evaluation of FSE/BSE reduced models across different transaction cost scenarios: zero transaction costs (Panel A), 12 basis points per month (Panel B), and 24 basis points per month (Panel C). Panels B and C exclusively examine two standard baseline factor models, FF5 and Q5. Annual rebalancing factors are denoted by an underline.
}
  \vspace{-0.2cm}
\end{table}

Table \ref{tab:cost} summarizes the FSE/BSE reduced models incorporating transaction cost effects. Key findings are highlighted below:
(1) Annually rebalanced factors (\texttt{SMB}, and \texttt{EPRD}) from the M8 model exhibit minimal sensitivity to transaction costs, suggesting low cost impact for annual rebalancing.
(2) Monthly rebalanced factors in the M8 model are more sensitive to transaction costs, leading to their replacement by annually rebalanced factors in some cases. 
For example, in Panel B's FF5$^{F+B}$ model, the monthly rebalanced \texttt{STR} and \texttt{ILR} are substituted with the annually rebalanced \texttt{CMA} and the monthly rebalanced \texttt{R5A}.
(3) As monthly transaction costs increase, annually rebalanced factors become more prevalent. In Panel C's FF5$^{F+B}$ model, 6 out of 8 factors (excluding \texttt{MKT}) are annually rebalanced, compared to 3 out of 7 factors in Panel B.
These findings underscore the need to account for transaction costs in factor model evaluation, suggesting that annual rebalancing factors as the most resilient.

\section{Conclusion} \label{sec:conclusion}

The asset pricing literature emphasizes the importance of factor models that minimize cross-sectional pricing errors. However, these models often neglect unselected candidate factors that could improve the performance of test assets. Incorporating these unselected factors into the test assets enhances the robustness of pricing tests.
To address these shortcomings, this paper proposes a framework for factor model selection and testing by (i) selecting the optimal model that spans the joint efficient frontier of test assets and all candidate factors, and (ii) testing pricing performance on both test assets and unselected candidate factors.

Our framework updates a baseline model (e.g., CAPM or Fama-French factor models) sequentially by adding or removing factors based on asset pricing tests. 
First, the forward stepwise evaluation (FSE) method sequentially adds factors from a pool of candidate factors to the baseline model. 
Second, the backward stepwise evaluation (BSE) method sequentially removes redundant factors from efficient but large models, mitigating the greedy nature of FSE against model overfitting.
Factor inclusion and exclusion are guided by 
model comparison 
metrics and asset pricing tests, ensuring optimal model selection. This framework enhances the model selection literature by offering a systematic approach tailored to asset pricing.

Analyzing the "factor zoo" in U.S. equities (1973–2021), our optimal 8-factor model (\texttt{MKT}, \texttt{REG}, \texttt{PEAD}, \texttt{HMLM}, \texttt{STR}, \texttt{ILR}, \texttt{SMB}, \texttt{EPRD}) effectively prices unselected candidate factors and is not rejected by asset pricing tests.
This framework demonstrates exceptional asset pricing performance and (out-of-sample) investment performance compared to alternative models. 
Furthermore, the stepwise evaluation based on mean-variance efficiency offers broader applications, including portfolio optimization and the evaluation of mutual fund performance.

\vspace{1cm}
\onehalfspacing{\small{\putbib}}
\end{bibunit}

\clearpage

\appendixpage
\appendix
\setcounter{page}{1}%

\renewcommand\thesection{\Alph{section}}
\renewcommand\thesubsection{\thesection.\arabic{subsection}}

\renewcommand\thefigure{A.\arabic{figure}}    
\setcounter{figure}{0}

\renewcommand\thetable{A.\arabic{table}}    
\setcounter{table}{0}

\renewcommand\theequation{A.\arabic{equation}}    
\setcounter{equation}{0}

\begin{bibunit}




\section{Additional Tables} \label{sec:description_data}

{\onehalfspacing
\begin{table}[H]
\caption{Description for the Factor Zoo}
\label{tab:factor_zoo}

\footnotesize{Notes: The table provides a comprehensive overview of the factors evaluated in the paper: monthly average returns and annualized Sharpe ratios from 1973 to 2021. }

\begin{center}
\scriptsize{

\begin{tabular}{lllccc}
\toprule
ID & Abbreviation & Long Description & Avg.Ret (\%) & Ann.SR (\%) \\ \hline
1  & MKT  & Excess market return                & 0.62 & 46.88  \\
2  & SMB  & Small minus big                     & 0.19 & 22.14  \\
3  & HML  & High minus low                      & 0.28 & 31.65  \\
4  & RMW  & Robust minus weak                   & 0.29 & 43.99  \\
5  & CMA  & Conservative minus aggressive       & 0.3  & 52.89  \\
6  & UMD  & Momentum on prior (2-12)            & 0.61 & 48.2   \\
7  & HMLM & HML Devil                           & 0.3  & 28.34  \\
8  & QMJ  & Quality minus junk                  & 0.39 & 56.76  \\
9  & BAB  & Betting against beta                & 0.86 & 85.99  \\
10 & LIQ  & Liquidity                           & 0.4  & 38.98  \\
11 & STR  & Short-term reversal on prior (1-1)  & 0.44 & 46.4   \\
12 & LTR  & Long-term reversal on prior (13-60) & 0.17 & 22.51  \\
13 & REG  & Expected growth                     & 0.8  & 134.49 \\
14 & VOL  & Return volatility                   & 0.43 & 18.08  \\
15 & SUE  & Standard unexpected earnings        & 0.44 & 43.83  \\
16 & IA   & Investment factor                   & 0.34 & 60.64  \\
17 & ROE  & Equity factor returns               & 0.53 & 69.16  \\
18 & IMD  & Intermediary                        & 1.02 & 52.02  \\
19 & NI   & Net share issues                    & 0.44 & 47.75  \\
20 & BETA & Market beta                         & 0.13 & 6.69   \\
21 & PEAD         & Post-earnings announcement drift                                & 0.61         & 110.74      \\
22 & FIN          & Financing factor                                                & 0.69         & 61.77       \\
23 & $S^{PLS}$    & Aligned investor sentiment index                                & 0.36         & 28.71       \\
24 & ABR          & CAR around earnings announcement                                & 0.71         & 74.36       \\
25 & ACI          & Abnormal corporate investment                                   & 0.23         & 25.06       \\
26 & ATO          & Assets turnover                                                 & 0.45         & 36.03       \\
27 & CEI          & Composite equity issuance                                       & 0.53         & 46.62       \\
28 & CIM          & Customer industries momentum                                    & 0.79         & 49.06       \\
29 & CLA          & Cash-based operating profits-to-lagged assets                   & 0.57         & 49.48       \\
30 & COP          & Operating cash flow-to-assets                                   & 0.65         & 54.47       \\
31 & CP           & Cash flow-to-price                                              & 0.35         & 24.99       \\
32 & CTO          & Capital turnover                                                & 0.34         & 31.76       \\
33 & DAC          & Discretionary accruals                                          & 0.43         & 47.31       \\
34 & dBE          & Changes in book equity                                          & 0.26         & 24.25       \\
35 & dCOA         & Changes in current operating assets                             & 0.22         & 22.67       \\
36 & dFIN         & Changes in net financial assets                                 & 0.22         & 26.51       \\
37 & dFNL         & Changes in financial liabilities                                & 0.25         & 33.45       \\
38 & dII          & Percent changes in investment relative to industry              & 0.24         & 28.61       \\
39 & dLNO         & Changes in long-term net operating assets                       & 0.31         & 32.37       \\
40 & dNCA         & Changes in non-current operating assets                         & 0.36         & 41.08       \\
41 & dNCO         & Changes in net non-current operating assets                     & 0.35         & 41.66       \\
42 & dNOA         & Changes in net operating assets                                 & 0.42         & 44.7        \\ 
\bottomrule
\end{tabular}}
\end{center}
\end{table}}

\begin{table}[H]
\begin{center}
    
{\onehalfspacing
\scriptsize{
\begin{tabular}[h]{lllccc}
\toprule
ID & Abbreviation & Long Description & Avg.Ret (\%) & Ann.SR (\%) \\ \hline
43 & DP           & Dividend yield                                                  & 0.21         & 12.69       \\
44 & dPIA         & Changes in PPE and inventory scaled by lagged assets            & 0.39         & 41.52       \\
45 & dROA         & 4-quarter changes in return on assets                           & 0.51         & 49.55       \\
46 & dROE         & 4-quarter changes in return on equity                           & 0.77         & 77.94       \\
47 & DUR          & Equity duration                                                 & 0.3          & 21.98       \\
48 & dWC          & Changes in net non-cash working capital                         & 0.4          & 43.56       \\
49 & EBP          & Enterprise book-to-price                                        & 0.27         & 18.52       \\
50 & EM           & Enterprise multiple                                             & 0.4          & 31.12       \\
51 & EP           & Earnings-to-price                                               & 0.29         & 21.21       \\
52 & EPRD         & Earnings predictability                                         & 0.68         & 54.13       \\
53 & ETL          & Earnings timeliness                                             & 0.2          & 21.99       \\
54 & ETR          & Effective tax rate                                              & 0.22         & 27.69       \\
55 & GPA          & Gross profits-to-assets                                         & 0.39         & 40.67       \\
56 & HS           & Industry concentration in sales                                 & 0.27         & 25.4        \\
57 & IG           & Investment growth                                               & 0.44         & 47.6        \\
58 & ILE          & Industry lead-lag effect in earnings surprises                  & 0.5          & 43.09       \\
59 & ILR          & Industry lead-lag effect in prior returns                       & 0.58         & 38.59       \\
60 & IM           & Industry momentum                                               & 0.64         & 36.36       \\
61 & IOCA         & Industry-adjusted organizational capital-to-assets              & 0.53         & 57.9        \\
62 & IR           & Intangible return                                               & 0.3          & 20.77       \\
63 & ISFF         & Idiosyncratic skewness based on FF3 factor model                & 0.3          & 43.06       \\
64 & ISQ          & Idiosyncratic skewness based on Q-factor model                  & 0.3          & 41.85       \\
65 & IVC          & Inventory changes                                               & 0.37         & 39.72       \\
66 & IVFF         & Idiosyncratic volatility based on FF3 factor model              & 0.5          & 23.91       \\
67 & IVG          & Inventory growth                                                & 0.29         & 32.94       \\
68 & IVQ          & Idiosyncratic volatility based on Q-factor model                & 0.44         & 21.4        \\
69 & NDF          & Net debt financing                                              & 0.24         & 30.57       \\
70 & NEI          & \# quarters with consecutive earnings increase                  & 0.26         & 32.07       \\
71 & NOA          & Net operating assets                                            & 0.53         & 58.97       \\
72 & NOP          & Net payout yield                                                & 0.51         & 42.13       \\
73 & OA           & Operating accruals                                              & 0.26         & 28.38       \\
74 & OCA          & Organizational capital-to-assets                                & 0.57         & 41.87       \\
75 & OCP          & Operating cash flow-to-price                                    & 0.54         & 37.14       \\
76 & OL           & Operating leverage                                              & 0.41         & 38.41       \\
77 & OLE          & Quarterly operating profits-to-lagged book equity               & 0.63         & 44.85       \\
78 & OP           & Payout yield                                                    & 0.36         & 28.36       \\
79 & OPA          & Operating profits-to-assets                                     & 0.5          & 40.58       \\
80 & OPE          & Operating profits-to-book equity                                & 0.27         & 19.48       \\
81 & 52W          & 52-week high                                                    & 0.48         & 23.7        \\
82 & PDA          & Percent discretionary accruals                                  & 0.47         & 53.46       \\
83 & POA          & Percent operating accruals                                      & 0.42         & 47.48       \\
84 & PTA          & Percent total accruals                                          & 0.38         & 39.7        \\
85 & R1A          & Seasonality, return in month t-12                               & 0.53         & 38.52       \\
86 & R5A          & Seasonality, years 2-5 lagged returns                           & 0.71         & 59.86       \\
87 & RER          & Industry-adjusted real estate ratio                             & 0.37         & 35.01       \\
88 & Resid        & 11-month residual momentum                                      & 0.53         & 46.42       \\
89 & ROA          & Return on assets                                                & 0.57         & 40.89       \\
90 & RS           & Revenue surprises                                               & 0.39         & 38.46       \\
91 & SG           & Quarterly sales growth                                          & 0.33         & 25.63       \\
92 & SIM          & Supplier industries momentum                                    & 0.74         & 47.17       \\
93 & SP           & Sales-to-price                                                  & 0.45         & 32.92       \\
94 & TA           & Total accruals                                                  & 0.26         & 28.8        \\
95 & TBI          & Quarterly tax income-to-book income                             & 0.18         & 22.84       \\
96 & TV           & Total volatility                                                & 0.31         & 13.49       \\
97 & VHP          & Roe-based intrinsic value-to-market                             & 0.26         & 19.53  \\ 
\bottomrule
\end{tabular}}}
\end{center}
\end{table}

\clearpage
\section{Stepwise Evaluation Method} \label{sec:method_detail}

Our stepwise evaluation framework includes Forward Stepwise Evaluation (FSE) and Backward Stepwise Evaluation (BSE). 
To expand the model to make it efficient, FSE sequentially shifts a candidate factor from the LHS unselected candidate factors to the RHS (inefficient) selected factors.
To reduce a large efficient model, BSE sequentially shifts a redundant factor from the RHS (efficient) to the LHS.

\subsection{Forward Stepwise Evaluation} \label{sec:FSE_detail}
Starting from a baseline model $\mathcal{M}_b \subset \mathcal{F} $,
in the $k$-th forward iteration, where $k = 1, \cdots, N-|\mathcal{M}_b|-1$, we are given $\mathcal{S}_{(k-1)}$, where $\mathcal{S}_{(k-1)} \subset \mathcal{F}$ denote the currently $(k-1)$-th selected factor model and $\mathcal{S}_{(0)} = \mathcal{M}_b$. For every $j \in \mathcal{F} \backslash  \mathcal{S}_{(k-1)} = \widetilde{\mathcal{S}}_{(k-1)}$, 
we construct a expanded factor model $\mathcal{M}_{j (k-1)} = \mathcal{S}_{(k-1)} \cup \{ j \} $,
we regress $\mathbf{F}_{\{ \widetilde{\mathcal{M}}_{j (k-1)} \} t }$ on $ \mathbf{F}_{\{\mathcal{M}_{j (k-1)} \} t}$ as in Eq. (\ref{eq: true_model}) and calculate the GRS value $ \mbox{GRS}\{\mathcal{M}_{j (k-1)}\}$ for the corresponding model $\mathcal{M}_{j (k-1)}$ in Eq. (\ref{eq:GRS_j}).
We see that the $k$-th factor added is
  \vspace{-0.2cm}
\begin{equation*}
a_{k} = \arg \min_{j \in \widetilde{\mathcal{S}}_{(k-1)}} \mbox{GRS}\{\mathcal{M}_{j (k-1)}\} \mbox{~or~} a_{k} = \arg \max_{j \in \widetilde{\mathcal{S}}_{(k-1)}} \SR^2\{\mathbf{F}_{(\mathcal{M}_{j (k-1)} ) }\}. 
  \vspace{-0.2cm}
\end{equation*}

Path algorithms can then be seen as providing us with an ordering of the factors $a_1, a_2, \cdots \in \{1, \cdots, N - |\mathcal{M}_b|\}$ along with a sequence of $k$ nested factor models contained in some upper model $\mathcal{S}_{(N-|\mathcal{M}_b|)}$:
  \vspace{-0.2cm}
\begin{equation}\label{eq: solution_path}
\mathcal{M}_b = \mathcal{S}_{(0)} \subset \mathcal{S}_{(1)} \subset \cdots \subset  \mathcal{S}_{(k)}  \cdots \subset \mathcal{S}_{(N-|\mathcal{M}_b|)} \mbox{~with~} \mathcal{S}_{(k)} = \{\mathcal{M}_b, a_1, \cdots, a_k\}.
  \vspace{-0.2cm}
\end{equation}

The $(k)$-th ordered hypothesis tests whether or not adding the $k$-th factor $a_k$ was informative.
We can test the vector of alphas from spanning regression by using the unselected $N- |\mathcal{M}_b| - k$ candidate factors on the selected model $\mathcal{S}_{(k)}$ in Eq. (\ref{eq: solution_path}).

As mentioned earlier, the GRS test requires that the number of test assets is much smaller than the number of observations. Even if the number of test assets is slightly larger, the power of the GRS test may be compromised.
For this reason, this paper uses a new HDA test from \cite{pesaran2023testing} in Eq. (\ref{eq: FSE}).

The simplest way is to keep rejecting until the first time that $\mbox{HDA}\{\mathcal{S}_{(k)}\}  < z_{\lambda} $,
  \vspace{-0.2cm}
\begin{equation*}
\widehat{k} = \min\{ k: \mbox{HDA}\{\mathcal{S}_{(k)}\} < z_{\lambda} \} \mbox{~and~} \mathcal{M}_b^F = \mathcal{S}_{(\widehat{k})},
  \vspace{-0.2cm}
\end{equation*}
where $z_{\lambda}$ is the threshold value of the HDA test based on its asymptotic distribution, and $\lambda$ is a pre-specified significance level. $\mathcal{M}_b^F$ refers to the expanded efficient model (the final model of FSE process) constructed based on the baseline model $\mathcal{M}_b$.
Motivated by this concept, we propose the FSE method to obtain an efficient factor model. The detailed Algorithm \ref{alg: FSE} is provided below.

\begin{algorithm}[h!]
\caption{Forward Stepwise Evaluation} \label{alg: FSE}
	\bigskip
	\textbf{Input} $N \times T$ factors return matrix $\mathbf{F}$, a baseline model $\mathcal{M}_b$ \\
	\textbf{Output} The expanded efficient model $\mathcal{M}_b^F$
	\begin{algorithmic}[1]
		\Procedure{FSE} {{$\mathbf{F}$, $\mathcal{M}_b$}}
		\State Initial $\mathcal{S}_{(0)} = \mathcal{M}_b$, $\widetilde{\mathcal{S}}_{(0)} = \mathcal{F} \backslash \mathcal{M}_b$
		\For{$k$ from $1$ to $N-|\mathcal{M}_b|-1$} \Comment{Loop over number of iterations}
		\For{ every $j \in  \widetilde{\mathcal{S}}_{(k-1)}$} \Comment{Loop over current all models}
		\State Construct a candidate factor model $$\mathcal{M}_{j (k-1)} = \mathcal{S}_{(k-1)} \cup \{ j \}. $$
		\State Regress $\mathbf{F}_{\{ \widetilde{\mathcal{M}}_{j (k-1)} \} t }$ on $ \mathbf{F}_{\{\mathcal{M}_{j (k-1)} \} t}$ as in Eq. (\ref{eq: true_model}).
		\State Calculate the GRS value $\mbox{GRS}_{j(k-1)} = \mbox{GRS}\{\mathcal{M}_{j (k-1)}\}$ in Eq. (\ref{eq:GRS_j}).
		\EndFor
		\State Find the best factor and minimizes GRS value $$a_k = \arg \min_{j \in \widetilde{\mathcal{S}}_{(k-1)}} \mbox{GRS}_{j(k -1)}.$$
		\State Update $\mathcal{S}_{(k)} = \mathcal{S}_{(k-1)} \cup \{a_k \} $, $\widetilde{\mathcal{S}}_{(k)} = \mathcal{F} \backslash \mathcal{S}_{(k)} $.
  \State  Regress $\mathbf{F}_{\{ \widetilde{\mathcal{S}}_{(k)}\} t }$ on $ \mathbf{F}_{\{\mathcal{S}_{(k)} \} t}$ as in Eq. (\ref{eq: true_model}).
		\State Calculate the HDA value $\mbox{HDA}\{\mathcal{S}_{(k)}\}$ in Eq. (\ref{eq: FSE})
		\If {$\mbox{HDA}\{\mathcal{S}_{(k)}\} < z_\lambda$} \Comment{Stop rule
		\State Stop
		\EndIf
		}
		\EndFor
		\State \textbf{return} $\mathcal{M}_b^F = \mathcal{S}_{(k)}$
		\EndProcedure
	\end{algorithmic}
\end{algorithm}
  \vspace{-0.2cm}

Note that it is unrealistic to require $P (\mathcal{T} = \mathcal{M}_b^F) \to 1$ under the FSE, because this is not guaranteed even in the fixed dimension situation.
However, it is indeed possible to have $\mathcal{T} \subset \mathcal{M}_b^F$.
Otherwise, at least one risk factor is completely missed by the selection result in $\mathcal{M}_b^F$.
We have the following theorem that shows the
screening consistency of our proposed FSE algorithm.

\begin{theorem}[Screening Consistency]  \label{theorem1}
Under model (\ref{eq: true_model}) and Assumptions \ref{assump1}-\ref{assump4} in Appendix \ref{sec:assumption}, suppose the threshold value of the HDA test satisfies  $z_{\lambda}^1  \lesssim T N^{-1/2- n_2 } $, we have
  \vspace{-0.2cm}
\begin{equation*}
P \big(\mathcal{T} \subset \mathcal{M}_b^F \big) \to 1,
  \vspace{-0.2cm}
\end{equation*}
as $(N, T) \to \infty$, and $|\mathcal{M}_b^F|  =O(N^{n_1+2n_2})$ with $n_1 + 2n_2 < 1/3$, where $n_1$ and $n_2$ are defined in Assumption \ref{assump1}.
\end{theorem}

Appendix \ref{sec:proof_th1} provides the proof of Theorem \ref{theorem1}. This theorem demonstrates that, given appropriate conditions, the FSE can detect all risk factors. 
As a result, the proposed approach enjoys the screening consistency property. It is important to control the model size moving forward.
With a high probability, FSE can detect the optimal model in significantly fewer steps than the total number of factors ($N$) with a complexity of $ O(N^{n_1+2n_2}) $. This suggests that overfitting is well-controlled.

\subsection{Backward Stepwise Evaluation} \label{sec:backward}
Forward stepwise regression is an algorithm commonly used to build a sequence of nested models, each incorporating additional factors. The BSE method is introduced to provide an exit mechanism to remove redundant factors.
Starting from an efficient baseline model $\mathcal{M}_b^F$,
in the $m$-th backward iteration, where $m = 1, \cdots, |{\mathcal{M}}_b^F|-1$, we are given ${\mathcal{S}}_{(m-1)}\subset {\mathcal{M}}^F_b$ denote the currently $(m-1)$-th included factor model and $\mathcal{S}_{(0)} = \mathcal{M}_b^F$. 
Then, for every $j \in  {\mathcal{S}}_{(m-1)} $,  we construct a reduced factor model ${\mathcal{M}}_{j (m-1)} = {\mathcal{S}}_{(m-1)} \backslash \{ j \}.$
we regress unselected candidate factors $\widetilde{\mathcal{M}}_{j (m-1)}$ on model ${\mathcal{M}}_{j (m-1)}$, 
and calculate the GRS value as $\mbox{GRS}\{{\mathcal{M}}_{j (m-1)}\}$.
For $j \in {\mathcal{S}}_{(m -1)} $, we see that the $(m)$-th factor removed is
  \vspace{-0.2cm}
\begin{equation*}
b_{m} = \arg \min_{j \in {\mathcal{S}}_{(m -1)}}  \mbox{GRS}\{ {\mathcal{M}}_{j (m-1)}\} \mbox{~or~} b_{m} = \arg \max_{j \in {\mathcal{S}}_{(m-1)}} \SR^2\{\mathbf{F}_{(\mathcal{M}_{j (m-1)}) }\}.
  \vspace{-0.2cm}
\end{equation*}

Path algorithms can then be seen as providing us with an ordering of the removed factors $b_1, b_2, \cdots \in \{1, \cdots, |{\mathcal{M}}_b^F|-1\}$ along with a sequence of $m$ nested models contained in the upper model ${\mathcal{M}}_b^F$:
  \vspace{-0.2cm}
\begin{equation}\label{eq: solution_path_BSE}
\mathcal{M}_b^F	= {\mathcal{S}}_{(0)} \supset {\mathcal{S}}_{(1)} \supset \cdots \supset  {\mathcal{S}}_{(m)}  \cdots \supset \emptyset, \mbox{~with~} {\mathcal{S}}_{(m)} =  \mathcal{M}_b^F \backslash \{ b_1, \cdots, b_m\}.
  \vspace{-0.2cm}
\end{equation}

We can test model $\mathcal{S}_{(m)}$ in (\ref{eq: solution_path_BSE}) using the unselected $N- |\mathcal{M}_b^F| + m$ candidate factors as test assets.
Similar to the FSE method, the stopping procedure is to keep not rejecting until the first time that $\mbox{HDA}\{ {\mathcal{S}}_{(m)} \} > z_\lambda$,
  \vspace{-0.2cm}
\begin{equation*}
\widehat{m} = \min\{ m: \mbox{HDA}\{ {\mathcal{S}}_{(m)} \} >  z_{\lambda} \} \mbox{~and~} \mathcal{M}_b^{F+B} = {\mathcal{S}}_{(\widehat{m}  -1)}.
  \vspace{-0.2cm}
\end{equation*}
Define the finally selected factor set as $\mathcal{M}_b^{F+B}$. Motivated by this concept, we propose the BSE method for selecting the optimal model. The detailed algorithm \ref{alg: BSE} is below.

\begin{algorithm}[h!]
	\caption{Backward Stepwise Evaluation}\label{alg: BSE}
	\bigskip
        \textbf{Input} $N \times T$ factors return matrix $\mathbf{F}$, an efficient model $\mathcal{M}_b^F$ \\
	\textbf{Output} The smallest efficient model $\mathcal{M}_b^{F+B}$
	\begin{algorithmic}[1]
		\Procedure{BSE} {{$\mathbf{F}$, $\mathcal{M}_b^F$}}
            \State Initial ${\mathcal{S}}_{(0)} = \mathcal{M}_b^F$, $\widetilde{\mathcal{S}}_{(0)} = \mathcal{F} \backslash \mathcal{M}_b^F$
		\For{$m$ from $1$ to $|\mathcal{M}_b^F| - 1$} \Comment{Loop over number of iterations}
		\For{ every $j \in  {\mathcal{S}}_{(m -1)}$} \Comment{Loop over current all models}
            \State Construct a candidate factor model $${\mathcal{M}}_{j (m-1)} = {\mathcal{S}}_{(m-1)} \backslash \{ j \}. $$
            \State Regress $\mathbf{F}_{\{ \widetilde{\mathcal{M}}_{j (m-1)} \} t } $ on $\mathbf{F}_{\{ {\mathcal{M}}_{j (m-1)} \} t } $ as in Eq. (\ref{eq: true_model}).
            \State Calculate the GRS value  $\mbox{GRS}_{j (m-1)} = \mbox{GRS}\{ {\mathcal{M}}_{j (m-1)}\}$ in Eq. (\ref{eq:GRS_j}).
		\EndFor
		\State Find the worst factor and minimizes GRS value $$b_m = \arg \min_{j \in {\mathcal{S}}_{(m -1)}} \mbox{GRS}_{j (m-1)}.$$
            \State Update ${\mathcal{S}}_{(m)} = {\mathcal{S}}_{(m-1)} \backslash \{b_m \} $, $\widetilde{\mathcal{S}}_{(m)} = \mathcal{F} \backslash \mathcal{S}_{(m)} $.
              \State  Regress $\mathbf{F}_{\{\widetilde{\mathcal{S}}_{(m)}\} t }$ on $ \mathbf{F}_{\{{\mathcal{S}}_{(m)}\} t}$ as in Eq. (\ref{eq: true_model}).
              \State Calculate the HDA value $\mbox{HDA}\{\mathcal{S}_{(m)}\}$ in Eq. (\ref{eq: FSE})
            \If {$\mbox{HDA} \{{\mathcal{S}}_{(m)}\} > z_\lambda$} \Comment{Stop rule
            \State Stop
		\EndIf
  }
		\EndFor
		\State \textbf{return} $\mathcal{M}_b^{F+B} = {\mathcal{S}}_{(m -1)}$
		\EndProcedure
	\end{algorithmic}
\end{algorithm}
  \vspace{-0.2cm}

\begin{theorem}[Selection Consistency]  \label{theorem2}
Under model (\ref{eq: true_model}) and Assumptions \ref{assump1}-\ref{assump4} in Appendix \ref{sec:assumption}, 
suppose the threshold value of the HDA test satisfies $\sqrt{N} \log N  \lesssim z_{\lambda}^2  \lesssim T N^{-1/2- n_2} $,
we have
  \vspace{-0.2cm}
\begin{equation*}
P \big(\mathcal{M}_b^{F+B} = \mathcal{T} \big) \to 1,
  \vspace{-0.2cm}
\end{equation*}
as $(N, T) \to \infty$.
\end{theorem}
Appendix \ref{sec:proof_th2} contains the proof of Theorem \ref{theorem2}. 
This theorem indicates that the BSE can consistently identify the optimal model as the sample size $(N, T)$ approaches infinity. It illustrates that, under suitable conditions, the BSE can consistently select all risk factors, thereby demonstrating the selection consistency property of the proposed approach.

\subsection{Illustration Examples}\label{sec:SE_Illustration}

\begin{figure}[h]
	\caption{Illustration Example of Stepwise Evaluation}
\label{fig:SE_Illustration}
\vspace{-0.4cm}
\begin{center}
	\includegraphics[ width=0.8\textwidth]{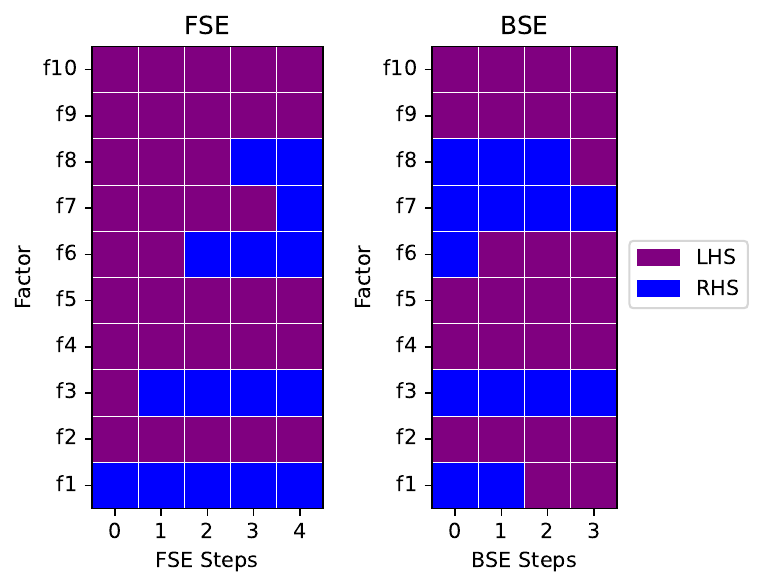}
 \end{center}
   \vspace{-0.2cm}
\footnotesize{
Notes: 
We provide illustration examples for the stepwise evaluation process using 10 candidate factors. 
The left graph shows the FSE process, and the right graph shows the BSE process. 
The horizontal axis indicates the steps FSE or BSE, while the vertical axis represents the factors. Purple squares on the left indicate unselected candidate factors as test assets, while blue squares on the right indicate those selected factors.
}
  \vspace{-0.2cm}
\end{figure}

Figure \ref{fig:SE_Illustration} provides illustration examples for the stepwise evaluation process using 10 candidate factors.
Consider an example of the baseline model with one pre-selected factor, $\{f_{1} \}$.
If the baseline model $\{f_{1}\}$ is rejected by the asset pricing test when pricing the unselected nine factors, we then shift one "potential" factor from the LHS to the RHS.
In step 1,  the baseline model (Step 0) is $\{f_{1}\}$ on the RHS. We can expand this model by adding another factor $\{f_{j} \}$ from the LHS and denote the expanded 2-factor model as $\{f_{1},f_{j}\}$ for each $ j = 2, \cdots, 10$.
Therefore, nine candidate non-nested expanded models require model comparisons. 
We calculate their GRS values in Eq. \eqref{eq:GRS_j} or $\SR^2$ for the expanded models $\{f_{1},f_{j}\}$, and define them as ($\mbox{GRS}_{2}, \mbox{GRS}_{3}, \cdots,\mbox{GRS}_{10}$) or ($\mbox{SR}^2_{2}, \mbox{SR}^2_{3},\cdots,\mbox{SR}^2_{10}$).
The selected expanded factor model has Min $\mbox{GRS}_j$ or Max $\mbox{SR}^2_j$ in step 1, and we suppose the most difficult-to-price factor is $f_{3}$, for example.
We need to test the model efficiency for $\{f_1, f_3\}$ to price the unselected eight factors.
If the asset pricing test is rejected, we add another factor.

In step 2,  the expanded baseline model is $\{f_1, f_3\}$. We can add another unselected candidate factor, $f_{j}$,  and denote the expanded 3-factor model as $\{f_{1}, f_{3},f_{j}\}$ for each $j = 2,4,\cdots, 10$.
We then calculate the GRS value or $\SR^2$ for $\{f_{1}, f_{3},f_{j}\}$. 
The best expanded factor model has Min $\mbox{GRS}_j$ or Max $\mbox{SR}^2_j$, and we suppose the best factor is $\{f_6\}$ and the expanded model is $\{f_{1}, f_{3}, f_{6}\}$ in step 2.
If the asset pricing test is rejected, we add another factor.
After adding $\{f_{8}\}$ and $\{f_{7}\}$ in steps 3 and 4, the asset pricing test is not rejected, and the final expanded model M$^{F} = \{f_1,f_3, f_6,f_8,f_7\}$ is efficient.

The figure on the right illustrates the BSE process. 
In step 1, the baseline model (step 0) is M$^{F} = \{f_1, f_3, f_6, f_8, f_7\}$, which is efficient but may contain redundant factors. 
We can remove one factor, $\{ f_{j} \}$, from the RHS to the LHS and denote the reduced model as $\{\mbox{M}^{F}\backslash f_{j}\}$  for each $j = 1,3,6,7,8$.
Therefore, five candidate non-nested reduced models require model comparisons. 
We then calculate their GRS value or $\SR^2$. 
The best reduced factor model has Min $\mbox{GRS}_j$ or Max $\mbox{SR}^2_j$ in step 1, and we suppose the worst low-contribution factor is $\{ f_{6} \}$, for example.
If the asset pricing test does not reject the reduced factor model $\{f_1, f_3, f_7, f_8\}$, we then proceed to remove $\{f_1\}$ in step 2. 
In step 3, suppose the reduced factor model is $\{f_3,f_7\}$ by removing $\{ f_8 \}$.
If the asset pricing test is rejected now, we can't remove $\{ f_8 \}$, and the final reduced model M$^{F+B} = \{f_3, f_7,f_8\}$.

\section{Simulation Evidence} \label{sec:simulation}

This section aims to show the finite sample performance of our stepwise evaluation framework through Monte Carlo experiments.
Section \ref{sec:DGP} discusses the Data Generating Process used for our simulations, while Section \ref{sec:Sim_result} presents the results.

\subsection{Data Generating Process} \label{sec:DGP}
In our simulation, $\textbf{F}_t$ comprises a large set of factors that includes $K_1$ risk factors, $\textbf{f}_{1,t}$, and $K_2$ unselected factors, $\textbf{f}_{2,t}$ (so the total dimension of $\textbf{F}_t$ is $K = K_1 + K_2 $). 
To simulate returns of factors and unselected factors, we employ the following simple steps:

(1) Risk factors: Simulate $\textbf{f}_{1,t}$ from a multivariate normal distribution with means and covariance matrix calibrated with the real data, that is,  $\textbf{f}_{1,t} \sim N_{K_1}(\pmb{\mu}_1, \pmb{\Sigma}_1)$. 

(2) Unselected factors: $\textbf{f}_{2,t}$ is a linear combination of $\textbf{f}_{1,t}$, that is, $\textbf{f}_{2,t} = \pmb{\beta}^\top \textbf{f}_{1,t} + \textbf{u}_t$,
where $\textbf{u}_t \sim N_{K_2}(0, \pmb{\Sigma}_2)$.

We set \cite{fama2015five} 5-factor model as the optimal model, $K_1 = 5$. All parameters of the simulation design are calibrated using the real data and fixed in the simulation.
We examine two different baseline models.
In Case 1, we have a one-factor baseline model that starts with the first factor in $\textbf{f}_{1,t}$ to see if the optimal model can be selected.
In Case 2, we have a 2-factor baseline model that includes the first factors in both $\textbf{f}_{1,t}$ and $\textbf{f}_{2,t}$ to assess if the optimal model can be selected and if the redundant factor can be removed when considering redundant factors in the baseline model.

To assess the accuracy of our framework in model selection, recall $\mathcal{T}$ represents the optimal model,
following \cite{wang2009forward} and \cite{lan2012bayesian}, we compute for each corresponding size 
  \vspace{-0.2cm}
\begin{equation*}
|\widehat{\mathcal{S}}| = M^{-1} \sum_{k} |\widehat{\mathcal{S}}_{(k)}|,
  \vspace{-0.2cm}
\end{equation*}
where $M$ is the number of simulations, and $\widehat{\mathcal{S}}_{(k)}$  is the selection result of the $k$-th simulation.

We evaluate the coverage probability. 
  \vspace{-0.2cm}
\begin{equation*}
\mbox{CP} (\%) = 100 \times M^{-1} \sum_{k} I(\widehat{\mathcal{S}}_{(k)} \supset \mathcal{T}),
  \vspace{-0.2cm}
\end{equation*}
which measures how likely all risk factors are discovered by one particular screening method.

We further compute the correct fit. 
  \vspace{-0.2cm}
\begin{equation*}
\mbox{CF} (\%) = 100 \times M^{-1} \sum_{k} I(\widehat{\mathcal{S}}_{(k)} = {\mathcal{T}}),
  \vspace{-0.2cm}
\end{equation*}
which measures how likely it is that all risk factors are accurately identified by one screening method. 

We further define the true rate to characterize the method's underfitting effect. 
  \vspace{-0.2cm}
\begin{equation*}
\mbox{TR} (\%) = 100 \times M^{-1} \sum_{k}|\widehat{\mathcal{S}}_{(k)} \cap \mathcal{T} | / | {\mathcal{T}}|.
  \vspace{-0.2cm}
\end{equation*}

To characterize the method's overfitting effect, we define the false rate. 
  \vspace{-0.2cm}
\begin{equation*}
\mbox{FR} (\%) = 100 \times M^{-1} \sum_{k}| \widehat{\mathcal{S}}_{(k)} \cap \widetilde{\mathcal{T}} | / |\widetilde{\mathcal{T}}|.
  \vspace{-0.2cm}
\end{equation*}

\subsection{Simulation Results} \label{sec:Sim_result}
To demonstrate the advantages of our proposed approach, we compare three methods of factor selection and stopping criteria applied in both FSE and BSE processes. The methods are as follows:
\begin{enumerate}   
     \item \textbf{FSE/BSE(HDA):} Our comprehensive approach leveraging the factor model SR$^2$ for selection and the HDA stopping criterion.
    \item \textbf{FSE/BSE(GRS):} A cases that uses factor model SR$^2$ for selection and applies the GRS criterion for stopping.
    \item \textbf{FSE/BSE(SR):} A simplified approach employing single-factor SR$^2$ selection coupled with the HDA stopping criterion.
\end{enumerate}

Table \ref{tab:simulation} summarizes the factor selection results of 1000 simulations, where we consider different numbers of observations ($T$ =  3000 and 600), and different signal-to-noise ratios $K_1$ = 5 risk factors and $K_2$ = 20 or 100 unselected factors. 

We can draw the following conclusions from the results of Case 1, as outlined in Panels A-B of Table \ref{tab:simulation}.
First, the results show that for a fixed $(K_1, K_2)$-specification, the metrics CP, CF, and TR of our methods steadily increase as the sample size $T$ increases, while FR steadily decreases. 
Our findings demonstrate that the factor model SR$^2$ consistently outperforms the single-factor SR$^2$, which tends to select larger models with many incorrectly identified factors. 
In Case 1, when $K_2 = 100$ and $T = 3000$, the false rate (FR) for our BSE(HDA) is remarkably low at 0.13\%, compared to the significantly higher false rate of 18.65\% observed for BSE(SR). This discrepancy arises because the single-factor SR$^2$ does not account for correlations between factors, a limitation consistent with prior theoretical analysis.
Additionally, our findings indicate that the selection criteria HDA test outperform the GRS test,
In Case 1, when $K_2$ is 100 and $T$ is 3000, the CP value of BSE(HDA) achieves 95.39\%, 
while the CP value of BSE(GRS) is only 50.4\%.

Overall, our BSE(HDA) method's finite sample performance is consistently competitive.
The results for Case 2 in Panels C to D of Table \ref{tab:simulation} resemble those of Case 1. 
The CF value of FSE(HDA) is 0\%, due to the baseline model including a redundant factor. In contrast, the CF value of BSE(HDA) is 89.08\%, indicating our method can successfully remove redundant factors from the baseline model.
Finally, we observe that the TR value of the BSE(HDA) method consistently remains close to 100\%, while the proportion of incorrect zeros diminishes as the sample period $T$ grows. These results highlight the advantages of BSE(HDA) in variable screening. However, further research is needed to evaluate its performance under varying conditions and refine the approach.

\begin{figure}[h]
  \vspace{-0.2cm}
\caption{Simulation Results for Factor Selections \label{fig:simulation_case1_2}}
\vspace{-0.4cm}
\begin{center}
\includegraphics[width=0.9\textwidth]{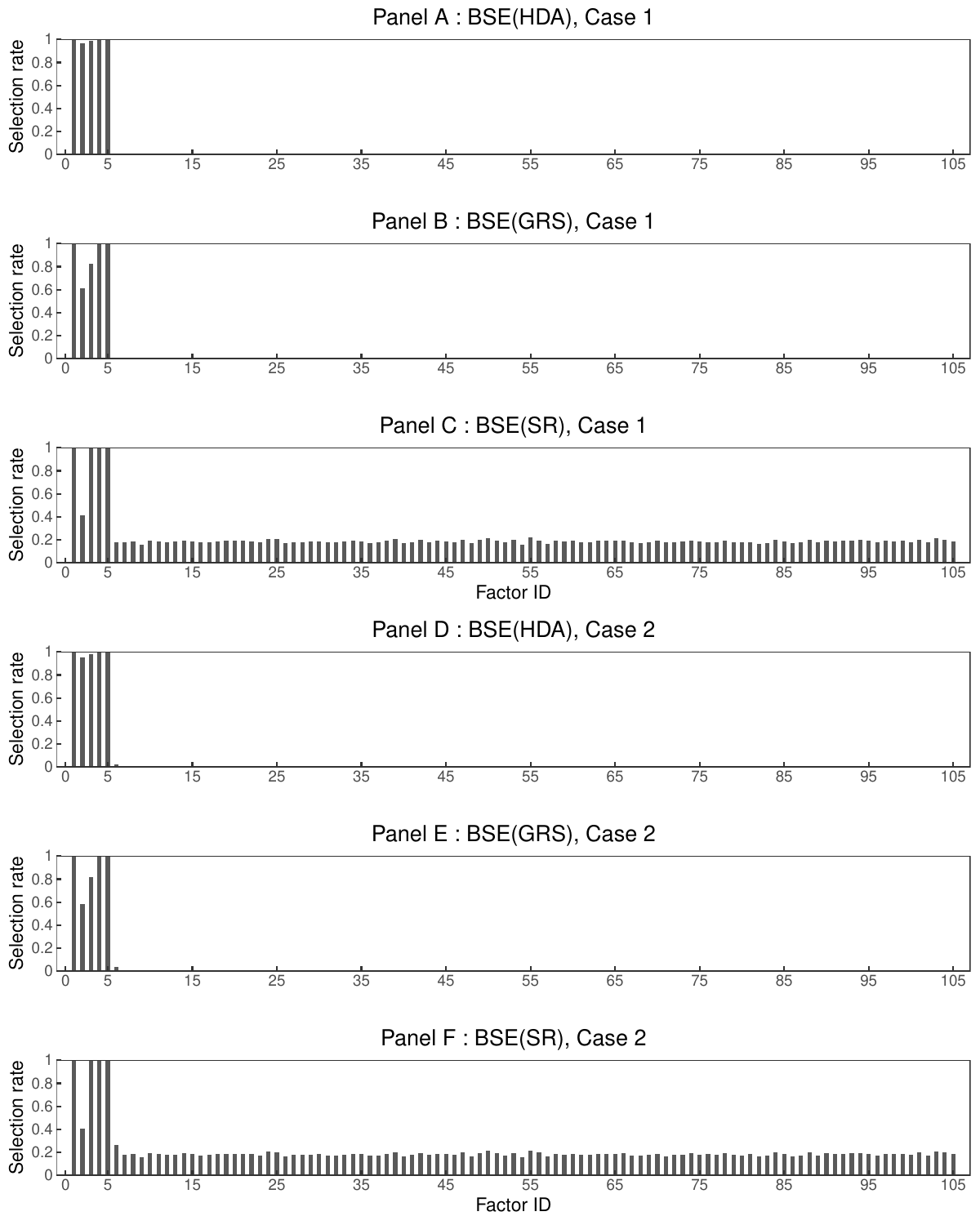}
\end{center}
  \vspace{-0.2cm}
\footnotesize{Notes: 
This figure illustrates the factor selection rates across 1,000 simulations using the BSE(HDA), BSE(GRS), and BSE(SR) methods for cases 1 and 2. The first five factors are risk factors, while the remaining 100 are unselected factors. Results are reported for $K_2 = 100$ and $T = 3000$.
}
  \vspace{-0.2cm}
\end{figure}

Figure \ref{fig:simulation_case1_2} summarizes the results of 1000 Monte Carlo simulations using BSE(HDA), BSE(GRS), and BSE(SR) methods under cases 1 and 2. Each bar in the figure represents a different simulated factor, identified by its ID from 1 to 105. We consider the case where $K_1=5$, $K_2=100$, and $T=3000$, and present the factor selection rates accordingly.
It is important to note that the first five factors are risk factors, and the remaining 100 are unselected factors. Therefore, if the final model selection could perfectly identify the correct control factors in all samples, the first five bars should have a selection rate of 100\%. In contrast, all other bars (corresponding to factors 6-105) should have a selection rate of 0\%.

In Panel A of Figure \ref{fig:simulation_case1_2}, the selection rate for the risk factors $\textbf{f}_{1,t}$ approaches 100\%, while rates for unselected factors remain near 0\%. This result highlights the advantages of the proposed BSE(HDA) in identifying the optimal model. 
Panel D further supports this conclusion, as the selection rate for the first factor of $\textbf{f}_{2,t}$ is nearly zero, demonstrating that BSE(HDA) effectively eliminates redundant factors.
In contrast, Panels B and E reveal that BSE(GRS) struggles to consistently select the risk factors $\textbf{f}_{1,t}$, with selection rates falling short of 100\%. Panels C and F indicate persistently high selection rates for other factors (6–105), underscoring the limitations of using single-factor $\SR^2$ as a criterion to remove redundancy.
We should not claim BSE (HDA) as the only effective method for variable screening. Our extensive simulation studies confirm that BSE(HDA) is a highly promising method compared to its competitors and can be beneficial in various applications.

\begin{table}[h!]
\caption{Simulation Results for Various Selection Methods}
\label{tab:simulation}
\vspace{-0.4cm}
\begin{center}
\footnotesize{
\begin{tabular}{lcccccccccccc}
\toprule
\\
& & \multicolumn{5}{c}{\underline{$(K_1, K_2)  = (5, 20)$}} & & \multicolumn{5}{c}{\underline{$(K_1, K_2)  = (5, 100)$}} \\  
\\
Method & & $|\widehat{\mathcal{S}}|$ & CP & CF & TR & FR & & $|\widehat{\mathcal{S}}|$ & CP & CF & TR & FR \\ \hline
\\
             &  & \multicolumn{11}{c}{\underline{Panel A: $T$ = 3000, Case   1} }                                              \\
             \\
FSE(HDA)                &  & 5.08         & 97.98  & 90.59 & 99.60 & 0.51  &  & 5.10         & 95.39  & 88.58  & 99.02 & 0.15  \\
FSE(GRS)                &  & 5.00         & 91.40  & 84.92 & 98.28 & 0.43  &  & 4.54         & 50.40  & 45.89  & 88.84 & 0.10  \\
FSE(SR)                 &  & 10.50        & 88.77  & 1.11  & 97.75 & 28.07 &  & 23.06        & 41.18  & 0.00   & 88.22 & 18.65 \\

BSE(HDA)                &  & 5.08         & 97.98  & 91.09 & 99.60 & 0.48  &  & 5.08         & 95.39  & 90.28  & 99.02 & 0.13  \\
BSE(GRS)                &  & 4.99         & 91.40  & 85.32 & 98.28 & 0.40  &  & 4.53         & 50.40  & 46.29  & 88.82 & 0.09  \\
BSE(SR)                 &  & 10.50        & 88.77  & 1.11  & 97.75 & 28.07 &  & 23.06        & 41.18  & 0.00   & 88.22 & 18.65 \\

\\
                        &  & \multicolumn{11}{c}{\underline{Panel B: $T$ = 600, Case   1}}                                                \\
                        \\
FSE(HDA)                &  & 3.49         & 10.22  & 8.50  & 63.60 & 1.53  &  & 3.40         & 3.71   & 3.11   & 50.82 & 0.86  \\
FSE(GRS)                &  & 3.43         & 4.96   & 3.54  & 63.00 & 1.41  &  & 2.70         & 0.90   & 0.60   & 42.14 & 0.59  \\
FSE(SR)                 &  & 6.12         & 19.84  & 0.51  & 68.62 & 13.45 &  & 10.19        & 5.01   & 0.00   & 52.38 & 7.57  \\
BSE(HDA)                &  & 3.46         & 9.82   & 8.70  & 63.46 & 1.44  &  & 3.33         & 3.61   & 3.31   & 50.38 & 0.81  \\
BSE(GRS)                &  & 3.38         & 4.45   & 3.85  & 62.67 & 1.26  &  & 2.66         & 0.90   & 0.60   & 41.76 & 0.57  \\
BSE(SR)                 &  & 6.10         & 19.43  & 0.40  & 68.32 & 13.44 &  & 9.98         & 4.71   & 0.00   & 49.86 & 7.48  \\

\\
                        &  & \multicolumn{11}{c}{\underline{Panel C: $T$ = 3000, Case   2}}                                               \\
                        \\
FSE(HDA)                &  & 6.08         & 97.06  & 0.00  & 99.39 & 5.54  &  & 6.09         & 94.09  & 0.00   & 98.76 & 1.15  \\
FSE(GRS)                &  & 5.99         & 90.08  & 0.00  & 98.02 & 5.43  &  & 5.51         & 48.10  & 0.00   & 88.28 & 1.10  \\
FSE(SR)                 &  & 11.19        & 88.26  & 0.00  & 97.63 & 31.52 &  & 23.62        & 40.98  & 0.00   & 88.18 & 19.21 \\
BSE(HDA)                &  & 5.07         & 96.86  & 90.28 & 99.35 & 0.53  &  & 5.08         & 93.89  & 89.08  & 98.70 & 0.15  \\
BSE(GRS)                &  & 5.00         & 89.57  & 83.50 & 97.91 & 0.50  &  & 4.53         & 47.29  & 43.29  & 88.12 & 0.13  \\
BSE(SR)                 &  & 10.46        & 87.75  & 1.11  & 97.53 & 27.93 &  & 22.82        & 40.48  & 0.00   & 88.08 & 18.42 \\

\\
                        &  & \multicolumn{11}{c}{\underline{Panel D: $T$ = 600, Case   2}}                                                \\
                        \\
FSE(HDA)                &  & 4.41         & 9.92   & 0.00  & 62.35 & 6.47  &  & 4.38         & 4.41   & 0.00   & 51.54 & 1.80  \\
FSE(GRS)                &  & 4.41         & 4.55   & 0.00  & 62.51 & 6.42  &  & 3.67         & 1.10   & 0.00   & 42.28 & 1.56  \\
FSE(SR)                 &  & 6.78         & 18.12  & 0.00  & 66.96 & 17.16 &  & 10.83        & 4.51   & 0.00   & 50.76 & 8.30  \\
BSE(HDA)                &  & 3.48         & 8.10   & 7.09  & 61.30 & 2.08  &  & 3.40         & 4.11   & 3.61   & 50.80 & 0.86 \\
BSE(GRS)                &  & 3.40         & 3.74   & 3.24  & 61.70 & 1.57  &  & 2.67         & 1.10   & 0.60   & 41.64 & 0.59  \\
BSE(SR)                 &  & 6.03         & 17.51  & 0.30  & 66.36 & 13.54 &  & 9.86         & 4.21   & 0.00   & 48.28 & 7.45  \\

\bottomrule
\end{tabular}}
\end{center}
 \vspace{-0.2cm}
\footnotesize{Notes: 
The table below summarizes model selection results from 1000 simulations. In Section \ref{sec:DGP}, we introduce five comparison criteria: model size ($|\widehat{\mathcal{S}}|$), coverage probability (CP), correct fit (CF), true rate (TR), and false rate (FR), to evaluate and compare different factor selection methods. Specifically, we present various FSE and BSE results for cases 1 and 2, including FSE/BSE(HDA), FSE/BSE(GRS), and FSE/BSE(SR).
}
 \vspace{-0.2cm}
\end{table}

\clearpage
\section{Proof of Theorems} \label{sec:proof_theorem}
\subsection{Notation}
We first introduce the notation used throughout.
The operators $\overset{p}{\rightarrow}$ and $\overset{d}{\rightarrow}$ denote convergence in probability and in distribution as $(N,T) \rightarrow \infty$, respectively.
For any arbitrary $n \times n$ symmetric matrix $\textbf{A}$,
we use $\lambda_{\min}(\textbf{A})$ and $\lambda_{\max}(\textbf{A})$  to denote the smallest and largest eigenvalues.
For a vector $\textbf{u} = (u_1, \cdots, u_n)^\top$, we denote $\| \textbf{u} \| = (
\sum_{i=1}^n u_i^2)^{1/2}$ as its $\mathbb{L}_2$ norm.
For any two sequences of positive numbers $\left\lbrace m_i\right\rbrace _{i=1}^{\infty}$ and $\left\lbrace n_i\right\rbrace _{i=1}^{\infty}$, we write $m_i = O(n_i)$ if there exists a constant $C_n$ such that $m_i/n_i \leq C_n$ for all $i$; we write $m_i = o(n_i)$ if $m_i/n_i \rightarrow 0$ as $i \rightarrow \infty$. 
Similarly, $m_i = O_p(n_i)$ if $m_i / n_i$ is stochastically bounded, and $m_i = o_p(n_i)$,
if $m_i / n_i \rightarrow_p 0$.
We also use the notation $a \lesssim b$ to denote $a \leq  \mathcal{K}_1 b$, and $a \gtrsim b$ to denote $a \geq   \mathcal{K}_2 b$, where $\mathcal{K}_1 > 0$ and $\mathcal{K}_2 > 0$ are finite constants holding asymptotically.
Let $\mathcal{S}$ denote a random candidate set, and define $\widetilde{\mathcal{S}} = \mathcal{S} \backslash \{j\}$ as the set obtained by removing element $j$ from $\mathcal{S}$.

\subsection{Technical Assumptions} \label{sec:assumption}
To gain theoretical insights,  we first introduce some notation.
Let $N$ be the total number of candidate factors, $f_{i,t}$ be the return of the $i$-th factor observed at time $t$ for $1 \leq i \leq N$ and $1\leq t\leq T$, $ \textbf{F}_t = (f_{1,t}, \cdots, f_{N,t})^\top$ 
with covariance matrix $\mbox{Cov}(\textbf{F}_t) =  \pmb{\Sigma} = (\sigma_{ij})_{i,j \in N}\in \mathbb{R}^{N \times N}$.
We use a generic notation $\mathcal{M} \subset \mathcal{F}$ to denote an arbitrary candidate model, where $\mathcal{F} = \{ 1, \cdots, N\} $ is the full model containing all candidate factors.
Define $\textbf{F}_{(\mathcal{M}) t} = \left\lbrace f_{i,t}: i \in \mathcal{M}\right\rbrace \in \mathbb{R}^{|\mathcal{M}|}$ and $\pmb{\Sigma}_{(\mathcal{M})} = (\sigma_{ij})_{i,j \in \mathcal{M}} \in \mathbb{R}^{|\mathcal{M}| \times |\mathcal{M}|}$ be the return vector of the factors $\mathcal{M}$ at time $t$ and its covariance matrix, respectively,
where $|\mathcal{M}|$ denotes the number of factors contained in model $\mathcal{M}$ (i.e., the model size).
Similarly, let
$\textbf{F}_{(\widetilde{\mathcal{M}}) t} = \left\lbrace f_{i,t}: i \in \widetilde{\mathcal{M}}\right\rbrace \in \mathbb{R}^{N-|\mathcal{M}|}$,
where $\widetilde{\mathcal{M}} = \{\mathcal{F} \backslash \mathcal{M} \} $ is the complement of $\mathcal{M}$.
For any candidate model $\mathcal{M}$, we consider the following regression form
 \vspace{-0.2cm}
\begin{equation}\label{eq:model_M}
\underbrace{\textbf{F}_{(\widetilde{\mathcal{M}})t}}_{\mbox{LHS}}  = \pmb{\alpha}_{(\widetilde{\mathcal{M}}) }  +  \pmb{\beta}_{(\widetilde{\mathcal{M}}) }^\top \underbrace{\textbf{F}_{(\mathcal{M})t} }_{\mbox{RHS}} + \pmb{\varepsilon}_{(\widetilde{\mathcal{M}})t},
\end{equation}
for $t = 1, \cdots, T$, where $\pmb{\alpha}_{(\widetilde{\mathcal{M}}) } = (\alpha_i, i \in \widetilde{\mathcal{M}}) \in \mathbb{R}^{|\widetilde{\mathcal{M}}|}$,  $\pmb{\beta}_{(\widetilde{\mathcal{M}}) } = (\pmb{\beta}_{(\widetilde{\mathcal{M}}) i}, i \in \widetilde{\mathcal{M}} ) \in \mathbb{R}^{|\mathcal{M}| \times |\widetilde{\mathcal{M}}|}$.
To prove the theoretical results, the following technical assumptions are needed.

%
\begin{assumption} \label{assump1}
\textbf{Rate:}

(i)  We assume that the optimal model $\mathcal{T}$ exists and is unique, and
$ |\mathcal{T}| = O(N^{n_1}) $ for some $0 < n_1 < 1/3$.

(ii) There exists a positive constant $n_2$, such that  $\min\{N, T\} \rightarrow \infty$ and $N^{(1 + n_2)} \log N = o(T)$ 
with $n_1 + 2 n_2 <1/3$.
\end{assumption}
Assumption \ref{assump1} (i) requires the optimal model to exist, and the number of risk factors tends to infinity but at a rate smaller than $N$.
This is reasonable, although there may be many candidate factors, only a small number of them capture systematic undiversifiable risk.  
Assumption \ref{assump1} (ii) allows the number of factors $N$ and the number of observations $T$ to be comparable.
This condition ensures our method remains valid in high-dimensional settings.
However, in current literature, Bayesian approaches \citep[e.g.,][]{chib2024winners, bryzgalova2022bayesian} are limited to a small number of factors.

\begin{assumption} \label{assump2}
\textbf{Factors and Covariance Matrix}: 

(i) We assume that $\textbf{F}_{(\mathcal{M}) t}$ is strictly stationary and independent of $\pmb{\varepsilon}_{(\widetilde{\mathcal{M}})t}$, and
there exist finite constants $c_{1}, c_{2}>0$, such that for each $s>0$, $\max _{1 \leq j \leq |\mathcal{M}|} P\left(\left|f_{j,t}\right|>s\right) \leq \exp \left\{-\left(s / c_{2}\right)^{c_{1}}\right\}$.

(ii) There exist finite $c_{3}, c_{4}, c_{5}>0$, such that $c_{3}^{-1}+c_{4}^{-1}>0.5$ and for any $T \in \mathcal{Z}^{+}$, the $\alpha$-mixing coefficient satisfies $\alpha(T) \leq \exp \left(-c_{5} T^{c_{4}}\right)$.

(iii) We assume that there exist two constants $0 < 2\tau_{\min} <2^{-1}\tau_{\max} < \infty$, such that $2\tau_{\min} < \lambda_{\min}(\pmb{\Sigma} ) \leq \lambda_{\max}(\pmb{\Sigma}) < 2^{-1}\tau_{\max}$.

\end{assumption}

Assumption \ref{assump2} is equivalent to Assumption 4.1 (ii-iv) of \cite{fan2015power} and posits that the candidate factors are approximately normally distributed. 
Assumption \ref{assump2} (i) primarily concerns the homoskedasticity relationship between factors and errors, a relationship routinely assumed in the literature on factor models.  
This assumption can be satisfied when $\textbf{F}_{(\mathcal{M})t} $ is multivariate normally distributed.
In Assumption \ref{assump2} (ii), the factors are assumed to be weakly correlated across $t$ and satisfy the strong mixing condition.
Note that we do allow the factors $\textbf{F}_{(\mathcal{M})t}$ to be weakly correlated across $t$, but satisfy the strong mixing condition.
Assumption \ref{assump2} (iii)  plays a role as the Sparse Riesz Condition (SRC), as defined by \cite{zhang2008sparsity} and discussed by \cite{wang2009forward}.
Assumption \ref{assump2} (iii) assumes that the covariance matrix of factor returns 
$\pmb{\Sigma}$ is a full-rank matrix, and the eigenvalues are bounded. 
A diagonal covariance matrix for latent factors introduced in \cite{lettau2020estimating} represents a special case satisfying this assumption.
Economically,  this implies that the covariance matrix controls the global correlation of factors placed on the LHS or RHS. This is equivalent to both $\pmb{\Sigma}_{(\mathcal{M})}$ and $\pmb{\Sigma}_{(\widetilde{\mathcal{M}})}$ being full-rank matrices.
Furthermore, we show in Lemma \ref{lemma1} that the sample covariance matrix is also bounded.

\begin{assumption} \label{assump3}
\textbf{Intercepts and Loadings:}

(i)
We assume that if $\mathcal{M} \not\supset \mathcal{T} $,  then $\min_{i \in \widetilde{\mathcal{M}} \cap \mathcal{T}} \alpha_{i}^2 \geq C_a N^{-n_2}$ for a positive constant $C_a$.
Conversely, if $\mathcal{M} \supset \mathcal{T}$, then 
$ \max_{i\in \widetilde{\mathcal{M}}} \alpha_{i}^2 = o( N^{-1/2} T^{-1})$.
The positive constant $n_2$ is defined in Assumption \ref{assump1}(ii).

(ii) There exists $c_{6}>0$, such that $\max _{1 \leq i \leq |\widetilde{\mathcal{M}}|}\left\|\pmb{\beta}_{(\widetilde{\mathcal{M}})i}\right\|<$ $c_{6}<\infty$.

\end{assumption}
Assumption \ref{assump3} (i) controls the magnitude of the pricing error $\alpha$.
Economically, we require that:
(i) For any factor in the optimal model $\mathcal{T}$, its $\alpha$ when regressed on the candidate model $\mathcal{M}$ should be large;
(ii) For any factor excluded from $\mathcal{T}$, its $\alpha$ when regressed on $\mathcal{M}$ should be small.
This assumption of a similar type is commonly assumed in the literature \citep[see, e.g.,][]{giglio2021thousands, feng2022high}. Assumption \ref{assump3} (ii)  measures the correlation between the LHS $\widetilde{\mathcal{M}}$ and the RHS $\mathcal{M}$.
We allow for correlation between the LHS and RHS sides, and the magnitude of the correlation is bounded.

\begin{assumption} \label{assump4}
\textbf{ Residuals:}

We assume that
$\{ \pmb{\varepsilon}_{(\widetilde{\mathcal{M}})1}, \cdots, \pmb{\varepsilon}_{(\widetilde{\mathcal{M}})T} \} $ are independent and distributed with $E\{\pmb{\varepsilon}_{(\widetilde{\mathcal{M}})t}\}=\mathbf{0}$ and $\mbox{Cov}\{\pmb{\varepsilon}_{(\widetilde{\mathcal{M}})t}\}=\pmb{\Sigma}_{\pmb{\varepsilon} }$, where $\pmb{\Sigma}_{\pmb{\varepsilon} }= (\sigma_{\pmb{\varepsilon},ij})_{i,j \in |\widetilde{\mathcal{M}}|}$ is a $|\widetilde{\mathcal{M}}| \times |\widetilde{\mathcal{M}}|$ positive definite matrix.
\end{assumption}

Assumption \ref{assump4} is equivalent to Assumption 4.1 (i) of \cite{fan2015power}. 
In Assumption \ref{assump4}, $\pmb{\varepsilon}_{(\widetilde{\mathcal{M}})t}$ is required to be serially uncorrelated across $t$. This is sensible since the time correlation is taken into account by the factors $\textbf{F}_{(\mathcal{M})t}$.

\subsection{Useful Lemmas}
The following lemmas are needed to prove Theorem \ref{theorem1}.
For convenience, we define $\widehat{\pmb{\Sigma}} = T^{-1}\sum_{t=1}^T (\textbf{F}_t - \Bar{\textbf{F}})(\textbf{F}_t - \Bar{\textbf{F}})^\top $, where $\Bar{\textbf{F}} = T^{-1} \sum_{t=1}^T \textbf{F}_t $, and write  $\widehat{\pmb{\Sigma}} = (\widehat{\sigma}_{ij})_{i,j \in N}$. In addition, define that $\widehat{\pmb{\Sigma}}_{(\mathcal{M})}$ and ${\pmb{\Sigma}}_{(\mathcal{M})}$ are the submatrices of $\widehat{\pmb{\Sigma}}$ and ${\pmb{\Sigma}}$ (corresponding to $\mathcal{M}$), respectively.

\begin{lemma} \label{lemma1}
Under Assumptions \ref{assump1}-\ref{assump4}, and $\omega = O( N^{n_1+2 n_2} )$. As $\min (T, N) \to \infty$, we have, with probability tending to one,
 \vspace{-0.2cm}
\begin{equation*}
\tau_{\min} \leq  \min_{|\mathcal{M}| \leq \omega} \lambda_{\min}\{ \widehat{\pmb{\Sigma}}_{(\mathcal{M})} \}  \leq  \min_{|\mathcal{M}| \leq \omega} \lambda_{\max} \{ \widehat{\pmb{\Sigma}}_{(\mathcal{M})} \} \leq \tau_{\max}.
 \vspace{-0.2cm}
\end{equation*}
\end{lemma} 
	
\noindent{\sc Proof:} Let $\textbf{r} = (r_1, \cdots, r_N)^\top \in \mathbb{R}^N$ be an arbitrary $N$-dimensional vector and $r_{(\mathcal{M})}$ be the subvector corresponding to $\mathcal{M}$. By Assumption \ref{assump2} (iii), we know immediately
 \vspace{-0.2cm}
\begin{equation*}
2\tau_{\min} \leq \min_{\mathcal{M} \subset \mathcal{F}} \inf_{\| r(\mathcal{M}) \|= 1 } \textbf{r}^\top_{(\mathcal{M})} {\pmb{\Sigma}}_{(\mathcal{M})} \textbf{r}_{(\mathcal{M})} \leq \max_{\mathcal{M} \subset \mathcal{F}} \sup_{\| r(\mathcal{M}) \|= 1 } \textbf{r}^\top_{(\mathcal{M})} {\pmb{\Sigma}}_{(\mathcal{M})} \textbf{r}_{(\mathcal{M})} \leq 2^{-1}\tau_{\max}.
 \vspace{-0.2cm}
\end{equation*}
Thus, the proof of Lemma \ref{lemma1} is implied by
 \vspace{-0.2cm}
\begin{equation} \label{A.1}
P\Big( \max_{|\mathcal{M}| \leq \omega} \sup_{\| r_{(\mathcal{M})}\| = 1} |r^\top_{(\mathcal{M})} \{ \widehat{\pmb{\Sigma}}_{(\mathcal{M})} - {\pmb{\Sigma}}_{(\mathcal{M})}\} r_{(\mathcal{M})}| > \eta \Big) \to 0,
 \vspace{-0.2cm}
\end{equation}
where $\eta > 0$ is an arbitrary positive number.
	
By the Bonferroni inequality, the left-hand side of Eq. (\ref{A.1}) can be bounded by
 \vspace{-0.2cm}
\begin{equation} \label{A.2}
\leq \sum_{|\mathcal{M}| \leq \omega} P\Big(  \sup_{\| r_{(\mathcal{M})}\| = 1} |r^\top_{(\mathcal{M})} \{ \widehat{\pmb{\Sigma}}_{(\mathcal{M})} - {\pmb{\Sigma}}_{(\mathcal{M})}\} r_{(\mathcal{M})}| > \eta \Big).
 \vspace{-0.2cm}
\end{equation}
Note that, for any $\mathcal{M}$ with $|\mathcal{M}|\leq \omega$, by the Cauchy-Schwarz inequality, we have
 \vspace{-0.2cm}
\begin{align*}
|r^\top_{(\mathcal{M})} \{ \widehat{\pmb{\Sigma}}_{(\mathcal{M})} - {\pmb{\Sigma}}_{(\mathcal{M})}\} r_{(\mathcal{M})}| 
& \leq \sum_{j_1, j_2 \in \mathcal{M}} |r_{j_1}| \times |r_{j_2}| \times |\widehat{\sigma}_{j_1j_2} - \sigma_{j_1j_2}| \\
& \leq \max_{1\leq j_1, j_2 \leq N} |\widehat{\sigma}_{j_1j_2} - \sigma_{j_1j_2}| \sum_{j_1, j_2 \in \mathcal{M}} |r_{j_1}| \times |r_{j_2}| \\
& = \max_{1\leq j_1, j_2 \leq N} |\widehat{\sigma}_{j_1j_2} - \sigma_{j_1j_2}| \Big( \sum_{j\in \mathcal{M}} |r_j|\Big)^2 \\
& \leq |\mathcal{M}| \max_{1\leq j_1, j_2 \leq N} |\widehat{\sigma}_{j_1j_2} - \sigma_{j_1j_2}| \\
& \leq \omega \max_{1\leq j_1, j_2 \leq N} |\widehat{\sigma}_{j_1j_2} - \sigma_{j_1j_2}|.
 \vspace{-0.2cm}
\end{align*}
	
By Assumptions \ref{assump2}-\ref{assump4}, and Lemma A3  of Theorem 1 in \cite{bickel2008covariance}, 
for ant $\eta >0$,
there exists positive constants $C^{'} > 0$ and $C^{*} > 0$,
such that
 \vspace{-0.2cm}
\[
P (|\widehat{\sigma}_{j_1j_2} - \sigma_{j_1j_2}| > \eta) \leq C^{'} \exp(-C^{*} T \eta^2 ).
 \vspace{-0.2cm}
\]
	
Employing the above result, as $T \to \infty$, the right-hand side of Eq. (\ref{A.2}) can be further bounded by
 \vspace{-0.2cm}
\begin{align*}
& \leq \sum_{|\mathcal{M}| \leq \omega} P\Big( \max_{1\leq j_1, j_2 \leq N} |\widehat{\sigma}_{j_1j_2} - \sigma_{j_1j_2}| > \frac{\eta}{\omega}\Big) \\
& \leq \sum_{|\mathcal{M}| \leq \omega} \sum_{1 \leq j_1, j_2 \leq N}  P\Big( |\widehat{\sigma}_{j_1j_2} - \sigma_{j_1j_2}| > \frac{\eta}{\omega}\Big)\\
&\leq  N^{(\omega+1)} \times N^2 \times C^{'} \exp(-C^{*} T \eta^2 \omega^{-2}) \\
& =C^{'} \exp  \Big\{ (\omega+3) \log N - C^{*} T \eta^2 \omega^{-2}  \Big\} \\
& =C^{'} \exp\Big\{ N^{(n_1 + 2 n_2)}\log N - C^{*} T \eta^2 N^{-2(n_1 + 2 n_2)} \Big\}.
 \vspace{-0.2cm}
\end{align*}

By Assumption \ref{assump1} (ii), we have $N^{3(n_1 + 2 n_2)} \log N = o(T)$ and $T N^{-2(n_1 + 2 n_2)} \to \infty$. Thus, the right-hand side of the above equation converges to 0 as $\min (T, N) \to \infty$.
This proves Eq. (\ref{A.1}) and completes the proof.
	
\begin{lemma} \label{lemma2}
Under model (\ref{eq:model_M}), Assumptions \ref{assump2}-\ref{assump4}, and Lemma \ref{lemma1}, we have
 \vspace{-0.2cm}
\begin{equation*}
\begin{split}
& \max_{1\le i,j\le |\widetilde{\mathcal{M}}|} \left| \widehat{\sigma}_{\pmb{\varepsilon},ij} - \sigma_{\pmb{\varepsilon},ij}\right| = O_p\left\lbrace\sqrt{\frac{\log(N)}{T}}\right\rbrace, \\
& \max_{i\in \widetilde{\mathcal{M}}} \left| \widehat{\alpha}_{i} - {\alpha}_{i} \right| = O_p\left\lbrace\sqrt{\frac{\log(N)}{T}}\right\rbrace,\\
& P\left\lbrace \frac{4}{9} \leq \frac{\widehat{\sigma}_{\pmb{\varepsilon},ii}}{\sigma_{\pmb{\varepsilon},ii}} \leq \frac{9}{4}, i = 1,\cdots,  |\widetilde{\mathcal{M}}| \right\rbrace \to 1.
  \end{split}
   \vspace{-0.2cm}
\end{equation*}

\end{lemma}
	
Lemma \ref{lemma2} is borrowed from Lemma E.2 and Proposition 4.1 in \cite{fan2015power}. We omit the detailed proof here to save space.

\subsection{Proof of Theorem \ref{theorem1}} \label{sec:proof_th1}

We prove the theorem in two steps.
In the first step, we prove that, with probability tending to one, FSE can detect all risk factors within $O(N^{n_1+2 n_2})$ steps.
In the second step, we show that the HDA test is to keep rejecting when $k \leq O(N^{n_1+2 n_2})$.

{\sc Step I.} To prove the desired conclusion, similar to \cite{wang2009forward},
we assume that no risk factor has been identified in the first $k$ steps. We then evaluate how likely it is that at least one risk factor will be discovered in the next step, that is, $(k + 1)$'s.
Let
 \vspace{-0.2cm}
\begin{equation}
\mbox{GRS}^{*}\{\mathcal{S}_{(k)}\}  = \frac{N - k}{T -N} \mbox{GRS}\{\mathcal{S}_{(k)}\} \mbox{~and~} \mbox{GRS}^{*}\{\mathcal{S}_{(k+1)}\}  = \frac{N -k-1}{T -N} \mbox{GRS}\{\mathcal{S}_{(k+1)}\} , \nonumber
 \vspace{-0.2cm}
\end{equation}
where $\mbox{GRS}\{\mathcal{S}_{(k)}\}$ and $\mbox{GRS}\{\mathcal{S}_{(k+1)}\}$ are the GRS values of $k$-th and $(k+1)$-th steps.

For any given subset $\mathcal{S}_{(k)}  \not\supset 
\mathcal{T}$ and $\mathcal{S}_{(k+1)} = \mathcal{S}_{(k)} \cup a_{k+1}$, it suffices to prove

 \vspace{-0.2cm}
\begin{equation} \label{eq: omega_k}
\Omega(k) = \mbox{GRS}^{*}\{\mathcal{S}_{(k)}\} - \mbox{GRS}^{*}\{\mathcal{S}_{(k+1)}\} \geq C_k {N^{ -n_2}}
 \vspace{-0.2cm}
\end{equation}
for a finite positive constant $C_k$ if $a_{k+1} \notin \mathcal{T}$ uniformly for any $ k \leq O(N^{2 n_2})$.
Accordingly, a risk factor is selected within at least every $O(N^{2 n_2})$ step.
Since $|\mathcal{T}| = O(N^{n_1})$ by Assumption \ref{assump1} (i), then all the risk factors are selected within $O(N^{n_1+2 n_2})$ steps.

We now demonstrate Eq. (\ref{eq: omega_k}). By the definition of GRS in Eq. (\ref{eq:GRS_j}) and the maximal quadratic
Sharpe ratios be bounded \citep{barillas2018comparing, avramov2023integrating},  we have
 \vspace{-0.2cm}
\begin{equation} \label{eq: omega_k1}
\begin{split}
\Omega{(k)} &= \Big( \frac{1 + \SR^2\{ \mathbf{F} \}  } {1+ \SR^2\{\mathbf{F}_{\mathcal{S}_{(k )}}\} } -1 \Big) - 	\Big( \frac{1 + \SR^2\{ \mathbf{F}\}  } {1+ \SR^2\{\mathbf{F}_{\mathcal{S}_{(k + 1)}} \} } -1 \Big) \\
& = \frac{(1 + \SR^2\{ \mathbf{F}\}) {(\SR^2\{\mathbf{F}_{\mathcal{S}_{(k + 1)}}\} - \SR^2\{\mathbf{F}_{\mathcal{S}_{(k)}}\} ) }} {(1+ \SR^2\{\mathbf{F}_{\mathcal{S}_{(k )}}\})  (1+ \SR^2\{\mathbf{F}_{\mathcal{S}_{(k + 1)}}\})} \\
		& = \frac{(1 + \SR^2 \{ \mathbf{F}\})}{(1+ \SR^2\{\mathbf{F}_{\mathcal{S}_{(k + 1)}}\})} \times \frac{ {(\widehat{\alpha}_{k+1})^2(\widehat{\sigma}_{\pmb{\varepsilon}, k+1})^{-1 } } } {(1+ \SR^2\{\mathbf{F}_{\mathcal{S}_{(k )}}\})} \\
		& \geq C_{\Omega} \times { \frac{(\widehat{\alpha}_{k+1})^2}{\widehat{\sigma}_{\pmb{\varepsilon}, k+1}}  }
\end{split}
 \vspace{-0.2cm}
\end{equation}
for a finite positive constant $C_{\Omega}$, where $\widehat{\alpha}_{k+1}$ is the estimated intercept in the spanning regression of $\mathbf{F}_{(k+1)}$ on $\mathbf{F}_{\mathcal{S}_{(k )}}$ and $\widehat \sigma_{\pmb{\varepsilon},k+1}$ is the estimated variance of the regression residuals.
$({\widehat{\alpha}_{k+1})^2 / \widehat \sigma_{\pmb{\varepsilon},k+1}}$ characterizes that the increase in the maximal $\SR^2$ when factor $k+1$ is added to the model $\mathcal{S}_{(k )}$, that is,
$({\widehat{\alpha}_{k+1})^2 / \widehat \sigma_{\pmb{\varepsilon},k+1}} = \SR^2\{\mathbf{F}_{\mathcal{S}_{(k + 1)}}\} - \SR^2\{\mathbf{F}_{\mathcal{S}_{(k)}}\} $; see, for example, \cite{barillas2017alpha} and \cite{fama2018choosing}.
	
We are assuming that $a_{k+1} \notin  \mathcal{T} $ and $\mathcal{S}_{(k )} \not\supset \mathcal{T}$.
By Assumption \ref{assump1} (ii) and Assumption \ref{assump3} (i),
we have $N^{(1 + n_2)} \log N = o(T)$ and $\min_{j\in \widetilde{\mathcal{S}}_{(k)}   \cap \mathcal{T}} ({\alpha}_{j})^2 \geq C_a N^{-n_2} $. This result,
together with Assumption \ref{assump4} and Lemma \ref{lemma2},
indicates that there exist finite positive constants $C_1$ and $C_2$, 
 \vspace{-0.2cm}
\begin{equation} \label{eq: omega_k2}
\begin{split}
 \frac{(\widehat{\alpha}_{k+1})^2}{\widehat{\sigma}_{\pmb{\varepsilon},k+1}} & \geq \max_{j\in \widetilde{\mathcal{S}}_{(k )}}\frac{(\widehat{\alpha}_{j})^2}{\widehat{\sigma}_{\pmb{\varepsilon},jj}} \\
	& \geq \max_{j\in \widetilde{\mathcal{S}}_{(k )}}\frac{ ({\alpha}_{j})^2}{2\widehat{\sigma}_{\pmb{\varepsilon},jj}} - \max_{j\in \widetilde{\mathcal{S}}_{(k )}}\frac{(\widehat{\alpha}_{j} - {\alpha}_{j})^2}{\widehat{\sigma}_{\pmb{\varepsilon},jj}} \\
	& \geq \frac{4}{9}\max_{j\in \widetilde{\mathcal{S}}_{(k )}}\frac{({\alpha}_{j})^2}{2{\sigma}_{\pmb{\varepsilon},jj}} - \frac{4}{9} \max_{j\in \widetilde{\mathcal{S}}_{(k )}}\frac{(\widehat{\alpha}_{j} - {\alpha}_{j})^2}{{\sigma}_{\pmb{\varepsilon},jj}} \\
	& \geq \frac{4}{9} \min_{j\in \widetilde{\mathcal{S}}_{(k )} \cap \mathcal{T}}\frac{({\alpha}_{j})^2}{2{\sigma}_{\pmb{\varepsilon},jj}} - \frac{4}{9} \max_{j\in \widetilde{\mathcal{S}}_{(k )}}\frac{(\widehat{\alpha}_{j} - {\alpha}_{j})^2}{{\sigma}_{\pmb{\varepsilon},jj}} \\
	& \geq  C_1 N^{ - n_2} - O \Big(\frac{\log N}{T} \Big)  \geq C_2 N^{ - n_2}.
\end{split}
 \vspace{-0.2cm}
\end{equation}
Combining Eqs. (\ref{eq: omega_k1}) and (\ref{eq: omega_k2}), with probability approaching one, we find 
$\Omega(k)  \geq C_k {N^{ -n_2}}$
uniformly for every $k \leq C_3 N^{2 n_2}$ for a finite positive constant $C_3$.

The above results
indicate that there exists a finite positive constant $C_4$ such that
 \vspace{-0.2cm}
\begin{equation*}
\begin{split}
\frac{N }{T-N } \mbox{GRS}\{\mathcal{S}_{(0)}\}  &=   \SR^2(\mathbf{F}) \\
	& > \frac{N }{T - N } \Big\{ \mbox{GRS}\{\mathcal{S}_{(0)}\} - \mbox{GRS}\{\mathcal{S}_{(1)}\} + \cdots + \mbox{GRS}\{\mathcal{S}_{(k  )}\} - \mbox{GRS}\{\mathcal{S}_{(k +1)}\} \Big\} \\
	& \geq  \mbox{GRS}^{*}\{\mathcal{S}_{(0)}\}  - \frac{N}{N-1 }  \mbox{GRS}^{*}\{\mathcal{S}_{(1)}\}
			+ \cdots \\
	& + \frac{N }{N-k }  \mbox{GRS}^{*}\{\mathcal{S}_{(k)}\}  - \frac{N}{N-k-1 }  \mbox{GRS}^{*}\{\mathcal{S}_{(k+1)}\} \\
	& \geq  \sum_{k=0}^{[C_3 N^{2 n_2}]}  \frac{N}{N-k }   \Big\{\mbox{GRS}^{*}\{\mathcal{S}_{(k)}\} - \mbox{GRS}^{*}\{\mathcal{S}_{(k+1)}\}  \Big\}  \\
	& - \sum_{k=0}^{{[C_3 N^{2 n_2}]}+1} \frac{1}{(N-k)(N-k-1) }  \mbox{GRS}^{*}(\mathcal{S}_{(k)})  \\
	& \geq 	\sum_{k=0}^{[C_3 N^{2 n_2}]} \frac{N}{N-k }   \Omega(k) - O\Big(\frac{k}{(N-k)(N-k-1) }\Big) \\
	& \geq   C_3  N^{2 n_2 } \times C_k N^{ - n_2} - o(1) \geq C_4 N^{n_2} \to \infty
 \end{split}
  \vspace{-0.2cm}
 \end{equation*}
as $N \to \infty$, where $[n]$ stands for the largest integer that is not larger than $n$. This contradicts the boundedness of $\SR^2(\mathbf{F})$, which indicating that it is impossible to have $\mathcal{S}_{(k)} \cap \mathcal{T} = \emptyset$ for every $1 \leq k \leq C_3 N^{2 n_2 }$.
Similar to the discussion in \cite{wang2009forward}, a risk factor is selected within at least every $O(N^{2 n_2})$ steps.
Since $|\mathcal{T}| = O(N^{n_1})$ in Assumption \ref{assump1} (i), then all the risk factors are selected within $O(N^{n_1+2 n_2})$ steps.

{\sc Step II.}
Define $k_{\min} = \min_{1 \leq k \leq N} \{k: \mathcal{T} \subset\mathcal{S}_{(k)} \}$. Based on the results of Step I, we know that, with probability tending to one, $k_{\min}$ is well defined and satisfies $k_{\min} = O(N^{n_1 + 2 n_2 })$ as $N \to \infty$.
Thus, the theorem conclusion follows if we can prove that the $\mbox{HDA}\{\mathcal{S}_{(k)}\} $ can be rejected when $0<k < k_{\min}$ with probability approaching one, that is $ \mathcal{S}_{(k)} \not\supset \mathcal{T}$.
To this end, it suffices to show that
 \vspace{-0.2cm}
\begin{equation} \label{eq: proof_HDA}
\inf_{0<k < k_{\min}} P \Big(  \mbox{HDA}\{\mathcal{S}_{(k)}\} \geq z_{\lambda}^1 \Big) \to 1,
 \vspace{-0.2cm}
\end{equation}
 where $z_{\lambda}^1 $ is the threshold value of the HDA test, indicating that the power of $\mbox{HDA}\{\mathcal{S}_{(k)}\}$ converges to one for the case $\mathcal{S}_{(k)}  \not\supset \mathcal{T}$.

Under Assumptions \ref{assump1}-\ref{assump4}, according to Theorem 6 in \cite{pesaran2023testing},
we have
 \vspace{-0.2cm}
\begin{equation*}
\mbox{HDA}\{\mathcal{S}_{(k)}\} \overset{d}{\longrightarrow} N(\phi^2_k/\sqrt{2},1 ),
 \vspace{-0.2cm}
\end{equation*}
where $\phi^2_k = \lim_{(N - k) \to \infty} \frac{T}{N - k} \sum_{j = 1}^{N-k} (\alpha_j)^2 / \sigma_{\pmb{\varepsilon}, jj}$.
Therefore, to prove Eq. (\ref{eq: proof_HDA}), it suffices to show $\phi^2_k  \gtrsim z_{\lambda}^1 $ when $0<k < k_{\min}$ with probability approaching one.

We are assuming that ${\mathcal{S}}_{(k)} \not\supset  \mathcal{T}$. 
By Assumption \ref{assump1} (ii) and Assumption \ref{assump3} (i),
we have $N^{(1 + n_2)} \log N = o(T)$ and $\min_{j\in \widetilde{\mathcal{S}}_{(k)}  \cap \mathcal{T}} ({\alpha}_{j})^2 \geq C_a N^{-n_2}$.
The above result, together with Assumption \ref{assump4} and Lemma \ref{lemma2}, indicating that
there exist finite positive constants $C_5$ and $C_6$, 
 \vspace{-0.2cm}
\begin{equation} \label{eq: omega_k2_1}
\begin{split}
 \sum_{ j\in \widetilde{\mathcal{S}}_{(k)} }\frac{(\widehat{\alpha}_{j})^2}{\widehat{\sigma}_{\pmb{\varepsilon},jj}} & \geq \sum_{j\in \widetilde{\mathcal{S}}_{(k)}  }  \frac{ ({\alpha}_{j})^2}{ 2 \widehat{\sigma}_{\pmb{\varepsilon},jj}} - \sum_{j\in \widetilde{\mathcal{S}}_{(k)}  }\frac{(\widehat{\alpha}_{j} - {\alpha}_{j})^2}{\widehat{\sigma}_{\pmb{\varepsilon},jj}} \\
	& \geq \frac{4}{9} \sum_{j\in \widetilde{\mathcal{S}}_{(k)}  }\frac{({\alpha}_{j})^2}{2{\sigma}_{\pmb{\varepsilon},jj}} - \frac{4}{9} \sum_{j\in \widetilde{\mathcal{S}}_{(k)} 
 }\frac{(\widehat{\alpha}_{j} - {\alpha}_{j})^2}{{\sigma}_{\pmb{\varepsilon},jj}} \\
	& \geq \frac{4}{9} \min_{j\in \widetilde{\mathcal{S}}_{(k)}  \cap \mathcal{T}}\frac{({\alpha}_{j})^2}{2{\sigma}_{\pmb{\varepsilon},jj}} - \frac{4}{9}  |\mathcal{S}_{(k)}| \max_{j\in \widetilde{\mathcal{S}}_{(k)} 
 }\frac{(\widehat{\alpha}_{j} - {\alpha}_{j})^2}{{\sigma}_{\pmb{\varepsilon},jj}} \\
	& \geq  C_5 N^{ - n_2} - O \Big(\frac{N \log N}{T} \Big)  \geq C_6 N^{ - n_2}.
\end{split}
 \vspace{-0.2cm}
\end{equation}

By Eq. (\ref{eq: omega_k2_1}) and theorem assumption that $z_{\lambda}^1  \lesssim T N^{-1/2- n_2} $, when $0<k < k_{\min}$, with probability approaching one, we have
 \vspace{-0.2cm}
\begin{equation} \label{eq: HDA_up}
\lim_{(N - k) \to \infty} \frac{T}{\sqrt{N - k}} 
\sum_{ j\in \widetilde{\mathcal{S}}_{(k)} }\frac{(\widehat{\alpha}_{j})^2}{\widehat{\sigma}_{\pmb{\varepsilon},jj}}  \gtrsim T N^{-1/2- n_2} \gtrsim z_{\lambda}^1,
 \vspace{-0.2cm}
\end{equation}
which means that $\mbox{HDA}\{\mathcal{S}_{(m)}\}$ can be rejected within $k_{\min}$ steps and needs to continue adding factors.
We have completed the entire proof by the result in Eq. (\ref{eq: HDA_up}).

\subsection{ Proof of Theorem \ref{theorem2}}  \label{sec:proof_th2}
To prove this theorem, we consider two cases: overfitting and underfitting.
Thus, the theorem conclusion follows if we can prove that the $\mbox{HDA}\{{\mathcal{S}}_{(m)}\}$  cannot be rejected when ${\mathcal{S}}_{(m)} \supset \mathcal{T} $ and the $\mbox{HDA}\{{\mathcal{S}}_{(m)} \} $ can be rejected when 
${\mathcal{S}}_{(m)} \not\supset \mathcal{T}$.
To this end, given $\sqrt{N} \log N \lesssim z_{\lambda}^2 \lesssim T N^{-1/2- n_2}$, it suffices to show that
 \vspace{-0.2cm}
\begin{equation} \label{eq: proof_HDn_2}
\inf_{{\mathcal{S}}_{(m)} \supset \mathcal{T}} P \Big( \mbox{HDA}\{{\mathcal{S}}_{(m)}\} \geq z_{\lambda}^2 \Big) \to 0 \mbox{~and~} \inf_{{\mathcal{S}}_{(m)} \not\supset \mathcal{T} } P \Big( \mbox{HDA}\{{\mathcal{S}}_{(m)}\}  \geq z_{\lambda}^2 \Big) \to 1.
 \vspace{-0.2cm}
\end{equation}
The above results indicate that the power of $\mbox{HDA}\{\mathcal{S}_{(m)}\}$ converges to zero
for the overfitting case, whereas the power of $\mbox{HDA}\{\mathcal{S}_{(m)}\}$ test converges to one for the underfitting case.

\noindent \textsc{CASE I} ({Overfitting}: $ {\mathcal{S}}_{(m)} \supset 
 \mathcal{T}$ ). We are assuming that  $ {\mathcal{S}}_{(m)} \supset \mathcal{T} $ and ${\mathcal{S}}_{(m)} \neq \mathcal{T}$ in Eq. (\ref{eq: proof_HDn_2}). By Assumption \ref{assump1} (ii), we have $\max_{j \in\widetilde{\mathcal{S}}_{(m)}} ({\alpha}_{j})^2 = o(N^{-1/2} T^{-1}) $.
The above result, together with Assumption \ref{assump4} and Lemma \ref{lemma2}, 
leads to 
 \vspace{-0.2cm}
\begin{equation} \label{eq: omega_k2_2}
\begin{split}
 \max_{j \in \widetilde{\mathcal{S}}_{(m)}} \frac{(\widehat{\alpha}_j)^2 }{\widehat{\sigma}_{\pmb{\varepsilon},jj}} 
	& \leq \max_{j \in\widetilde{\mathcal{S}}_{(m)}} 2  \frac{ ({\alpha}_{j})^2}{\widehat{\sigma}_{\pmb{\varepsilon},jj}} + 2 \max_{j \in\widetilde{\mathcal{S}}_{(m)}} \frac{(\widehat{\alpha}_{j} - {\alpha}_{j})^2}{\widehat{\sigma}_{\pmb{\varepsilon},jj}} \\
	& \leq 2\times \frac{9}{4} \max_{j \in\widetilde{\mathcal{S}}_{(m)}} \frac{({\alpha}_{j})^2}{{\sigma}_{\pmb{\varepsilon},jj}} + 2\times \frac{9}{4}  \max_{j \in\widetilde{\mathcal{S}}_{(m)}} \frac{(\widehat{\alpha}_{j} - {\alpha}_{j})^2}{{\sigma}_{\pmb{\varepsilon},jj}}  \\
 	&=  o(N^{-1/2} T^{-1})  + O \Big(\frac{\log N}{T} \Big) \\
  &= O \Big(\frac{\log N}{T} \Big). 
\end{split}
 \vspace{-0.2cm}
\end{equation}

Subsequently, using the result in Eq. (\ref{eq: omega_k2_2}), we have
 \vspace{-0.2cm}
\begin{equation} \label{eq: HDA_down_1}
\begin{split}
\lim_{(N - m) \to \infty} \frac{T}{\sqrt{N - m}} \sum_{j \in\widetilde{\mathcal{S}}_{(m)}} \frac{(\widehat{\alpha}_j)^2 }{\widehat{\sigma}_{\pmb{\varepsilon},jj}} & \leq \lim_{(N - m) \to \infty}   T \sqrt{N - m} \max_{j \in\widetilde{\mathcal{S}}_{(m)}}\frac{(\widehat{\alpha}_{j})^2}{\widehat{\sigma}_{\pmb{\varepsilon},jj}} \\
& \lesssim \sqrt{N} \log N  \lesssim z_{\lambda}^2,
\end{split}
 \vspace{-0.2cm}
\end{equation}
which means that in the case of overfitting,  with probability approaching one, $\mbox{HDA}\{\mathcal{S}_{(m)}\}$ cannot be rejected and needs to continue removing factors.

\noindent \textsc{CASE II} ({Underfitting}: $\widetilde{\mathcal{S}}_{(m)} \not\supset \mathcal{T}$). We are assuming that $\widetilde{\mathcal{S}}_{(m)} \not\supset \mathcal{T}$ in Eq. (\ref{eq: proof_HDn_2}), according to the proof of Theorem \ref{theorem1}, we have, 
 \vspace{-0.2cm}
\begin{equation} \label{eq: HDA_up_1}
\lim_{(N - m) \to \infty} \frac{T}{\sqrt{N -m}} \sum_{j \in\widetilde{\mathcal{S}}_{(m)}} \frac{(\widehat{\alpha}_j)^2 }{\widehat\sigma_{\pmb{\varepsilon},jj}} \gtrsim  T  N^{-1/2- n_2}  \gtrsim z_{\lambda}^2, 
 \vspace{-0.2cm}
\end{equation}
  which means that in the case of underfitting, with probability approaching one, $\mbox{HDA} \{\mathcal{S}_{(m)}\}$ can be rejected and needs to continue adding factors. 
Combining the results in Eqs. (\ref{eq: HDA_down_1}) and (\ref{eq: HDA_up_1}), we have completed the entire proof.

\section{Additional Empirical Results} \label{sec:add_empirical}

\subsection{Pairwise Factor Model Comparison}

Table \ref{tab:model_comparison_sr} presents the results of pairwise equality tests for $\SR^2$ values, comparing both nested and non-nested models to evaluate their relative performance.\footnote{Recent studies explore pairwise model comparisons. \cite{li2023comparing} assess CAPM, FF5, FF6, BS6, and Q5 models, finding Q5 best spans the investment opportunity set, excluding transaction costs. \cite{kan2024sample} analyze models using INS and OOS Sharpe ratios, also identifying Q5 as superior.}
We use GRS statistics in the case of nested models to test whether the factors excluded from a smaller model have zero alphas when regressed on the smaller model.
When the models are non-nested, we test the difference in $\SR^2$ for non-nested models using the statistics from \cite{barillas2020model}.
The models are arranged in a grid from left to right and top to bottom in order of increasing $\SR^2$ values.
The table displays the differences between the sample $\SR^2$ values for various pairs of models.

\begin{table}[!h]
\caption{
Pairwise Tests of Equality of Squared Sharpe Ratios}
\label{tab:model_comparison_sr}
\vspace{-0.4cm}
\begin{center}
\footnotesize{
\setlength{\tabcolsep}{0.2em}{
\begin{tabular}{lccccccccccc}
\toprule
        & FF3         & FF5          & RRA          & FF6          & FGX2020      & KNS2020      & RP-PCA       & BS6          & DHS3         & Q5           & M8           \\ \hline
CAPM    & 0.02$^{**}$ & 0.08$^{***}$ & 0.10$^{***}$ & 0.11$^{***}$ & 0.13$^{***}$ & 0.16$^{***}$ & 0.17$^{***}$ & 0.18$^{***}$ & 0.21$^{***}$ & 0.33$^{***}$ & 0.60$^{***}$ \\
FF3     &             & 0.07$^{***}$ & 0.08$^{***}$ & 0.10$^{***}$ & 0.11$^{***}$ & 0.15$^{***}$ & 0.15$^{***}$ & 0.16$^{***}$ & 0.20$^{***}$ & 0.31$^{***}$ & 0.59$^{***}$ \\
FF5     &             &              & 0.01         & 0.03$^{***}$ & 0.04         & 0.08$^{*}$   & 0.08$^{**}$  & 0.09$^{***}$ & 0.13$^{***}$ & 0.24$^{***}$ & 0.52$^{***}$ \\
RRA     &             &              &              & 0.02         & 0.03         & 0.07$^{*}$   & 0.07$^{***}$ & 0.08$^{***}$ & 0.12$^{***}$ & 0.23$^{***}$ & 0.51$^{***}$ \\
FF6     &             &              &              &              & 0.02         & 0.05         & 0.05$^{**}$  & 0.07$^{***}$ & 0.10$^{***}$ & 0.22$^{***}$ & 0.49$^{***}$ \\
FGX2020 &             &              &              &              &              & 0.04         & 0.04         & 0.05         & 0.08$^{**}$  & 0.20$^{***}$ & 0.48$^{***}$ \\
KNS2020 &             &              &              &              &              &              & 0            & 0.01         & 0.05         & 0.17$^{***}$ & 0.44$^{***}$ \\
RP-PCA  &             &              &              &              &              &              &              & 0.01         & 0.05         & 0.16$^{***}$ & 0.44$^{***}$ \\
BS6     &             &              &              &              &              &              &              &              & 0.03         & 0.15$^{***}$ & 0.43$^{***}$ \\
DHS3    &             &              &              &              &              &              &              &              &              & 0.12$^{**}$  & 0.39$^{***}$ \\
Q5      &             &              &              &              &              &              &              &              &              &              & 0.27$^{***}$ \\
\bottomrule
\end{tabular}
}}
\end{center}
\footnotesize{Notes: 
Using pairwise equality tests, we compare the squared Sharpe ratios of 7 asset pricing models, 2 PCA models, 2 Lasso-based models, and our 8-factor model.
The models are arranged in a grid from left to right and top to bottom in order of increasing $\SR^2$ values.
The table presents the differences between the sample squared Sharpe ratios for various pairs of models.
We use GRS statistics in the case of nested models to test whether the factors excluded from a smaller model have zero alphas when regressed on the smaller model.
When the models are non-nested, we test the difference in $\SR^2$ for non-nested models using the statistics from \cite{barillas2020model}.
We also report the significance of the test statistic,
$^*$, $^{**}$, and $^{***}$ indicate significance as the 10\%, 5\%, and 1\% levels, respectively.
}
\vspace{-0.2cm}
\end{table}

The main empirical findings suggest that the CAPM and FF3 models are less effective than the other models, and this difference is statistically significant.
When FF5 adds \texttt{CMA} and \texttt{RMW} factors to the FF3 model, it performs significantly the same as models RRA, FGX2020, and KNS2020.
Second, RP-PCA significantly outperforms the first five models, performs equally to models FGX2020, KNS2020, BS6, and DHS3, and is outperformed by models Q5 and our M8.
Third, we find that the Q5 model outperforms many other models. Specifically, the Q5 model significantly outperforms the other models at the 1\% level, except for the M8, which outperforms the DHS3 with a 5\% significance level.
Finally, all the results confirm that our M8 model outperforms other models and is statistically significant at the 1\% level. The magnitude increases of $\SR^2$ values, ranging from 0.27 for Q5 to 0.60 for CAPM.
We conclude that the M8 model best spans the investment opportunity set.

\subsection{Stepwise Evaluation for Test Assets}
To further assess the pricing power of candidate factors, we perform a robustness check by expanding the set of basis portfolio test assets. 
Given the high dimensionality of test assets, we mitigate complexity by constructing 20 RP-PCA factors \citep{lettau2020estimating}, which parsimoniously represent the test assets and approximate the joint efficient frontier. Specifically, we analyze two sets of underlying assets: the top 20 RP-PCA factors derived from 285 bivariate-sorted portfolios (Top 20 RP-PCA) and their combination with 97 additional factors (Factor97 + Top 20 RP-PCA).

\begin{table}[!h]
\caption{
Stepwise Evaluation for Test Assets
}
\label{tab:test_asset_SE}
\vspace{-0.5cm}
\begin{center}
\footnotesize{
\setlength{\tabcolsep}{1em}{
\begin{tabular}{cc}
\toprule
Top 20 RP-PCA & Factor97+Top 20 RP-PCA \\ \hline
RP-PCA3       & REG                    \\
RP-PCA12      & RP-PCA1                \\
RP-PCA8       & RP-PCA6                \\
RP-PCA6       & PEAD                   \\
RP-PCA7       & RP-PCA7                \\
RP-PCA1       & HMLM                   \\
RP-PCA4       & ILR                    \\
RP-PCA20      & RP-PCA8  \\                        
\bottomrule
\end{tabular}
}}
\end{center}
\vspace{-0.3cm}
\end{table}

Table \ref{tab:test_asset_SE} presents the reduced FSE/BSE models under test asset choices. 
Key findings include:
(1) Substituting RP-PCA factors for the 285 test assets reveals that the selection order of the Top 20 RP-PCA factors deviates from traditional eigenvalue rankings, as our criterion prioritizes Sharpe ratios. Moreover, using only 20 RP-PCA factors necessitates selecting more factors for convergence due to lower correlations among PCA components.
(2)
When considering model selection starting from 97 observable factors and 20 RP-PCA factors, we find that an eight-factor model was ultimately selected. Among these, four factors are from the observable factors, and the other four are from the RP-PCA factors.

\subsection{Stepwise Evaluation for Different Sentiment Levels}

We examine how factor selection varies across market sentiment regimes, contrasting high- and low-sentiment periods. Following \cite{huang2015investor}, a month is classified as high sentiment if the sentiment level ($S^{PLS}$) in the prior month exceeds its median value for the sample period; otherwise, it is classified as low sentiment.

\begin{table}[!h]
\caption{Reduced Factor Models for Different Sentiment Levels}
\label{tab:I1}
\vspace{-0.4cm}
\begin{center}
\footnotesize{\setlength{\tabcolsep}{1em}{
\begin{tabular}{ccccccc}
\toprule							
CAPM$^{F+B}$ & FF3$^{F+B}$ & FF5$^{F+B}$ & FF6$^{F+B}$ & Q5$^{F+B}$ & DHS3$^{F+B}$ & BS6$^{F+B}$ \\ \hline
\\
\multicolumn{7}{c}{\underline{Panel A: High Sentiment Periods}}                                \\
\\
MKT          & MKT         & MKT         & MKT         & MKT        & MKT          & MKT         \\
REG          & SMB         & SMB         & SMB         & SMB        & PEAD         & SMB         \\
\underline{$S^{PLS}$}           & HML         & HML         & HML         & REG        & REG          & UMD         \\
STR          & REG         & REG         & UMD         & \underline{$S^{PLS}$}         & HMLM         & HMLM        \\
SIM          & \underline{$S^{PLS}$}          & \underline{$S^{PLS}$}          & REG         & STR        & STR          & REG         \\
             &             & HMLM        & \underline{$S^{PLS}$}          & ILR        & ILR          & EPRD        \\
             &             & EPRD        & HMLM        &            & SMB          & HML         \\
             &             & UMD         & EPRD        &            & EPRD         & \underline{$S^{PLS}$}          \\           
\\
\multicolumn{7}{c}{\underline{Panel B: Low Sentiment Periods}}                                        \\
\\
MKT          & MKT         & MKT         & MKT         & MKT        & MKT          & MKT         \\
REG          & SMB         & SMB         & SMB         & IA         & PEAD         & SMB         \\
PEAD         & HML         & CMA         & CMA         & REG        & REG          & ROE         \\
STR          & REG         & REG         & REG         & PEAD       & STR          & HMLM        \\
OCP          & PEAD        & PEAD        & PEAD        & STR        & OCP          & REG         \\
             &             &             &             & CIM        &              & PEAD        \\
\bottomrule
\end{tabular}
}}
\end{center}

\footnotesize{Notes: 
The table presents reduced models based on high- and low-sentiment periods across seven baseline models. 
Panel A provides a detailed analysis of the results during high-sentiment periods, while Panel B examines the outcomes observed during low-sentiment periods. 
The sentiment factor is marked with underlining.
}
  \vspace{-0.2cm}
  
\end{table}

Table \ref{tab:I1} presents the FSE/BSE reduced models across sentiment regimes. Two key findings emerge: 
First, high-sentiment periods consistently feature more factors across all specifications, reflecting greater model complexity. For instance, the FF6$^{F+B}$ model selects eight factors during high sentiment versus five in low sentiment. 
Second, the factor $S^{PLS}$ demonstrates sensitivity to sentiment, appearing in six models during high-sentiment periods but in none during low-sentiment periods.
This corroborates \cite{huang2015investor}, who show $S^{PLS}$'s predictive power is concentrated in high-sentiment periods.
These results underscore the crucial role of market sentiment in factor selection, affecting both model complexity and factor relevance.

\vspace{1cm}
\onehalfspacing{\small{\putbib}}
\end{bibunit}

\end{document}